\documentclass[onecolumn,authoryear]{els-mrw} 

\usepackage{amsmath,amssymb,amsfonts,amsthm,makeidx,graphicx}
\usepackage{txfonts}
\usepackage{helvet}

\begin{document}

\chapter{Global analyses of helicity-dependent parton distribution functions}
\label{chap1}

\author[1,2]{Emanuele R. Nocera}

\address[1]{\orgname{Universit\`a degli Studi di Torino}, \orgdiv{Dipartimento di Fisica}, \orgaddress{Via Pietro Giuria 1, 10125 Torino, Italy}}
\address[2]{\orgname{INFN}, \orgdiv{Sezione di Torino}, \orgaddress{Via Pietro Giuria 1, 10125 Torino, Italy}}

%\articletag{First version}

\maketitle

\begin{glossary}[Glossary]

\term{Axial anomaly}
The quantum-mechanical breaking of the classical conservation law for the
flavour-singlet axial current. In polarised deep-inelastic scattering, it
affects the separation of the singlet contribution into quark and gluon
helicity components.

\term{Axial charge}
A quantity defined by the forward proton matrix element of an axial current.
The flavour-singlet axial charge $a_0$ is related to the first moment of the
quark-singlet helicity distribution, with a relation that depends on the
factorisation scheme.

\term{Coefficient function}
A perturbatively calculable function describing the short-distance interaction
between the external probe and the partons. In a factorisation formula,
coefficient functions are convoluted with PDFs and, where appropriate,
fragmentation functions.

\term{Collinear factorisation}
The leading-power separation of a hard-scattering observable into
process-dependent short-distance coefficients and universal long-distance
parton distributions or fragmentation functions.

\term{Factorisation scheme}
A convention specifying how finite contributions are distributed between
perturbative coefficient functions and nonperturbative PDFs or fragmentation
functions. Physical observables are scheme independent, whereas PDFs and
their moments are not.

\term{Fragmentation function}
A nonperturbative function $D_i^h(z,\mu^2)$ describing the probability density
for a parton $i$ to produce an observed hadron $h$ carrying a fraction $z$ of
the parton momentum.

\term{Global QCD analysis}
A simultaneous determination of PDFs from a broad collection of experimental
measurements, using QCD evolution and factorisation to relate all observables
to a common set of parton distributions.

\term{Helicity-dependent parton distribution}
The difference between the densities of partons whose helicity is aligned and
anti-aligned with the helicity of the parent nucleon. Helicity PDFs are also
called polarised PDFs.

\term{Longitudinal spin asymmetry}
A ratio of polarised to unpolarised cross sections constructed from scattering
rates with different longitudinal helicity configurations of the incoming
particles.

\term{Parton distribution function}
A universal nonperturbative function $f_i(x,\mu^2)$ describing the density of
partons of species $i$ carrying a fraction $x$ of the longitudinal momentum
of a fast-moving hadron.

\term{Parton evolution}
The dependence of PDFs or fragmentation functions on the factorisation scale,
governed in perturbative QCD by the DGLAP evolution equations.

\end{glossary}

\begin{glossary}[Nomenclature]
\begin{tabular}{@{}lp{34pc}@{}}
CC       & Charged current \\
DIS      & Deep-inelastic scattering \\
DGLAP    & Dokshitzer--Gribov--Lipatov--Altarelli--Parisi \\
EIC      & Electron--Ion Collider \\
FF       & Fragmentation function \\
FFN      & Fixed-flavour-number scheme \\
GM-VFN   & General-mass variable-flavour-number scheme \\
IR       & Infrared \\
LO       & Leading order \\
MAP      & Maximum a posteriori \\
MHOU     & Missing-higher-order uncertainty \\
NC       & Neutral current \\
NLO      & Next-to-leading order \\
NNLO     & Next-to-next-to-leading order \\
N$^3$LO  & Next-to-next-to-next-to-leading order \\
OPE      & Operator-product expansion \\
PDF      & Parton distribution function \\
QCD      & Quantum chromodynamics \\
QED      & Quantum electrodynamics \\
RHIC     & Relativistic Heavy Ion Collider \\
SIDIS    & Semi-inclusive deep-inelastic scattering \\
UV       & Ultraviolet \\
VFN      & Variable-flavour-number scheme \\
ZM-VFN   & Zero-mass variable-flavour-number scheme \\   
\end{tabular}
\end{glossary}

\begin{abstract}[Abstract]
  Helicity-dependent, or polarised, parton distribution functions describe the
  longitudinal polarisation of quarks, antiquarks, and gluons inside a
  polarised nucleon. They encode how the spin of a fast-moving proton or
  neutron is shared among its partonic constituents and are therefore central
  to the modern understanding of nucleon spin structure in Quantum
  Chromodynamics. Helicity PDFs are extracted from global QCD analyses of
  polarised hard-scattering data, including inclusive deep-inelastic scattering,
  semi-inclusive deep-inelastic scattering, and polarised proton--proton
  collisions. This article reviews the definition and physical interpretation
  of polarised PDFs, the perturbative QCD framework used to determine
  them, the experimental observables that constrain the different parton
  flavours, and the statistical methodology of global analyses. The state of
  the art in polarised PDF determination is presented, together with the impact
  expected from future measurements at the Electron--Ion Collider.
\end{abstract}

\section{Introduction}
\label{sec:introduction}

A century after its introduction in quantum
mechanics~\citep{Uhlenbeck:1925pqz,Goudschmidt:1926ea}, the concept of spin
remains a compelling manifestation of the fundamental theoretical principles
that govern Nature. The subject of this review article is the proton. Like the
electron, it is a particle of spin one half; unlike the electron, however, it
is not elementary: it is a relativistic bound state of quarks and gluons,
continuously interacting, radiating and recombining according to the laws of
quantum chromodynamics (QCD), the field theory that describes the strong
interaction. The apparently simple statement that the proton has spin $1/2$
therefore conceals a much deeper question: how does this angular momentum
emerge from the spin and orbital motion of its constituents?

This question has accompanied the development of our understanding of proton
structure over the past century. That the proton has spin $1/2$ and that it has
an internal structure are facts that emerged in the early days of quantum
mechanics. The former property was inferred from the rotational statistics and
specific heat of molecular hydrogen~\citep{Dennison:1927}; the latter
was deduced from the measurement of the anomalous magnetic
moment~\citep{Frisch:1933a,Estermann:1933a,Estermann:1933b}, which suggested
that the proton did not behave like a pointlike Dirac particle. A change in
perspective came only in the early 1960s, when,
from a theoretical point of view, the quark model~\citep{Gell-Mann:1962yej,
  Gell-Mann:1964ewy,Zweig:1964ruk,Zweig:1964jf} was introduced to bring order
into the large array of strongly interacting particles observed in experiment,
by recognising that the known hadrons, including the proton, could be
associated with representations of the special unitary ${\rm SU}(3)$ group.
This led to the concept of quarks as the building blocks of hadrons. From an
experimental point of view, intense investigations of the proton structure
followed at SLAC in the late 1960s, when the first deep-inelastic scattering
(DIS) events were recorded~\citep{Friedman:1972sy}. In these experiments, the
proton was probed through high-energy lepton scattering, revealing that it
contained quasi-free, point-like constituents. In particular, the
unexpectedly weak dependence of the measured cross section on the
momentum-transfer scale --- the so-called Bjorken
scaling~\citep{Bjorken:1968dy} --- found a natural interpretation in
Feynman's parton model~\citep{Feynman:1969ej}. In this picture, a fast-moving
proton is described as a collection of elementary constituents carrying
definite fractions $x$ of its longitudinal momentum. The cross section is then
an incoherent sum of partonic cross sections, each weighted by the probability
that a parton $f$ carries the momentum fraction $x$. The corresponding
probability density is the parton distribution function (PDF). Early evidence
for gluons came from the observation that quarks accounted for
only about $50\%$ of the proton momentum~\citep{deGroot:1979ugm,
  deGroot:1978feq}. The subsequent formulation
of QCD~\citep{Gross:1973ju,Gross:1973zrg,Weinberg:1973un}, and the discovery of
asymptotic freedom~\citep{Gross:1973id}, identified these partons
with quarks and gluons and explained why they behave approximately as free
particles when probed at short distances, while predicting logarithmic
violations of exact scaling.

Once the partonic structure of the proton had been established, it became
natural to investigate its spin dependence. The SLAC E80 experiment reported
in 1976 the first measurement of the DIS asymmetry in the scattering of
longitudinally polarised electrons from longitudinally polarised
protons~\citep{Alguard:1976bm}. The later E130
experiment~\citep{Baum:1983ha} extended these measurements to higher energies
and momentum transfers and improved their precision. These pioneering
experiments established polarised DIS as a probe of the contribution of quark
helicities to the proton spin. The measured asymmetries were positive and
sizeable and, within their relatively limited precision and kinematic
coverage, were broadly consistent with contemporary quark--parton models.
The quantity relevant to this interpretation is the flavour-singlet axial
charge $a_0$, defined by the forward proton matrix element of the 
flavour-singlet axial current,
\begin{equation}
  \langle P,s | J_5^{\mu}(\mu^2) |P,s\rangle
  =
  2M s^\mu a_0(\mu^2)\,,
  \qquad
  J_{5}^{\mu}
  =
  \sum_{q=1}^{n_f}\bar q\gamma^\mu\gamma_5 q\,,
  \label{eq:a0}
\end{equation}
where $P$ and $M$ are the proton momentum and mass, and $s^\mu$ is its covariant
spin vector, normalised such that $s^2=-1$ and $P\cdot s=0$. In the naive
quark-parton model, $a_0$ is identified with the total quark and antiquark
helicity and would equal unity if the proton spin were entirely carried
by the helicities of its quark constituents.

Unexpectedly, in 1989 the significantly more precise EMC experiment at
CERN~\citep{EuropeanMuon:1989yki} measured the first moment of the proton
structure function $g_1$ and inferred a value of $a_0$ that was compatible
with zero within uncertainties and significantly smaller than the naive
expectation of unity. This evidence gave rise to the so-called proton-spin
crisis~\citep{Leader:1988vd}. It led to the realisation that, first,
relativistic effects reduce the quark-model expectation for the total quark
helicity to a value closer to $0.6$~\citep{Jaffe:1989jz}, and, second, that
in QCD the interpretation of $a_0$ in terms of separate quark and gluon
contributions is affected by the axial anomaly~\citep{Altarelli:1988nr}.
In particular, the separation of the singlet contribution into quark and
gluon helicities depends on the factorisation scheme. In the commonly used
$\overline{\mathrm{MS}}$ scheme, the singlet axial charge is identified with
the first moment of the renormalised singlet quark distribution. In an
Adler--Bardeen-type scheme, instead, the anomalous gluon contribution is
displayed explicitly:
\begin{equation}
  \begin{aligned}
    a_0(\mu^2)
    &=
    \Delta\Sigma_{\overline{\mathrm{MS}}}(\mu^2)\,,
    && \overline{\mathrm{MS}}\ \text{scheme}\,,
    \\
    a_0(\mu^2)
    &=
    \Delta\Sigma_{\mathrm{AB}}
    -
    n_f\frac{\alpha_s(\mu^2)}{2\pi}
    \Delta G_{\mathrm{AB}}(\mu^2)\,,
    && \text{Adler--Bardeen-type scheme}\,.
  \end{aligned}
  \label{eq:a0_QCD}
\end{equation}
In the latter convention, $\Delta\Sigma_{\mathrm{AB}}$ is chosen to be
independent of the scale. The physical first moment of $g_1$ is independent
of the factorisation scheme, although the attribution of its singlet
contribution to quark and gluon helicities is not.
The quark-singlet and gluon helicity moments in Eq.~\eqref{eq:a0_QCD} are
defined as
\begin{equation}
  \Delta\Sigma(\mu^2)
  =
  \sum_q\int_0^1 dx\,
  \left[
    \Delta f_q(x,\mu^2)+\Delta f_{\bar q}(x,\mu^2)
  \right]\,,
  \qquad
  \Delta G(\mu^2)
  =
  \int_0^1 dx\,\Delta f_g(x,\mu^2)\,,
  \label{eq:moments}
\end{equation}
where factorisation-scheme labels have been suppressed. The functions
$\Delta f_q$, $\Delta f_{\bar q}$ , and $\Delta f_g$ are the polarised PDFs.
In the parton-model interpretation, they are the differences between the
number densities of partons $f$ carrying a fraction $x$ of the longitudinal
proton momentum with helicity aligned ($\uparrow$) or anti-aligned
($\downarrow$) with that of the parent proton:
\begin{equation}
  \Delta f_i(x,\mu^2)
  =
  f_i^\uparrow(x,\mu^2)-f_i^\downarrow(x,\mu^2)\,,
  \qquad
  f_i=f_q,f_{\bar q},f_g\,.
  \label{eq:pol_pdfs}
\end{equation}
In QCD, polarised PDFs depend on the factorisation scale $\mu^2$ through QCD
evolution. In addition, the renormalised flavour-singlet axial current has a
nonzero anomalous dimension induced by the axial anomaly, which starts at
two-loop order. Consequently,
$\Delta\Sigma_{\overline{\mathrm{MS}}}(\mu^2)$ evolves with the scale,
whereas in an Adler--Bardeen-type scheme the corresponding scale dependence
is carried by the explicit gluonic term in Eq.~\eqref{eq:a0_QCD}. At leading
logarithmic accuracy, $\Delta G(\mu^2)$ grows asymptotically as
$1/\alpha_s(\mu^2)$, so that the product
$\alpha_s(\mu^2)\Delta G(\mu^2)$ need not vanish at large scales. This
observation historically motivated the possibility that a large gluon
helicity contribution could reconcile a small value of $a_0$ with a larger
scale-independent quark helicity contribution. 

The proton-spin decomposition is, however, more involved than the separation
of quark and gluon helicities alone~\citep{Ji:2020ena}. Quarks and gluons may
also carry orbital angular momentum. In the canonical Jaffe--Manohar
decomposition~\citep{Jaffe:1989jz}, the proton angular momentum is written as
\begin{equation}
  \frac{1}{2}
  =
  \frac{1}{2}\Delta\Sigma(\mu^2)
  +
  \Delta G(\mu^2)
  +
  \mathcal{L}^{\mathrm{JM}}_q(\mu^2)
  +
  \mathcal{L}^{\mathrm{JM}}_g(\mu^2)\,.
  \label{eq:Jaffe_and_Manohar}
\end{equation}
This decomposition is not unique. Alternative decompositions organise the
spin and orbital contributions differently, and the precise operator
definitions, gauge properties, and experimental accessibility of the
individual terms are theoretically delicate
\citep{Leader:2013jra,Ji:2020ena}. Within a specified decomposition and
factorisation scheme, knowledge of $\Delta\Sigma(\mu^2)$ and
$\Delta G(\mu^2)$ constrains the total orbital contribution,
\begin{equation}
  \mathcal{L}^{\mathrm{JM}}_q(\mu^2)
  +
  \mathcal{L}^{\mathrm{JM}}_g(\mu^2)
  =
  \frac{1}{2}
  -
  \frac{1}{2}\Delta\Sigma(\mu^2)
  -
  \Delta G(\mu^2)\,,
  \label{eq:total_orbital}
\end{equation}
but does not by itself permit a separate determination of the quark and gluon
orbital angular momenta. An accurate determination of $\Delta\Sigma(\mu^2)$ and
$\Delta G(\mu^2)$ may therefore set a bound on the sum
$\mathcal{L}^{\mathrm{JM}}_q(\mu^2)+\mathcal{L}^{\mathrm{JM}}_g(\mu^2)$.

Since the pioneering results of the EMC experiment, the spin structure of
the proton, and in particular the determination of each term in
Eq.~\eqref{eq:Jaffe_and_Manohar}, has been the subject of intense theoretical
and experimental investigations, see~\citep{Aidala:2012mv} for a review.
The Electron--Ion Collider (EIC)~\citep{AbdulKhalek:2021gbh} will be the next
major stage of this programme.
The goal of this article is to review the determination of polarised PDFs and
the current knowledge of their moments $\Delta\Sigma(\mu^2)$ and
$\Delta G(\mu^2)$. In Sect.~\ref{sec:theoretical_input}, longitudinal spin
asymmetries are presented from a theoretical point of view. The processes of
interest in which these are measured are described, and predictions based on
QCD factorisation are discussed together with the status of perturbative
computations and general theoretical constraints.
In Sect.~\ref{sec:experimental_input}, current and future experimental
measurements of longitudinal spin asymmetries are reviewed, highlighting their
constraining power on different PDFs in different kinematic regions. In
Sect.~\ref{sec:methodological_input}, the problem of determining polarised PDFs
from experimental data is presented from a methodological point of view,
using a Bayesian framework. Specifically, the focus is on several aspects of
the problem, which include parametrisation, optimisation, uncertainty
representation, and uncertainty validation. In Sect.~\ref{sec:state_of_the_art},
the current state of the art in polarised PDF determination is assessed by
comparing the most recent sets of polarised PDFs. General conclusions
concerning $\Delta\Sigma(\mu^2)$ and $\Delta G(\mu^2)$ are drawn. Finally,
Sect.~\ref{sec:conclusions} presents the summary and outlook.

\section{Theoretical input}
\label{sec:theoretical_input}

Longitudinal spin asymmetries are the physical observables currently used
to determine polarised PDFs. They are defined as the ratio of polarised to
unpolarised hadronic cross sections, where the former are constructed
analogously to the latter but retain the dependence on the polarisation of the
initial states, be they leptonic or hadronic. In this respect, longitudinal
spin asymmetries are categorised as single or double, depending on whether one
or both initial scattering particles are polarised. This section presents
longitudinal spin asymmetries from a theoretical point of view. First, the
processes that are currently measured in experiments are described. Second,
the corresponding factorisation formul\ae{} are introduced. Third, the status
of the perturbative computations entering them is reviewed. Fourth,
general theoretical constraints on them are discussed.

\subsection{Spin asymmetries}
\label{subsec:spin_asymmetries}

Single- and double-longitudinal spin asymmetries have been measured in the
following processes.
\begin{description}
\item[Deep-inelastic scattering (DIS)]: $\ell N \to \ell^\prime X$.
  In this process, a beam of leptons $\ell$ and a nucleon $N$ interact
  inclusively, {\it i.e.}~only the four-momentum of the outgoing lepton
  $\ell^\prime$ is measured. The longitudinal double-spin asymmetry is defined as
  \begin{equation}
    A_{LL}^i=\frac{d^2\Delta\sigma^i}{d^2\sigma^i}\,,
    \label{eq:A_LL_DIS}
  \end{equation}
  where $d^2\Delta\sigma^i$ and $d^2\sigma^i$ are the polarised and unpolarised
  DIS cross sections, respectively.  The index $i$ denotes whether the
  process is neutral-current (NC) or charged-current (CC), that is, whether
  the electroweak interaction proceeds through the exchange of an electrically
  neutral photon and a $Z$ boson (NC) or through the exchange of electrically
  charged $W$ bosons (CC). Using charge and momentum conservation and
  interaction symmetries, one can parametrise them in terms of polarised and
  unpolarised DIS structure functions~\citep{Anselmino:1993tc}
  \begin{align}
    \frac{d^2\Delta\sigma^i}{dxdy}
    & = \frac{4\pi\alpha^2}{xyQ^2}
    \eta^i
    \left[\pm Y_- xg_1^i + Y_+ x g_5^i + (1-y) g_L^i \right]\,,
    \label{eq:DIS_SF_pol}\\
    \frac{d^2\sigma^i}{dxdy}
    & = \frac{2\pi\alpha^2}{xyQ^2}
    \eta^i
    \left[Y_+ 2x F_1^i \mp Y_- x F_3^i + (1-y) 2 F_L^i \right]\,,
    \label{eq:DIS_SF_unp}
  \end{align}
  where the kinematic variables are the nucleon momentum fraction
  $x=Q^2/(2q\cdot P)$, the inelasticity $y=(q\cdot P)/(k\cdot P)$, and the
  exchanged-boson virtuality $Q^2=-q^2$, with $k$, $P$, and $q$ the lepton,
  nucleon, and transferred four-momenta, respectively. The polarised cross
  section is defined as the difference $\Delta\sigma^i=\sigma^i(\lambda_N=-1,\lambda_\ell)-\sigma^i(\lambda_N=+1,\lambda_\ell)$, where $\lambda_N$
  ($\lambda_\ell$) is the nucleon (lepton)
  helicity, which can take the values $\pm 1$ according to its orientation
  parallel ($+$) or antiparallel ($-$) to the beam direction. The electroweak
  factors $\eta^i$ are, for NC and CC processes, $\eta^{\rm NC}=1$ and
  $\eta^{\rm CC}=(1\pm \lambda_\ell)^2$.
  In Eqs.~\eqref{eq:DIS_SF_pol}--\eqref{eq:DIS_SF_unp}, $\alpha$ is the
  Quantum Electrodynamics (QED) coupling, $Y_\pm=1\pm (1-y)^2$ are kinematic
  factors and the $\mp$ and $\pm$ signs, respectively, refer to the charge
  of the incoming lepton. Note that corrections
  proportional to $M^2/Q^2$, where $M$ is the nucleon mass, have been neglected
  in Eqs.~\eqref{eq:DIS_SF_pol}--\eqref{eq:DIS_SF_unp} because in typical
  experiments, $Q^2\gg M^2$. The NC structure functions, which follow from
  $\gamma$ and $Z$-boson exchange, are
  \begin{align}
    g_{1}^{\rm NC} & = \eta_\gamma g_{1}^\gamma
    - (g_V^e\pm\lambda_\ell g_A^e)\eta_{\gamma Z}g_{1}^{\gamma Z}
    + ((g_V^e)^2+(g_A^e)^2\pm2\lambda_\ell g_V^e g_A^e)\eta_Z g_{1}^Z\,,\\
    g_{5,L}^{\rm NC} & = -(g_A^e\pm\lambda_\ell g_V^e)\eta_{\gamma Z}g_{5,L}^{\gamma Z}
    + [2g_V^e g_A^e\pm\lambda_\ell((g_V^e)^2+(g_A^e)^2)]\eta_Z g_{5,L}^Z\,,\\
    F_{1,L}^{\rm NC} & = \eta_\gamma F_{1,L}^\gamma
    - (g_V^e\pm\lambda_\ell g_A^e)\eta_{\gamma Z}F_{1,L}^{\gamma Z}
    + ((g_V^e)^2+(g_A^e)^2\pm2\lambda_\ell g_V^e g_A^e)\eta_Z F_{1,L}^Z\,,\\
    F_3^{\rm NC} & = -(g_A^e\pm\lambda_\ell g_V^e)\eta_{\gamma Z}F_3^{\gamma Z}
    + [2g_V^e g_A^e\pm\lambda_\ell((g_V^e)^2+(g_A^e)^2)]\eta_ZF_3^Z\,,
    \label{eq:NC_DIS}
  \end{align} 
  where $g_V^e=-1/2+2\sin^2\theta_W$ and $g_A^e=-1/2$, with $\theta_W$ the
  Weinberg angle. The CC structure functions, which follow
  from $W$-boson exchange only, are
  \begin{equation}
    g_{1,5,L}^{\rm CC}=\eta_W g_{1,5,L}^{W}\,,
    \qquad\qquad
    F_{1,3,L}^{\rm CC}=\eta_W F_{1,3,L}^{W}\,.
    \label{eq:CC_DIS}
  \end{equation}
  The NC and CC couplings are
  \begin{equation}
    \eta_\gamma=1
    \qquad
    \eta_{\gamma Z}
    =
    \left(\frac{G_FM_Z^2}{2\sqrt{2}\pi\alpha}\frac{Q^2}{Q^2+M_Z^2}\right)
    \qquad
    \eta_Z=\eta_{\gamma Z}^2
    \qquad
    \eta_W=\frac{1}{2}
    \left(
    \frac{G_FM_W^2}{4\pi\alpha}\frac{Q^2}{Q^2+M_W^2}
    \right)^2\,.
  \end{equation}
  In the NC case, specifically when the interaction
  is purely electromagnetic and the exchanged particle is a photon, the
  longitudinal double-spin asymmetry, Eq.~\eqref{eq:A_LL_DIS}, is often
  expressed in terms of the virtual photoabsorption asymmetry $A_1$. This is
  defined as $A_1=(\sigma_{1/2}^T-\sigma_{3/2}^T)/(\sigma_{1/2}^T+\sigma_{3/2}^T)$,
  where $\sigma_{1/2}^T$ and $\sigma_{3/2}^T$ are cross sections for the
  scattering of virtual transversely polarised photons (corresponding to
  longitudinal lepton polarisation) with helicity of the photon-nucleon system
  equal to $1/2$ and $3/2$, respectively. In the limit $Q^2\gg M^2$,
  $A_{LL}^{{\rm NC},\gamma}=DA_1$, where $D$ is the so-called depolarisation factor.
  It can then be shown that
  $A_1\approx g_1^\gamma/F_1^\gamma$~\citep{Roberts:1990ww}.

\item[Semi-inclusive DIS (SIDIS)]: $\ell N \to\ell^\prime h X$. In this process,
  a beam of leptons $\ell$ and a nucleon $N$ interact semi-inclusively: in
  addition to the four-momentum of the outgoing lepton $\ell^\prime$, the
  four-momentum of one identified hadron $h$ is measured. The
  longitudinal double-spin asymmetry is defined as in the case of
  DIS, Eq.~\eqref{eq:A_LL_DIS},
  \begin{equation}
    A_{LL}^{i,h}=\frac{d^3\Delta\sigma^{i,h}}{d^3\sigma^{i,h}}\,,
    \label{eq:A_LL_SIDIS}
  \end{equation}
  where the superscript $h$ denotes the semi-inclusive production of a hadron
  $h$. The polarised and unpolarised SIDIS cross sections
  $d^3\Delta\sigma^{i,h}$ and $d^3\sigma^{i,h}$ are now triple differential in the
  additional variable $z=P\cdot p_h/P\cdot q$, where $p_h$ is the four-momentum
  of the identified hadron $h$. The parametrisation of these cross sections
  in terms of SIDIS structure functions can be derived in the same way as for
  DIS, and reads~\citep{deFlorian:2012wk}
  \begin{align}
    \frac{d^3\Delta\sigma^{i,h}}{dxdydz}
    & = \frac{8\pi\alpha^2}{xyQ^2}
    \eta^i
    \left[\pm Y_- x g_1^{i,h} + Y_+ x g_5^{i,h} + (1-y) g_L^{i,h}\right]\,,
    \label{eq:SIDIS_SF_pol}\\
    \frac{d^3\sigma^{i,h}}{dxdydz}
    & = \frac{2\pi\alpha^2}{xyQ^2}
    \eta^i
    \left[Y_+ 2x F_1^{i,h} \mp Y_- x F_3^{i,h} - (1-y) 2F_L^{i,h}  \right]\,,
    \label{eq:SIDIS_SF_unp}
  \end{align}
  where the coefficients $\eta^i$ and $Y_\pm$ are as in
  Eqs.~\eqref{eq:DIS_SF_pol}--\eqref{eq:DIS_SF_unp}. Also in this case, one can
  define CC and NC structure functions with the same form as
  Eqs.~\eqref{eq:NC_DIS}--\eqref{eq:CC_DIS}, and a photoabsorption asymmetry
  for the NC process mediated by a photon as
  $A_1^h\approx g_1^{h,\gamma}/F_1^{h,\gamma}$.
  
\item[Particle production in proton--proton ($pp$) collisions]: $pp\to HX$. In
  this process, a proton beam interacts with another proton beam;
  the final state is inclusive with respect to the particle $H$, {\it i.e.}~no
  measurements of the other reaction products are made. The particle $H$ can be
  an electroweak boson, reconstructed from its decays, a hadron, such as a pion
  or a kaon, or one or two reconstructed jets, defined by a suitable jet
  algorithm. The measured longitudinal-spin asymmetry can be single or double,
  depending on whether one ($\rightarrow$) or both ($\rightleftarrows$)
  initial-state protons are polarised in the process:
  \begin{equation}
    A_{L}^H=\frac{d\Delta\sigma^{\rightarrow, H}}{d\sigma^H}
    \qquad \qquad
    A_{LL}^H=\frac{d\Delta\sigma^{\rightleftarrows, H}}{d\sigma^H}\,.
    \label{eq:A_LL_pp}
  \end{equation}
  The polarised and unpolarised cross sections may be single or
  multi-differential in one or more kinematic variables that depend on the
  process. These could be, for instance, the rapidity of the electroweak boson,
  the transverse momentum of the leading jet, or the transverse momentum of the
  produced hadron. In contrast to DIS and SIDIS, cross sections are not
  parametrised in terms of structure functions: because there are two protons
  in the initial state, there is no leptonic contribution that can be computed
  separately from the hadronic cross section.
  
\end{description}

\subsection{Factorisation}
\label{subsec:factorisation}

The polarised and unpolarised cross sections that enter the definition of the
longitudinal spin asymmetries described in Sect.~\ref{subsec:spin_asymmetries},
Eqs.~\eqref{eq:A_LL_DIS}, \eqref{eq:A_LL_SIDIS}, and \eqref{eq:A_LL_pp},
are inclusive with respect to the full final state or to all but a small number
of identified final-state particles or jets. They therefore obey collinear
factorisation theorems~\citep{Collins:1989gx,Collins:2011zzd}. Within this
framework, in the hard-scattering limit, {\it i.e.}~when the momentum transfer
in lepton--nucleon or nucleon--nucleon scattering processes is sufficiently
large to probe wavelengths shorter than the QCD confinement scale, the strong
interaction involves approximately collinear quarks and gluons extracted from
the incoming hadrons. The hadronic cross section can then be written as a
convolution of two components. The first is computable as a perturbative series
in the QCD coupling $\alpha_s$ and describes the hard interaction of free
partons through process-dependent partonic cross sections. The second describes
the nonperturbative momentum distributions of partons through universal PDFs and
must be determined as explained in Sect.~\ref{sec:methodological_input}.

Partonic cross sections and PDFs describe, respectively, the short- and
long-distance components of the scattering process. Their separation
introduces the factorisation scale $\mu_F$, at which initial-state collinear
singularities are absorbed into the PDFs. Ultraviolet renormalisation
introduces instead the renormalisation scale $\mu_R$, at which the strong
coupling $\alpha_s(\mu_R^2)$ is evaluated. The dependence of $\alpha_s$ on
$\mu_R^2$ is governed by the QCD beta function, while the dependence of the
PDFs on $\mu_F^2$ is governed by the DGLAP evolution
equations~\citep{Gribov:1972ri,Lipatov:1974qm,Altarelli:1977zs,
Dokshitzer:1977sg}. Physical cross sections are independent of these
unphysical scales when computed to all perturbative orders, whereas a
residual scale dependence remains at any finite order. In phenomenological
applications one often sets $\mu_R^2=\mu_F^2=\mu^2$, although the two scales
are conceptually distinct. Throughout this review, unless stated otherwise,
this choice is adopted, with $\mu^2$ the characteristic hard scale of the
process. In Eqs.~\eqref{eq:a0_QCD}--\eqref{eq:Jaffe_and_Manohar}, $\mu^2$ should
instead be understood as the common scale at which the renormalised and
factorised quark, gluon, and angular-momentum contributions are defined.

In a flavour basis containing the $n_f$ quark distributions, the $n_f$
antiquark distributions, and the gluon distribution, the DGLAP equations form
a system of $2n_f+1$ coupled integro-differential equations. For polarised and
unpolarised PDFs they read, respectively,
\begin{equation}
  \frac{\partial \Delta f_i(x,\mu^2)}
       {\partial \ln\mu^2}
  =
  \sum_{j=1}^{2n_f+1}
  \int_x^1\frac{dz}{z}\,
  \Delta P_{ij}\left(z,\alpha_s(\mu^2)\right)
  \Delta f_j\left(\frac{x}{z},\mu^2\right)\,,
  \qquad
  \frac{\partial f_i(x,\mu^2)}
       {\partial \ln\mu^2}
  =
  \sum_{j=1}^{2n_f+1}
  \int_x^1\frac{dz}{z}\,
  P_{ij}\left(z,\alpha_s(\mu^2)\right)
  f_j\left(\frac{x}{z},\mu^2\right)\,,
  \label{eq:DGLAP}
\end{equation}
where the indices $i$ and $j$ run over all active quarks, antiquarks, and the
gluon. The polarised and unpolarised splitting functions,
$\Delta P_{ij}$ and $P_{ij}$, are perturbatively calculable.
Flavour symmetry allows the evolution equations to be reorganised into
singlet and non-singlet sectors. The singlet distributions evolve in a coupled
system with the corresponding gluon distributions. The remaining $2n_f-1$
independent quark combinations, including valence distributions and flavour
differences, form the non-singlet sector and evolve independently of the gluon,
although distinct non-singlet combinations may be governed by different
splitting functions beyond leading order.

The perturbative calculation of coefficient functions and partonic cross
sections gives rise to ultraviolet, soft, and collinear singularities.
Ultraviolet singularities are removed by renormalising the parameters and
fields of the theory. For sufficiently inclusive, infrared-safe observables,
soft and final-state collinear singularities cancel between real-emission and
virtual contributions. Initial-state collinear singularities are instead
absorbed into the PDFs at the scale $\mu_F$. When an identified hadron is
observed, the corresponding final-state collinear singularities are absorbed
into fragmentation functions at a fragmentation scale $\mu_D$, which may
be chosen such that $\mu_D^2=\mu^2$. After
renormalisation and mass factorisation, the perturbative coefficients are
finite, but retain a residual dependence on $\mu_R$, $\mu_F$, and, where
relevant, $\mu_D$. Renormalisation and factorisation also require the choice
of a subtraction scheme; the most commonly adopted one is the modified
minimal subtraction scheme
($\overline{\rm MS}$)~\citep{tHooft:1972tcz,tHooft:1973mfk,Bardeen:1978yd},
in which the poles of dimensional regularisation are subtracted together with
a standard set of accompanying constants.

With these conventions, upon adopting the common scale choice
$\mu_F^2=\mu_R^2=\mu^2$, where $\mu^2$ is identified with the
characteristic DIS scale $Q^2$, the polarised and unpolarised DIS structure
functions factorise as
\begin{align}
  g_k^i(x,\mu^2)
  &=
  \sum_j
  \int_x^1\frac{dx^\prime}{x^\prime}\,
  \Delta C_{k,j}^i
  \left(
    \frac{x}{x^\prime},
    \alpha_s(\mu^2)
  \right)
  \Delta f_j(x^\prime,\mu^2)\,,
  \label{eq:g_DIS_factorisation}
\end{align}
\begin{align}
  F_l^i(x,\mu^2)
  &=
  \sum_j
  \int_x^1\frac{dx^\prime}{x^\prime}\,
  C_{l,j}^i
  \left(
    \frac{x}{x^\prime},
    \alpha_s(\mu^2)
  \right)
  f_j(x^\prime,\mu^2)\,.
  \label{eq:F_DIS_factorisation}
\end{align}
where $i={\rm NC},{\rm CC}$, $k=1,5,L$, $l=1,3,L$, $\Delta C_{k,j}^i$ and
$C_{l,j}^i$ are the polarised and unpolarised DIS coefficient functions,
$\Delta f_j$ and $f_j$ are the polarised and unpolarised PDFs, and the sum runs
over all active quarks, antiquarks, and the gluon. Similarly, the polarised
and unpolarised SIDIS structure functions factorise as
\begin{align}
  g_k^{i,h}(x,z,\mu^2)
  &=
  \sum_{a,b}
  \int_x^1\frac{dx^\prime}{x^\prime}
  \int_z^1\frac{dz^\prime}{z^\prime}\,
  \Delta C_{k,ab}^{i}
  \left(
    \frac{x}{x^\prime},
    \frac{z}{z^\prime},
    \alpha_s(\mu^2)
  \right)
  \Delta f_a(x^\prime,\mu^2)
  D_b^h(z^\prime,\mu^2)\,,
  \label{eq:g_SIDIS_factorisation}
  \\
  F_l^{i,h}(x,z,\mu^2)
  &=
  \sum_{a,b}
  \int_x^1\frac{dx^\prime}{x^\prime}
  \int_z^1\frac{dz^\prime}{z^\prime}\,
  C_{l,ab}^{i}
  \left(
    \frac{x}{x^\prime},
    \frac{z}{z^\prime},
    \alpha_s(\mu^2)
  \right)
  f_a(x^\prime,\mu^2)
  D_b^h(z^\prime,\mu^2)\,.
  \label{eq:F_SIDIS_factorisation}
\end{align}
where the indices $i$, $k$, and $l$ are defined as in 
Eqs.~\eqref{eq:g_DIS_factorisation}--\eqref{eq:F_DIS_factorisation},
$\Delta C_{k,ab}^i$ and $C_{l,ab}^i$ are the polarised and unpolarised
SIDIS coefficient functions, $\Delta f_a$ and $f_a$ are the same polarised and
unpolarised PDFs as in
Eqs.~\eqref{eq:g_DIS_factorisation}--\eqref{eq:F_DIS_factorisation}, and
$D_b^h$ is the fragmentation function (FF) of the parton $b$ into a
hadron $h$. The sum runs over the initial- and final-state partons $a$ and $b$.
There is an additional dependence on the fragmentation scale $\mu_D$, which
is the final-state analogue of the factorisation scale $\mu_F$, and there are
two convolutions.
The dependence on the fragmentation scale $\mu_D$ is perturbative, and is
determined from DGLAP equations for FFs of the same form as
Eq.~\eqref{eq:DGLAP}, but with different expressions for the splitting
functions, denoted by $P_{ij}^T$, which are timelike. The common scale choice is
$\mu_F^2=\mu_D^2=\mu_R^2=\mu^2$, where $\mu^2$ is
the characteristic SIDIS scale $Q^2$.
Finally, the single-polarised and unpolarised cross sections for
$pp$ collisions, for a fully inclusive final state, read
\begin{align}
  d\Delta\sigma^{\rightarrow,\mathrm{incl.}}(\tau,\mu^2)
  &=
  \sum_{a,b}
  \int_{x_1}^1\frac{dx^\prime}{x^\prime}
  \int_{x_2}^1\frac{dx^{\prime\prime}}{x^{\prime\prime}}\,
  d\Delta\hat{\sigma}_{ab}^{\rightarrow,\mathrm{incl.}}
  \left(
    \frac{x_1}{x^\prime},
    \frac{x_2}{x^{\prime\prime}},
    \alpha_s(\mu^2)
  \right)
  \Delta f_a(x^\prime,\mu^2)
  f_b(x^{\prime\prime},\mu^2)\,,
  \label{eq:pol_pp_factorisation_incl}
  \\
  d\sigma^{\mathrm{incl.}}(\tau,\mu^2)
  &=
  \sum_{a,b}
  \int_{x_1}^1\frac{dx^\prime}{x^\prime}
  \int_{x_2}^1\frac{dx^{\prime\prime}}{x^{\prime\prime}}\,
  d\hat{\sigma}_{ab}^{\mathrm{incl.}}
  \left(
    \frac{x_1}{x^\prime},
    \frac{x_2}{x^{\prime\prime}},
    \alpha_s(\mu^2)
  \right)
  f_a(x^\prime,\mu^2)
  f_b(x^{\prime\prime},\mu^2)\,.
  \label{eq:unp_pp_factorisation_incl}
\end{align}
where $\tau$ is a dimensionless scaling variable, {\it e.g.}~the ratio of the
invariant mass of the final state to the centre-of-mass energy (that can be
expressed as a function of the momentum fractions of each proton, $x_1$ and
$x_2$), and $\Delta\hat\sigma_{ab}^{\rightarrow,incl.}$ and
$\hat\sigma_{ab}^{incl.}$ are the polarised and unpolarised partonic cross
sections. If the process is semi-inclusive, for instance when a hadron $h$ is
measured in the final state, the cross sections read
\begin{align}
  d\Delta\sigma^{\rightarrow,h}(\tau,\mu^2)
  &=
  \sum_{a,b,c}
  \int_{x_1}^1\frac{dx^\prime}{x^\prime}
  \int_{x_2}^1\frac{dx^{\prime\prime}}{x^{\prime\prime}}
  \int_z^1\frac{dz^\prime}{z^\prime}\,
  d\Delta\hat{\sigma}_{ab\to c}^{\rightarrow}
  \left(
    \frac{x_1}{x^\prime},
    \frac{x_2}{x^{\prime\prime}},
    \frac{z}{z^\prime},
    \alpha_s(\mu^2)
  \right)
  \Delta f_a(x^\prime,\mu^2)
  f_b(x^{\prime\prime},\mu^2)
  D_c^h(z^\prime,\mu^2)\,,
  \label{eq:pol_pp_factorisation_h}
  \\
  d\sigma^h(\tau,\mu^2)
  &=
  \sum_{a,b,c}
  \int_{x_1}^1\frac{dx^\prime}{x^\prime}
  \int_{x_2}^1\frac{dx^{\prime\prime}}{x^{\prime\prime}}
  \int_z^1\frac{dz^\prime}{z^\prime}\,
  d\hat{\sigma}_{ab\to c}
  \left(
    \frac{x_1}{x^\prime},
    \frac{x_2}{x^{\prime\prime}},
    \frac{z}{z^\prime},
    \alpha_s(\mu^2)
  \right)
  f_a(x^\prime,\mu^2)
  f_b(x^{\prime\prime},\mu^2)
  D_c^h(z^\prime,\mu^2)\,.
  \label{eq:unp_pp_factorisation_h}
\end{align}
where the polarised and unpolarised partonic cross sections are now
$\Delta\hat\sigma_{ab\to c}^{\rightarrow}$ and 
$\hat\sigma_{ab\to c}$, the sum runs over all
active partons $a$, $b$, and $c$, and there are three
convolutions, including one with the fragmentation function $D_c^h$.
The double-polarised cross sections $d\Delta\sigma^{\leftrightarrows,incl.}$ and
$d\Delta\sigma^{\leftrightarrows,h}$ can be obtained from
Eqs.~\eqref{eq:pol_pp_factorisation_incl} and~\eqref{eq:pol_pp_factorisation_h}
by replacing the partonic cross sections
$\Delta\hat\sigma_{ab}^{\rightarrow,incl.}$ and
$\Delta\hat\sigma_{ab\to c}^{\rightarrow}$ with their counterparts
$\Delta\hat\sigma_{ab}^{\leftrightarrows,incl.}$ and
$\Delta\hat\sigma_{ab\to c}^{\leftrightarrows}$, and the unpolarised PDF $f_b$ with
its polarised counterpart $\Delta f_b$.
As usual, $\mu_F^2=\mu_D^2=\mu_R^2=\mu^2$, where $\mu^2$ is the characteristic
scale of the process, whose precise definition is process dependent.

The formal proof of the factorisation formul\ae,
Eqs.~\eqref{eq:g_DIS_factorisation}--\eqref{eq:unp_pp_factorisation_h},
is highly non-trivial. It requires the identification of all leading momentum
regions, the separation of hard, collinear, and soft dynamics, the use of gauge
invariance and Ward identities to reorganise longitudinally polarised gluons
into Wilson lines, and the demonstration that potentially
factorisation-breaking Glauber exchanges either cancel or can be avoided by
suitable contour deformations~\citep{Collins:2011zzd,Collins:1985ue,
  Collins:1988ig}. These arguments are largely insensitive to the polarisation
of the incoming hadrons, except for the spin projections that define the
relevant parton densities and partonic cross sections. For inclusive DIS, the
factorised structure admits a particularly transparent derivation through the
light-cone operator product expansion (OPE): Mellin moments of the polarised
structure functions are related to forward proton matrix elements of local
twist-two axial quark and gluon operators, while their complete $x$-dependence
is encoded in gauge-invariant bilocal light-ray operators. In a commonly used
convention, for a proton state of momentum $P$ and positive helicity and for a
light-like separation $\xi^\mu=(0,\xi^-,\boldsymbol{0}_T)$, the quark and gluon
helicity distributions are defined by (in light-cone coordinates)
\begin{equation}
  \Delta q(x,\mu^2)
  =
  \frac{1}{4\pi}
  \int d\xi^-
  e^{-ixP^+\xi^-}
  \langle P,+ |
  \bar\psi(\xi^-)\,
  \gamma^+\gamma_5\,
  \mathcal{W}_F(\xi^-,0)\,
  \psi(0)
  \left|P,+\right\rangle\,,
\end{equation}
and
\begin{equation}
  \Delta g(x,\mu^2)=
  \frac{i}{4\pi xP^+}
  \int d\xi^-
  e^{-ixP^+\xi^-}
  \left\langle P,+\right|
  F_a^{+\alpha}(\xi^-)\,
  \mathcal{W}_A^{ab}(\xi^-,0)\,
  \widetilde F^{+}_{\alpha b}(0)
  \left|P,+\right\rangle\,,
\end{equation}
up to conventional choices for the overall normalisation and Fourier-transform
signs. Here $\mathcal{W}_F$ and $\mathcal{W}_A$ are straight light-like Wilson
lines in the fundamental and adjoint representations, respectively, and
$\widetilde F^{\mu\nu}=\tfrac12\epsilon^{\mu\nu\rho\sigma}F_{\rho\sigma}$.
These operator definitions make the universality of polarised PDFs precise and
provide the continuum quantities targeted by first-principles lattice-QCD
calculations: the traditional OPE approach determines their Mellin moments from
local twist-two operators, whereas modern quasi- and pseudo-PDF methods
calculate related Euclidean bilocal correlators and perturbatively match them
onto the corresponding light-cone
distributions~\citep{Lin:2017snn,Constantinou:2020hdm}.

\subsection{Perturbative computations}
\label{subsec:perturbative_computations}

As mentioned, the splitting functions entering DGLAP equations,
Eq.~\eqref{eq:DGLAP}, and all the coefficient functions and partonic cross
sections in
Eqs.~\eqref{eq:g_DIS_factorisation}--\eqref{eq:unp_pp_factorisation_h} can
be computed as a perturbative series in the strong coupling $\alpha_s$
\begin{equation}
  \mathcal{P}=\sum_{n=0}^\infty
  \left(\frac{\alpha_s}{2\pi}\right)^{n+1} \mathcal{P}^{(n)}\,,
  \qquad
  \mathcal{C}=\sum_{n=0}^\infty
  \left(\frac{\alpha_s}{2\pi}\right)^n \mathcal{C}^{(n)}\,,
  \label{eq:pert_expansion}
\end{equation}
where, for splitting functions, $\mathcal{P}=\Delta P_{ff^\prime}$,
$P_{ff^\prime}, P^T_{ff^\prime}$ (for polarised PDFs, unpolarised PDFs, and FFs,
respectively) and, for coefficient functions and partonic matrix
elements, $\mathcal{C}=\Delta C_{k,f}^i$, $C_{j,f}^i$, $\Delta C_{k,f,f^\prime}^{i,h}$,
$C_{j,f,f^\prime}^{i,h},\Delta\hat\sigma_{ff^\prime}^{\rightarrow,incl.}$,
$\hat\sigma_{ff^\prime}^{incl.}$,
$\Delta\hat\sigma_{ff^\prime f^{\prime\prime}}^{\rightarrow,h}$,
$\hat\sigma_{ff^\prime f^{\prime\prime}}^h$,
$\Delta\hat\sigma_{ff^\prime}^{\leftrightarrows,incl.}$,
$\Delta\hat\sigma_{ff^\prime f^{\prime\prime}}^{\leftrightarrows,h}$.

Beyond leading order (LO), the coefficients $\mathcal{P}^{(k)}$ and
$\mathcal{C}^{(k)}$ contain logarithms of ratios of the hard, factorisation,
fragmentation, and renormalisation scales, while
both coefficient and splitting functions contain logarithmically enhanced
terms near the kinematic endpoints. DGLAP evolution resums the collinear
logarithms associated with scale evolution. Additional resummations may
be required for powers of $\ln(1/x)$ at small $x$,
see~\citep{JAMCollaborationSmall-xAnalysisGroup:2025tfa}
and references therein, and of $\ln (1-x)$ at large $x$,
see~\citep{Anderle:2013lka} and references therein.

If the lowest non-vanishing contribution to a given cross section is of order
$\alpha_s^p$, a calculation through N$^k$LO includes terms up to
$\alpha_s^{p+k}$, while its residual dependence on the factorisation,
fragmentation, and renormalisation scales is formally of order
$\alpha_s^{p+k+1}$. This dependence therefore generally decreases as higher
perturbative orders are included, although this expectation may be altered
when new partonic channels or kinematic configurations first appear. Scale
variations provide a conventional estimate of missing higher-order
uncertainties, rather than a rigorous uncertainty determination. The scales
$\mu_R$, $\mu_F$ and, where relevant, $\mu_D$
are commonly varied independently by factors of $2$ and $1/2$ about their
central values, generally assumed to be equal to $\mu$, using prescribed
combinations that avoid excessively large
ratios between them; where the next perturbative order is known, such
prescriptions can be tested against the actual higher-order correction
(see~\citep{NNPDF:2019ubu} for details).

Significant progress has been made to compute the higher-order coefficients
$\mathcal{P}^{(n)}$ and $\mathcal{C}^{(n)}$ in recent years. Before reviewing
this progress, it must be noted that all factorisation formul\ae,
Eqs.~\eqref{eq:g_DIS_factorisation}--\eqref{eq:unp_pp_factorisation_h},
and evolution equations, Eq.~\eqref{eq:DGLAP}, include sums over the number of
active flavours $n_f$, that is over the number of partons that are kinematically
accessible at an energy scale $\mu^2$. Concerning splitting functions, this
requires appropriate matching conditions for PDF evolution across
the threshold. For coefficient functions, decoupling
arguments~\citep{Appelquist:1974tg} imply that heavy-quark contributions to
any process are power-suppressed at scales below 
threshold for their production~\citep{Collins:1978wz}. Therefore, when 
expressing predictions for processes at different scales in terms of the same
PDF set, it is necessary to use a variable-flavour number (VFN) scheme in which
different numbers of active flavours are adopted consistently. In the vicinity
of the threshold for heavy quark production, the quark mass $m$ cannot be
ignored. This can be accounted for in a general-mass VFN (GM-VFN) scheme that
interpolates, in a model-dependent way, between the fixed-flavour number (FFN)
scheme near  production threshold and the asymptotic result of the zero-mass
VFN (ZM-VFN). In the FFN scheme, heavy quark mass effects are built into the 
partonic cross sections, but terms proportional to large logarithms of the form 
$\ln(Q/m)$ are not resummed; in the ZM-VFN scheme, heavy quark mass effects are 
ignored, but $\ln(Q/m)$ terms are instead resummed into the heavy quark PDFs. 
Various schemes have been worked out in the literature
(see~\citep{SM:2010nsa} for a review), however only the so-called FONLL
scheme~\citep{Forte:2010ta} has been extended to polarised DIS
structure functions up to next-to-next-to-leading order
(NNLO)~\citep{Hekhorn:2024tqm}. No extension to SIDIS
is available. Quark mass effects are expected to be small in current DIS and
SIDIS measurements; however, they may be sizeable in future EIC
measurements. 

The current status of perturbative computations is as follows.
Concerning splitting functions, NNLO corrections are consistently known for
the polarised~\citep{Moch:2014sna,
  Moch:2015usa,Blumlein:2021enk,Blumlein:2021ryt,Bierenbaum:2022biv},
unpolarised~\citep{Moch:2004pa,Vogt:2004mw},
and time-like~\citep{Mitov:2006ic,Moch:2007tx,Almasy:2011eq,Chen:2020uvt} cases.
In the unpolarised case, an increasing set of Mellin moments and the small- and
large-$x$ behaviour of splitting functions are known at
next-to-next-to-next-to-leading order (N$^3$LO),
see~\citep{Cridge:2024icl} and references therein. Heavy-flavour matching
conditions are known through NNLO for polarised PDFs~\citep{Bierenbaum:2022biv},
at N$^3$LO~\citep{Bierenbaum:2009mv,Ablinger:2023ahe,Kawamura:2012cr} for
unpolarised PDFs, and at next-to-leading-order (NLO) for
FFs~\citep{Cacciari:2005ry}.
Concerning coefficient functions, for polarised DIS, the massless polarised
structure function $g_1$ has been known at NNLO for a long
time~\citep{Zijlstra:1993sh}. Very recently, N$^3$LO corrections have been
computed~\citep{Blumlein:2022gpp}, as have NNLO
massive contributions~\citep{Hekhorn:2018ywm}, their asymptotic
limit~\citep{Behring:2015zaa,Ablinger:2019etw,Behring:2021asx,Blumlein:2021xlc,
  Bierenbaum:2022biv,Ablinger:2022wbb,Ablinger:2023ahe,Ablinger:2024xtt},
and the NNLO parity-violating massless polarised structure
functions~\citep{Borsa:2022irn}. For polarised SIDIS, the massless polarised
structure function $g_1^h$ has recently been determined through NNLO: first
using an approximation based on the threshold resummation
formalism~\citep{Abele:2021nyo}, then exactly using various analytical
methods~\citep{Bonino:2024wgg,Goyal:2024tmo,Goyal:2024emo}; NNLO corrections
to polarised massless parity-violating SIDIS structure functions have also
followed~\citep{Bonino:2025bqa,Goyal:2026ccx}. Finally, NNLO corrections have
been obtained for $W$-boson production in polarised proton-proton
collisions~\citep{Boughezal:2021wjw}. Higher-order corrections to
single-inclusive jet, dijet, and single-hadron production in polarised
$pp$ collisions are known only up to
NLO~\citep{Jager:2004jh,Jager:2002xm,deFlorian:2002az}. For all the
aforementioned processes, the perturbative accuracy of the corresponding
unpolarised cross sections, needed for the computation of spin asymmetries,
is known consistently: NNLO for DIS~\citep{vanNeerven:1991nn,Zijlstra:1991qc,
  Zijlstra:1992kj,Zijlstra:1992qd},
SIDIS~\citep{Bonino:2024qbh,Bonino:2025tnf,Bonino:2025qta,Goyal:2023zdi,
  Goyal:2024emo,Goyal:2026ccx}, and inclusive gauge-boson production in
$pp$ collisions~\citep{Anastasiou:2003yy,Anastasiou:2003ds};
NLO for single-inclusive jet, dijet, and single-hadron production in 
$pp$ collisions~\citep{Jager:2004jh,Jager:2002xm,deFlorian:2002az}. 
Additional results at N$^3$LO are known for
DIS~\citep{Vermaseren:2005qc,Blumlein:2022gpp} and for inclusive
gauge boson production in $pp$ collisions~\citep{Chen:2022cgv},
and at NNLO for single-inclusive jet and
dijet~\citep{Currie:2016bfm,Currie:2017eqf,Chen:2022tpk}
and single-hadron~\citep{Czakon:2025yti} production in $pp$ collisions.

%-------------------------------------------------------------------------------
\begin{table}[!t]
  \renewcommand{\arraystretch}{1.4}
  \TBL{\caption{The leading-order factorisation of DIS and SIDIS structure
      functions, and of $pp$ collision cross sections in terms of PDF and FF
      combinations, for each of the observables defined in
      Eqs.~\eqref{eq:g_DIS_factorisation}--\eqref{eq:unp_pp_factorisation_h}
      entering the spin asymmetries introduced in
      Eqs.~\eqref{eq:A_LL_DIS}, \eqref{eq:A_LL_SIDIS}, and~\eqref{eq:A_LL_pp}.}
    \label{tab:factorisation}}
  \begin{tabular*}{\textwidth}{@{\extracolsep{\fill}}@{}lll@{}}
\toprule
\multicolumn{1}{@{}l}{\TCH{Process}} &
\multicolumn{1}{l}{\TCH{Observable}} &
\multicolumn{1}{l}{\TCH{Leading PDF/FF dependence at Born level}}\\
\colrule
DIS
& $\left[F_1^\gamma,F_1^{\gamma Z},F_1^Z\right]$
& $\displaystyle
  \frac{1}{2}\sum_q
  \left[e_q^2,\,2e_qg_V^q,\,(g_V^q)^2+(g_A^q)^2\right]
  \left(f_q+f_{\bar q}\right)$
\\
& $\left[F_3^\gamma,F_3^{\gamma Z},F_3^Z\right]$
& $\displaystyle
  \sum_q
  \left[0,\,2e_qg_A^q,\,2g_V^qg_A^q\right]
  \left(f_q-f_{\bar q}\right)$
\\
& $\left[F_1^{W^-},F_1^{W^+}\right]$
& $\left[
  f_u+f_{\bar d}+f_{\bar s}+f_c+\ldots,\,
  f_d+f_{\bar u}+f_{\bar c}+f_s+\ldots
  \right]$
\\
& $\left[F_3^{W^-},F_3^{W^+}\right]$
& $\displaystyle
  2\left[
  f_u-f_{\bar d}-f_{\bar s}+f_c+\ldots,\,
  f_d-f_{\bar u}-f_{\bar c}+f_s+\ldots
  \right]$
\\
& $\left[g_1^\gamma,g_1^{\gamma Z},g_1^Z\right]$
& $\displaystyle
  \frac{1}{2}\sum_q
  \left[e_q^2,\,2e_qg_V^q,\,(g_V^q)^2+(g_A^q)^2\right]
  \left(\Delta f_q+\Delta f_{\bar q}\right)$
\\
& $\left[g_5^\gamma,g_5^{\gamma Z},g_5^Z\right]$
& $\displaystyle
  \sum_q
  \left[0,\,e_qg_A^q,\,g_V^qg_A^q\right]
  \left(\Delta f_{\bar q}-\Delta f_q\right)$
\\
& $\left[g_1^{W^-},g_1^{W^+}\right]$
& $\left[
  \Delta f_u+\Delta f_{\bar d}+\Delta f_{\bar s}+\Delta f_c+\ldots,\,
  \Delta f_d+\Delta f_{\bar u}+\Delta f_{\bar c}+\Delta f_s+\ldots
  \right]$
\\
& $\left[g_5^{W^-},g_5^{W^+}\right]$
& $\left[
  -\Delta f_u+\Delta f_{\bar d}+\Delta f_{\bar s}-\Delta f_c+\ldots,\,
  -\Delta f_d+\Delta f_{\bar u}+\Delta f_{\bar c}-\Delta f_s+\ldots
  \right]$
\\
\colrule
SIDIS
& $\left[F_1^{\gamma,h},F_1^{\gamma Z,h},F_1^{Z,h}\right]$
& $\displaystyle
  \frac{1}{2}\sum_q
  \left[e_q^2,\,2e_qg_V^q,\,(g_V^q)^2+(g_A^q)^2\right]
  \left(f_qD_q^h+f_{\bar q}D_{\bar q}^h\right)$
\\
& $\left[F_3^{\gamma,h},F_3^{\gamma Z,h},F_3^{Z,h}\right]$
& $\displaystyle
  \sum_q
  \left[0,\,2e_qg_A^q,\,2g_V^qg_A^q\right]
  \left(f_qD_q^h-f_{\bar q}D_{\bar q}^h\right)$
\\
& $\left[F_1^{W^-,h},F_1^{W^+,h}\right]$
& $\left[
  f_uD_d^h+f_{\bar d}D_{\bar u}^h
  +f_{\bar s}D_{\bar c}^h+f_cD_s^h+\ldots,\,
  f_dD_u^h+f_{\bar u}D_{\bar d}^h
  +f_{\bar c}D_{\bar s}^h+f_sD_c^h+\ldots
  \right]$
\\
& $\left[F_3^{W^-,h},F_3^{W^+,h}\right]$
& $\displaystyle
  2\left[
  f_uD_d^h-f_{\bar d}D_{\bar u}^h
  -f_{\bar s}D_{\bar c}^h+f_cD_s^h+\ldots,\,
  f_dD_u^h-f_{\bar u}D_{\bar d}^h
  -f_{\bar c}D_{\bar s}^h+f_sD_c^h+\ldots
  \right]$
\\
& $\left[g_1^{\gamma,h},g_1^{\gamma Z,h},g_1^{Z,h}\right]$
& $\displaystyle
  \frac{1}{2}\sum_q
  \left[e_q^2,\,2e_qg_V^q,\,(g_V^q)^2+(g_A^q)^2\right]
  \left(\Delta f_qD_q^h+\Delta f_{\bar q}D_{\bar q}^h\right)$
\\
& $\left[g_5^{\gamma,h},g_5^{\gamma Z,h},g_5^{Z,h}\right]$
& $\displaystyle
  \sum_q
  \left[0,\,e_qg_A^q,\,g_V^qg_A^q\right]
  \left(\Delta f_{\bar q}D_{\bar q}^h-\Delta f_qD_q^h\right)$
\\
& $\left[g_1^{W^-,h},g_1^{W^+,h}\right]$
& $\left[
  \Delta f_uD_d^h+\Delta f_{\bar d}D_{\bar u}^h
  +\Delta f_{\bar s}D_{\bar c}^h+\Delta f_cD_s^h+\ldots,\,
  \Delta f_dD_u^h+\Delta f_{\bar u}D_{\bar d}^h
  +\Delta f_{\bar c}D_{\bar s}^h+\Delta f_sD_c^h+\ldots
  \right]$
\\
& $\left[g_5^{W^-,h},g_5^{W^+,h}\right]$
& $\left[
  -\Delta f_uD_d^h+\Delta f_{\bar d}D_{\bar u}^h
  +\Delta f_{\bar s}D_{\bar c}^h-\Delta f_cD_s^h+\ldots,\,
  -\Delta f_dD_u^h+\Delta f_{\bar u}D_{\bar d}^h
  +\Delta f_{\bar c}D_{\bar s}^h-\Delta f_sD_c^h+\ldots
  \right]$
\\
\colrule
proton--proton
& $\left[
  d\sigma^{\mathrm{jet}},
  d\Delta\sigma^{\rightleftarrows,\mathrm{jet}}
  \right]$
& $\displaystyle
  \left[
  \sum_{a,b}f_a f_b\,
  d\hat\sigma_{ab\to\mathrm{jet}+X},\,
  \sum_{a,b}\Delta f_a\Delta f_b\,
  d\Delta\hat\sigma_{ab\to\mathrm{jet}+X}
  \right],
  \quad a,b=q,\bar q,g$
\\
& $\left[
  d\sigma^{W^-},
  d\Delta\sigma^{\rightarrow,W^-},
  d\Delta\sigma^{\rightleftarrows,W^-}
  \right]$
& $\displaystyle
  \left[
  f_{d,1}f_{\bar u,2}+f_{\bar u,1}f_{d,2},\,
  -\Delta f_{d,1}f_{\bar u,2}
  +\Delta f_{\bar u,1}f_{d,2},\,
  -\Delta f_{d,1}\Delta f_{\bar u,2}
  -\Delta f_{\bar u,1}\Delta f_{d,2}
  \right]+\ldots$
\\
& $\left[
  d\sigma^{W^+},
  d\Delta\sigma^{\rightarrow,W^+},
  d\Delta\sigma^{\rightleftarrows,W^+}
  \right]$
& $\displaystyle
  \left[
  f_{u,1}f_{\bar d,2}+f_{\bar d,1}f_{u,2},\,
  -\Delta f_{u,1}f_{\bar d,2}
  +\Delta f_{\bar d,1}f_{u,2},\,
  -\Delta f_{u,1}\Delta f_{\bar d,2}
  -\Delta f_{\bar d,1}\Delta f_{u,2}
  \right]+\ldots$
\\
& $\left[
  d\sigma^h,
  d\Delta\sigma^{\rightleftarrows,h}
  \right]$
& $\displaystyle
  \left[
  \sum_{a,b,c}f_a f_b\,
  d\hat\sigma_{ab\to c+X}D_c^h,\,
  \sum_{a,b,c}\Delta f_a\Delta f_b\,
  d\Delta\hat\sigma_{ab\to c+X}D_c^h
  \right]$
\\
\botrule
\end{tabular*}

\end{table}
%-------------------------------------------------------------------------------

Given the perturbative expansions of the splitting functions, coefficient
functions, and partonic cross sections in Eq.~\eqref{eq:pert_expansion}, it is
useful to examine the factorisation formul\ae\
in Eqs.~\eqref{eq:g_DIS_factorisation}--\eqref{eq:unp_pp_factorisation_h} at
the lowest non-vanishing perturbative order for each process. This order is
not uniformly $\mathcal{O}(\alpha_s^0)$: inclusive DIS, SIDIS, and
electroweak-boson production begin at $\mathcal{O}(\alpha_s^0)$, whereas jet
and single-hadron production in $pp$ collisions begin at
$\mathcal{O}(\alpha_s^2)$. The corresponding Born-level expressions make the
leading flavour and partonic-channel sensitivity of each observable
particularly transparent. Parton evolution is not neglected in this
description: PDFs and FFs remain evaluated at the relevant factorisation
scales and may be evolved to those scales, while higher-order hard-scattering
corrections and the additional channels that they generate are omitted.

This exercise is summarised in Table~\ref{tab:factorisation}, where the
leading combinations of PDFs and FFs contributing to the polarised and
unpolarised DIS and SIDIS structure functions, and to single-inclusive jet,
$W$-boson, and single-inclusive hadron production in $pp$ collisions, are
displayed. The quark couplings are defined as
\begin{equation}
  g_V^q=\pm\frac{1}{2}-2e_q\sin^2\theta_W\,,
  \qquad
  g_A^q=\pm\frac{1}{2}\,,
\end{equation}
where the $\pm$ sign refers to whether $q$ is a $u$- or $d$-type quark.
In the charged-current DIS and SIDIS expressions, CKM factors and
off-diagonal flavour transitions are left implicit. In the proton--proton
expressions, $f_{a,1}$ and $f_{a,2}$ denote the PDF of parton $a$ in the first
and second proton, respectively; the first proton is taken to be polarised in
the single-spin cross sections. The $W^\pm$ expressions refer to production
differential in the boson kinematics. When the decay lepton is observed
instead, the two partonic contributions acquire different angular weights.
The jet and identified-hadron cross sections receive contributions from all
quark--quark, quark--gluon, and gluon--gluon channels, with the $qg$ and $gg$
channels providing the principal sensitivity to the gluon helicity
distribution in RHIC kinematics. Longitudinal single-spin jet and hadron
cross sections vanish in parity-conserving QCD; nonzero contributions can
arise from electroweak interactions.

Three remarks are in order. First, the longitudinal DIS and SIDIS structure
functions $F_L$ and $g_L$ do not appear in Table~\ref{tab:factorisation},
because they vanish at Born level in the massless approximation and receive
their first nonzero contributions at $\mathcal{O}(\alpha_s)$. By contrast, the
purely electromagnetic structure functions $F_3^\gamma$ and $g_5^\gamma$,
together with their semi-inclusive counterparts $F_3^{h,\gamma}$ and
$g_5^{h,\gamma}$, vanish because the electromagnetic interaction conserves
parity. Second, the parity-conserving electromagnetic DIS structure functions
$F_1^\gamma$ and $g_1^\gamma$ probe charge-weighted combinations of
$f_q+f_{\bar q}$ and $\Delta f_q+\Delta f_{\bar q}$, respectively, and therefore do
not separate quark from antiquark distributions. Such a separation can instead
be obtained from parity-violating inclusive DIS, from SIDIS (where quark and
antiquark PDFs are weighted by generally different flavour-dependent
fragmentation functions) and from electroweak-boson production in $pp$
collisions. In the SIDIS case, this enhanced flavour sensitivity comes at the
price of introducing additional nonperturbative information and uncertainties
through the FFs. Finally, direct sensitivity to the gluon helicity distribution
at the lowest non-vanishing perturbative order is provided primarily by jet,
dijet, and single-hadron production in $pp$ collisions, through channels such
as $qg$ and $gg$ scattering. In DIS and SIDIS, the gluon enters instead through
higher-order coefficient functions and through QCD evolution. These
considerations illustrate why a determination of all quark, antiquark, and
gluon polarised PDFs requires the combination of a complementary set of
observables from several different processes.

\subsection{Theoretical constraints}
\label{subsec:theoretical_constraints}

In global determinations of polarised PDFs, the analysis of spin asymmetries is
typically accompanied by theoretical constraints. These belong to three main
classes, which are discussed in turn.

The first class of theoretical constraints follows from the fact that, for any
process $p$, the unpolarised cross section $d\sigma^p$ ought to be positive,
a fact that, by definition, provides a bound on the corresponding polarised
cross sections $d\Delta\sigma^p$ for the same process
\begin{equation}
  |d\Delta\sigma^p|\leq d\sigma^p\,.
  \label{eq:positivity_bound}
\end{equation}
Depending on the process $p$, different positivity bounds can be put on spin
asymmetries starting from Eq.~\eqref{eq:positivity_bound}. In the case of
neutral-current (photon-induced) DIS and SIDIS, neglecting power corrections,
the relationship between cross sections and structure functions,
Eqs.~\eqref{eq:DIS_SF_pol}--\eqref{eq:DIS_SF_unp}
and~\eqref{eq:SIDIS_SF_pol}-\eqref{eq:SIDIS_SF_unp}, immediately implies that
\begin{equation}
  |g_1^\gamma(x,\mu^2)|\leq F_1^\gamma(x,\mu^2)
  \qquad
  |g_1^{\gamma,h}(x,\mu^2)|\leq F_1^{\gamma, h}(x,\mu^2)\,.
  \label{eq:positivity_SFs}
\end{equation}
At LO, Eq.~\eqref{eq:positivity_SFs} constrains charge-weighted combinations of
quark and antiquark helicity distributions. Stronger flavour-by-flavour
inequalities follow from the probabilistic interpretation of the individual LO
parton densities, or equivalently from their definition in terms of an
appropriate set of physical processes. In particular, a boundary condition on
the polarised gluon PDF can be derived~\citep{Altarelli:1998gn} by considering
Higgs boson production in unpolarised and polarised $pp$ collisions.
Therefore the very general bound
\begin{equation}
  |\Delta f_i(x,\mu^2)|\leq f_i(x,\mu^2)\,
  \label{eq:PDF_positivity_bound}
\end{equation}
holds for $i$ denoting any quark, antiquark, or the gluon. Beyond LO,
Eq.~\eqref{eq:PDF_positivity_bound} should be replaced by more complicated
relationships between PDFs and NLO coefficient functions, that become
scheme-dependent~\citep{Altarelli:1998gn}. In comparison to the LO bound,
however, the NLO bound is modified only at small-to-intermediate
values of $x$~\citep{Forte:1998kd}, where its impact on polarised PDFs is
phenomenologically small~\citep{Ball:2013lla}.

Other positivity bounds hold for spin asymmetries measured in single-particle
inclusive production in $pp$ collisions, for both
parity-conserving~\citep{Soffer:2003qj} and
parity-violating~\citep{Kang:2011qz} processes. A particularly relevant bound
among these is
\begin{equation}
  1\pm A_{\rm LL}(y)\geq |A_{\rm L}(y)\pm A_{\rm L}(-y)|\,
  \label{eq:ALL_positivity_bound}
\end{equation}
where $A_{\rm LL}$ and $A_{\rm L}$ are the double- and single-spin asymmetries for
inclusive gauge boson production in polarised $pp$ collisions,
evaluated at forward ($y$) and backward ($-y$) rapidities of the gauge boson.

The second class of theoretical constraints follows from SU$_f$ symmetry
relations. These are constraints that can be imposed on polarised PDFs using
an established relationship between weak baryon decays and the non-singlet
combination of PDF moments. A first constraint follows from the Bjorken
sum rule~\citep{Bjorken:1966jh,Bjorken:1969mm} which relates the difference
between the lowest moments of the proton and neutron structure functions $g_1$
to the isovector axial charge $g_A$ assuming exact SU$_f$(2) symmetry.
A second constraint follows from relating the non-singlet combination of PDFs
to the octet axial charge $a_8$, extracted from weak hyperon decays assuming
exact SU$_f$(3) symmetry. In terms of PDF moments,
\begin{equation}
  \int_0^1 dx\,(\Delta f_u^+-\Delta f_d^+) = g_A
  \qquad
  \int_0^1 dx\,(\Delta f_u^++\Delta f_d^+-2\Delta f_s^+) = a_8\,
  \label{eq:pol_moments}
\end{equation}
where $\Delta f_q^+=\Delta f_q+\Delta f_{\bar q}$, with $q=u,d,s$. The second
relationship constrains
the strange polarisation, which is typically poorly determined by polarised DIS;
SIDIS, in particular with kaons in the final state, and gauge boson production
in $pp$ collisions are significantly more sensitive to the strange
polarisation, hence these processes can be used to test possible violations
of SU$_f$(3). While the Bjorken sum rule is experimentally verified at the
few-percent level~\citep{Ethier:2017zbq}, there is some uncertainty surrounding
the level of SU$_f$(3) symmetry breaking in the value of
$a_8$~\citep{Jaffe:1989jz,Bass:2009ed}. For this reason, the corresponding
theoretical constraint is often relaxed in modern polarised PDF determinations.

The third class of theoretical constraints concerns general properties of
PDFs. The first is smoothness: in practical parametrisations, PDFs are
assumed to be sufficiently regular on the interval $x\in (0,1)$, apart from
possible nodes and from the power-like or logarithmic behaviour associated
with the endpoints. The second is the behaviour at the endpoint $x=1$:
for a composite hadron, PDFs are expected to vanish as $x\to 1$, reflecting the
suppression of configurations in which one parton carries essentially all of
the hadron momentum. Perturbative counting arguments motivate power-law
behaviour, although they do not uniquely determine the exponent used in a
phenomenological parametrisation. The third is integrability: when
integrated quark helicities are identified with axial-current matrix elements,
the corresponding first moments must be finite after renormalisation.
Convergence of the gluon moment is instead an additional assumption about
the small-$x$ behaviour of the gluon helicity distribution. More generally, the
existence of full first moments must be checked against, or imposed through
assumptions on, the unmeasured small-$x$ region.

\section{Experimental input}
\label{sec:experimental_input}

\begin{figure}[!t]
  \centering
  \includegraphics[width=\textwidth]{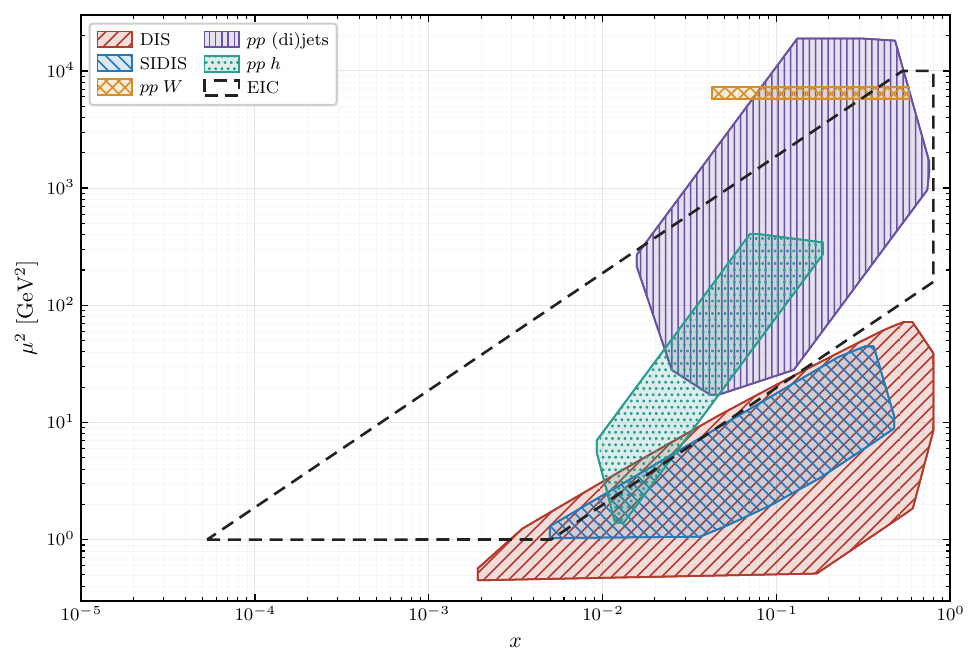}\\
  \caption{The kinematic coverage, in the $(x,\mu^2)$ plane (where $\mu^2$ is
    the process-dependent characteristic scale) of all current
    experimental measurements of longitudinal spin observables relevant for the
    determination of polarised PDFs, see Sect.~\ref{subsec:spin_asymmetries}.
    Approximate coverage regions are grouped by process: DIS, SIDIS, and
    particle production in $pp$ collisions. The displayed values of $x$ and
    $\mu^2$ are inferred from the kinematic variables of each measurement using
    LO QCD. The kinematic region covered by future DIS and SIDIS EIC
    measurements is outlined by a dashed contour.}
  \label{fig:kinematics}
\end{figure}

The determination of polarised PDFs benefits from a corpus of experimental
measurements that have been collected for almost fifty years. These correspond
to the longitudinal spin observables discussed in
Sect.~\ref{subsec:spin_asymmetries} for DIS, SIDIS, and particle production in
$pp$ collisions. Table~\ref{tab:facilities} summarises the principal
experimental facilities that have operated to date or are planned for the
future, separated according to whether they use fixed-target or collider
kinematics. Their main features are listed, specifically the
name of the experiments, the data-taking period, the nature of the scattering
particles, the measured observable, the beam or centre-of-mass energy, and the
kinematic coverage in $x$ and $\mu^2$. This last feature is also displayed in
Fig.~\ref{fig:kinematics}. The displayed values of $x$ and $\mu^2$ are inferred
from the kinematic variables of each measurement using LO QCD. 

%-------------------------------------------------------------------------------
\begin{table}[!t]
  \renewcommand{\arraystretch}{1.4}
  \TBL{\caption{A summary of the experimental facilities that have measured (or
      will measure) longitudinal spin observables relevant to the determination
      of polarised PDFs, see Sect.~\ref{subsec:spin_asymmetries}. They are
      separated according to whether they operate in fixed-target or collider
      mode. Their main features are listed, specifically the
      name of the experiments, the data-taking period, the nature of the
      scattering particles, the measured asymmetry, the beam or centre-of-mass
      energy, and the $(x,\mu^2)$ kinematic coverage.}
    \label{tab:facilities}}
  \begin{tabular*}{\textwidth}{@{\extracolsep{\fill}}@{}lllllcccc@{}}
\toprule
\multicolumn{9}{c}{\TCH{Fixed-target mode}}\\
\colrule
\multicolumn{1}{@{}l}{\TCH{Facility}} &
\multicolumn{1}{l}{\TCH{Experiments}} &
\multicolumn{1}{l}{\TCH{Operation}} &
\multicolumn{1}{l}{\TCH{Beam}} &
\multicolumn{1}{l}{\TCH{Target}} &
\multicolumn{1}{c}{\TCH{Observable}} &
\multicolumn{1}{c}{\TCH{$E_b$ [GeV]}} &
\multicolumn{1}{c}{\TCH{$x$}} &
\multicolumn{1}{c}{\TCH{$\mu^2$ [GeV$^2$]}} \\
\colrule
SLAC
& E80, E130
& 1976--1983
& $e^-$
& $p$
& $A_1$
& $\lesssim 23$
& $0.1-0.6$
& $1-10$ \\
& E142, E143
& 1992--1993
& $e^-$
& $p$, $d$, $n$
& $g_1^\gamma$
& $\lesssim 30$
& $0.03-0.8$
& $1-10$ \\
& E154, E155
& 1995--1999
& $e^-$
& $p$, $d$, $n$
& $g_1^\gamma$, $g_1^\gamma/F_1^\gamma$
& $\lesssim 50$
& $0.01-0.8$
& $1-35$ \\
\colrule
CERN
& EMC
& 1985
& $\mu^+$
& $p$
& $g_1^\gamma$
& $100$, $190$
& $0.01-0.5$
& $1-30$ \\
& SMC
& 1992--1996
& $\mu^+$
& $p$, $d$
& $g_1^\gamma$
& $100$, $190$
& $0.004-0.5$
& $1-60$ \\
& COMPASS
& 2002--2012
& $\mu^+$
& $p$, $d$
& $g_1^\gamma$, $g_1^{\pi^\pm/K^\pm,\gamma}/F_1^{\pi^\pm/K^\pm,\gamma}$
& $160, 200$
& $0.003-0.6$
& $1-70$ \\
\colrule
DESY
& HERMES
& 1995--2007
& $e^+$, $e^-$
& $p$, $d$, $n$
& $g_1^\gamma$, $g_1^{\pi^\pm/K^\pm,\gamma}/F_1^{\pi^\pm/K^\pm,\gamma}$
& $\sim 30$
& $0.02-0.7$
& $1-15$ \\
\colrule
JLab6
& Hall A/B
& 1999--2012
& $e^-$
& $p, d$
& $g_1^\gamma$, $g_1^\gamma/F_1^\gamma$
& $\lesssim 6$
& $0.05-0.6$
& $1.0-1.5$\\
JLab12
& Hall A/B/C
& 2014--
& $e^-$
& $p, d$
& $g_1^\gamma$, $g_1^\gamma/F_1^\gamma$
& $\lesssim 12$
& $0.05-0.8$
& $1.0-10$\\
\colrule
\multicolumn{9}{c}{\TCH{Collider mode}}\\
\colrule
\multicolumn{1}{@{}l}{\TCH{Facility}} &
\multicolumn{1}{l}{\TCH{Experiments}} &
\multicolumn{1}{l}{\TCH{Operation}} &
\multicolumn{1}{l}{\TCH{Beam 1}} &
\multicolumn{1}{l}{\TCH{Beam 2}} &
\multicolumn{1}{c}{\TCH{Observable}} &
\multicolumn{1}{c}{\TCH{$E_b$ [GeV]}} &
\multicolumn{1}{c}{\TCH{$x$}} &
\multicolumn{1}{c}{\TCH{$\mu^2$ [GeV$^2$]}} \\
\colrule
RHIC
& STAR
& 2002--2025
& $p$
& $p$
& $A_L^{W^\pm}$, $A_{LL}^{1-jet}$, $A_{LL}^{2-jet}$
& $2\times (31-255)$
& $\sim 0.02-0.4$
& $1-10^4$ \\
& PHENIX
& 2002--2025
& $p$
& $p$
&  $A_L^{W^\pm}$, $A_{LL}^{1-jet}$, $A_{LL}^{\pi^0}$
& $2\times (31-255)$
& $\sim 0.02-0.4$
& $1-10^4$ \\
\colrule
EIC
& ePIC, \dots
& 2030s
& $e^-$, $e^+$ (opt.)
& $p,d,{}^3\mathrm{He}$
& $g_{1,5}^{\gamma,\gamma Z, Z, W^\pm}$, $g_{1,5}^{h,\gamma,\gamma Z, Z, W^\pm}$
& $18\times 275$
& $\sim 10^{-4}-0.8$
& $1-10^4$ \\
\botrule
\end{tabular*}

\end{table}
%-------------------------------------------------------------------------------

%-------------------------------------------------------------------------------
%-------------------------------------------------------------------------------

As is apparent from Table~\ref{tab:facilities} and Fig.~\ref{fig:kinematics},
the bulk of the experimental information comes from the DIS structure function
$g_1^\gamma$, or, equivalently, from the asymmetry
$A_1\approx g_1^\gamma/F_1^\gamma$. This was measured by a number of experiments,
at SLAC, by the E80~\citep{Alguard:1976bm}, E130~\citep{Baum:1983ha},
E142~\citep{E142:1996thl}, E143~\citep{E143:1998hbs},
E154~\citep{E154:1997xfa}, E155~\citep{E155:2000qdr} experiments, at CERN,
by the EMC~\citep{EuropeanMuon:1989yki},
SMC~\citep{SpinMuon:1998eqa,SpinMuon:1999udj} and
COMPASS~\citep{COMPASS:2015mhb,COMPASS:2016jwv} experiments, at DESY, by the
HERMES~\citep{HERMES:1997hjr,HERMES:2006jyl} experiment, and
at JLab, by various experiments in Hall A, B, and
C~\citep{JeffersonLabHallA:2016neg,JeffersonLabHallA:2004tea,CLAS:2014qtg,
  CLAS:2006ozz}.
As discussed in Sect.~\ref{subsec:perturbative_computations}, these
measurements give access only to the sum of polarised quarks and antiquarks.
Because some measurements use proton targets and others use deuteron or neutron
targets, it is possible to disentangle the sums of up and down 
polarised quarks and antiquarks, using isospin symmetry to relate the PDFs in
the proton to those in the neutron. They do not, however, separate quarks from
antiquarks of the same flavour. Sensitivity to the strange-quark and gluon
distributions is also limited. Strange quarks and antiquarks contribute only a
small fraction of the cross section, so their helicity distribution is often
constrained by imposing the theoretical relation in Eq.~\eqref{eq:pol_moments}.
The gluon enters the structure function either as a
higher-order contribution to the DIS coefficient functions or through DGLAP
evolution. In both cases, sensitivity to the gluon is suppressed by a power of
the strong coupling.

Other measurements may overcome these limitations. On the one hand,
measurements of SIDIS structure functions, performed by
COMPASS~\citep{COMPASS:2010hwr} and HERMES~\citep{HERMES:2018awh} experiments,
and of gauge boson production asymmetries in polarised $pp$ collisions,
performed by the STAR~\citep{STAR:2014afm,STAR:2018fty} and
PHENIX~\citep{PHENIX:2010aru,PHENIX:2015ade} experiments, provide access to
separate quark and antiquark distributions, including strange quarks and
antiquarks (in particular when measuring kaons in the SIDIS final state).
Particle production in SIDIS, however, requires knowledge of FFs and
therefore introduces an additional source of nonperturbative uncertainty,
whereas gauge boson production in polarised $pp$ collisions only
requires knowledge of polarised and unpolarised PDFs. On the other hand,
measurements of single-inclusive jet, dijet, and single-inclusive hadron
production in polarised $pp$ collisions, performed by the
STAR~\citep{STAR:2012hth,STAR:2014wox,STAR:2016kpm,STAR:2019yqm,STAR:2021mqa,
  STAR:2021mfd,STAR:2018iyz}
and PHENIX~\citep{PHENIX:2007kqm,PHENIX:2008swq,PHENIX:2008sgl,PHENIX:2014gbf,
  PHENIX:2015fxo} experiments,
provide direct access to the gluon PDF, which enters
the corresponding asymmetries as a LO contribution, see
Table~\ref{tab:factorisation}. Although less numerous, these measurements are
an essential complement to the NC, photon-mediated DIS
structure functions.

Overall, the available measurements amount to roughly one thousand data points,
but they cover only a limited area in the $(x,\mu^2)$
plane. In particular, no measurements exist below $x\sim 0.004$, a fact that
renders the determination of polarised PDFs at small values of $x$ particularly
difficult. Furthermore, DIS and SIDIS measurements lie at relatively small
values of $Q^2$ and at relatively large values of $x$. Kinematic cuts are
typically imposed on these data to ensure the validity of
the factorisation framework, thereby further limiting the amount of experimental
information included. To ensure the applicability of perturbative QCD, only
measurements with $Q^2>1$~GeV$^2$ are usually considered. Because leading-twist
factorisation formul\ae~are used, see
Sect.~\ref{subsec:perturbative_computations}, a lower cut of a few GeV$^2$ on
the final-state invariant mass squared $W^2=M^2 + (1-x)/x Q^2$ (where $M$ is the
proton mass) is also typically imposed
to limit contamination from higher-twist corrections in the
computation of the physical observables. 

In the future, the EIC~\citep{AbdulKhalek:2021gbh} will measure a broad set of
DIS and SIDIS structure functions, including electroweak ones, over an extended
kinematic range, see Table~\ref{tab:facilities} and Fig.~\ref{fig:kinematics}.
This programme will allow quark and antiquark flavours to be mapped precisely,
in particular through parity-violating structure functions, and will constrain
the gluon PDF through the extended lever arm in $Q^2$ evolution. Until then,
polarised PDF determinations will remain limited by the available
experimental information.

\section{Methodological input}
\label{sec:methodological_input}

Determining (polarised) PDFs through a global analysis that compares the
theoretical predictions of Sect.~\ref{sec:theoretical_input} to the
experimental measurements of Sect.~\ref{sec:experimental_input} is, in
statistical terms, an inverse problem~\citep{Stuart_2010}. In this section,
the problem is formulated in Bayesian statistical language, together with the
standard treatment of several aspects of its solution, including
parametrisation, optimisation, uncertainty representation, and validation.

\subsection{Bayesian inference}
\label{sec:Bayesian_inference}

The goal of PDF determination is to infer a set of functions with
support in $x\in [0,1]$
\begin{equation}
  \Delta f(x,\mu^2), \qquad f=g,u,\bar u,d,\bar d,s,\bar s,\dots
\end{equation}
that, through the forward map provided by QCD factorisation of physical
observables,
Eqs.~\eqref{eq:g_DIS_factorisation}--\eqref{eq:unp_pp_factorisation_h},
give rise to the measured data. The problem is ill-posed because PDFs are
infinite-dimensional objects, whereas the number of data points is finite.
The standard practical solution to this problem is to approximate PDFs in terms
of a finite-dimensional parametrisation,
\begin{equation}
  \Delta f(x,\mu_0^2) \longrightarrow \Delta f(x,\mu_0^2;\theta)\,
\end{equation}
where $\theta$ denotes the set of parameters. Note that PDFs are typically
parametrised at an initial scale $\mu_0^2$, the dependence at higher scales
being uniquely determined by DGLAP evolution equations.

The inverse problem is addressed by inferring the parameters $\theta$ from the
data $D$. In Bayesian statistical language, this amounts to determining the
posterior probability density $p(\theta\mid D)$, obtained from Bayes' theorem
\begin{equation}
  p(\theta | D) = \frac{p(\theta)p(D | \theta)}{p(D)}\,,
\end{equation}
where $p(\theta)$ is the prior probability density, $p(D\mid\theta)$ is the
likelihood, and $p(D)$ is the evidence, or marginalised likelihood,
\begin{equation}
  p(D)=\int d\theta p(D | \theta) p(\theta)\,.
\end{equation}
The posterior distribution $p(\theta\mid D)$ should be understood as a
finite-dimensional representation of the more formal posterior distribution
$p(\Delta f\mid D)$ in function space. The prior $p(\theta)$ encodes
assumptions that make the reconstruction well defined: smoothness, endpoint
behaviour, integrability, positivity and sum rules, as discussed in
Sect.~\ref{subsec:theoretical_constraints}. The likelihood encodes the
agreement between theoretical predictions and data, including correlations. 
A point estimate of $\theta$ can be obtained by maximising the posterior
probability. This defines the maximum a posteriori (MAP) estimator
\begin{equation}
  \theta_{\rm MAP}
  = \operatorname*{arg\,max}_{\theta} p(\theta|D)
  = \operatorname*{arg\,max}_{\theta} p(D|\theta)p(\theta)\,,
  \label{eq:MAP_estimator}
\end{equation}
where the evidence $p(D)$ has been omitted in the last equality because it is
independent of $\theta$. Equivalently, the MAP estimator minimises the
negative log-posterior,
\begin{equation}
  \theta_{\rm MAP}
  =
  \operatorname*{arg\,min}_{\theta}
  \left[
    -\log p(D|\theta) - \log p(\theta)
  \right]\,.
  \label{eq:MAPmin}
\end{equation}
This expression makes explicit the relation between Bayesian inference and
regularised optimisation: the likelihood measures the agreement between the
model prediction and the data, while the prior acts as a regularising term.
Finally, given a physical observable $\mathcal{O}$ that depends on the PDF
parameters $\theta$, its posterior expectation value and variance are
\begin{align}
  \mathbb{E}[\mathcal{O}\mid D]
  &=
  \int_{\Theta} d^n\theta\,
  p(\theta\mid D)\,
  \mathcal{O}(\theta)\,,
  \nonumber\\
  \mathbb{V}[\mathcal{O}\mid D]
  &=
  \int_{\Theta} d^n\theta\,
  p(\theta\mid D)
  \left[
    \mathcal{O}(\theta)
    -
    \mathbb{E}[\mathcal{O}\mid D]
  \right]^2\,,
  \label{eq:posterior_moments}
\end{align}
where the integration extends over the full parameter space $\Theta$.
The MAP estimator $\theta_{\rm MAP}$ defined in
Eq.~\eqref{eq:MAP_estimator} is instead a single point estimate. The
corresponding prediction $\mathcal{O}(\theta_{\rm MAP})$ does not in general
coincide with the posterior expectation value $\mathbb{E}[\mathcal{O}\mid D]$
and, by itself, does not quantify the uncertainty on $\mathcal{O}$. Although
this general strategy appears relatively simple in principle, its
practical realisation entails a number of choices concerning the
parametrisation, optimisation, uncertainty representation, and validation
that are not merely technical steps, but an essential part of the definition of
the inference problem. Each of these aspects is discussed next.

\subsection{Parametrisation}
\label{subsec:parametrisation}

Polarised proton PDFs are usually parametrised as
\begin{equation}
  \Delta f (x,\mu_0^2)
  =
  \mathcal{N} x^{\alpha_f}(1-x)^{\beta_f}\mathcal{I}(x; \theta)\,,
  \label{eq:parametrisation}
\end{equation}
at an input scale $\mu_0^2$. The power-law factors $x^{\alpha_f}$ and
$(1-x)^{\beta_f}$ describe the small- and large-$x$ behaviour of PDFs close to
the endpoints $x\to 0$ and $x\to 1$, and are motivated by general QCD
arguments: Regge theory~\citep{Regge:1959mz} and Brodsky--Farrar counting
rules~\citep{Brodsky:1973kr}. While various models can predict approximate
values for $\alpha_f$ and $\beta_f$~\citep{Nocera:2014uea}, they are treated
as part of the prior or as additional parameters fitted to the data in global
QCD analyses. The factor $\mathcal{N}$ is an overall normalisation used to
implement sum rules, Eq.~\eqref{eq:pol_moments}, when they are imposed. 
The function $\mathcal{I}(x;\theta)$ provides an interpolation between the
small- and large-$x$ regions. Typically, it is chosen to have a simple
polynomial form,
{\it e.g.}~$\mathcal{I}(x;\theta)=(1+\gamma\sqrt x + \delta x + \dots)$, where
$\theta=(\gamma, \delta, \dots)$ denotes parameters to be determined by the fit.
The function $\mathcal{I}(x;\theta)$ can also be represented by the output of
a neural network, thereby reducing dependence on a specific functional form. In
this context, neural networks act as flexible function parametrisations.

The input scale $\mu_0^2$ is chosen at the lowest possible value that
renders perturbative QCD reliable, typically $\mu_0^2=1$~GeV$^2$. The number of
independent PDFs that can be parametrised depends on the
measurements included in the fit. As discussed in
Sect.~\ref{sec:experimental_input}, photon-mediated NC DIS structure functions
on proton targets directly constrain one charge-weighted combination
of quark and antiquark helicity distributions, with indirect sensitivity
to the gluon distribution through scaling violations and higher-order
coefficient functions. The availability of deuteron and neutron
measurements provides additional independent flavour combinations,
while sum rules supply moment constraints involving the strange distribution.
Inclusion of SIDIS and electroweak measurements in $pp$ collisions finally
extends the parametrisation basis to six independent distributions, because of
their sensitivity to up- and down-quark and antiquark distributions separately.
A typical set of fitted functions is therefore
$\Delta f = \Delta u, \Delta\bar u, \Delta d, \Delta\bar d, \Delta s+\Delta\bar s, \Delta g$,
which may be recast in the so-called evolution basis, in terms of singlet
and non-singlet distributions that simplify the DGLAP evolution equations.
The available measurements allow, in principle, a separation of
$\Delta s$ and $\Delta\bar s$ distributions, which also evolve differently at
NNLO and beyond. However, they are typically required to be the same at the
initial scale $\mu_0$, because the measurements are practically insensitive to
any asymmetry between the two. Polarised PDFs for heavy quarks are generated
perturbatively through gluon splitting and are therefore not parametrised. A
nonperturbative (intrinsic) component is not forbidden by theory; however, it
is too small to be detected with currently available measurements.

\subsection{Optimisation}
\label{subsec:optimisation}

The optimal set of parameters $\theta$ is obtained by minimising the negative
log-posterior in Eq.~\eqref{eq:MAPmin}. For a Gaussian likelihood and a flat
prior in the fitted region, this reduces, up to an additive constant, to
minimising a $\chi^2$. Given a set of  $N_{\rm dat}$ measurements
$D_i$, and a set of corresponding theoretical  predictions $T_i(\theta)$,
the $\chi^2$ is defined as
\begin{equation}
  \chi^2({\theta})
  = 
  \sum_{i=1}^{N_{\rm dat}}\sum_{j=1}^{N_{\rm dat}} 
  \left[D_i-T_i({\theta})\right] 
       {\rm cov}_{ij}^{-1} 
       \left[D_j-T_j({\theta})
  \right]\,,
  \label{eq:chi2cov}
\end{equation}
where the covariance matrix ${\rm cov}_{ij}$ is the sum of the experimental,
${\rm cov}_{ij}^{\rm exp}$, and theoretical, ${\rm cov}_{ij}^{\rm th}$, covariance
matrices
\begin{equation}
  {\rm cov}_{ij} = {\rm cov}_{ij}^{\rm exp} + {\rm cov}_{ij}^{\rm th}\,.
\end{equation}
The experimental covariance matrix may be written as
\begin{equation}
  {\rm cov}_{ij}^{\rm exp}
  =
  \delta_{ij}\sigma_i^{\rm unc}\sigma_j^{\rm unc}
  +
  \sum_{k=1}^{n_{\rm corr}}\sigma_{k,i}^{\rm corr}\sigma_{k,j}^{\rm corr}
  \,,
  \label{eq:covariance}
\end{equation}
in terms of uncorrelated ($\sigma_i^{\rm unc}$) and $n_{\rm corr}$ correlated
($\sigma_{k,i}^{\rm corr}$) uncertainties.
If $\mathrm{cov}^{\rm th}=0$, or if the theoretical covariance is
represented by an additional set of nuisance parameters, the $\chi^2$
can equivalently be written as~\citep{Pumplin:2002vw}
\begin{equation}
  \chi^2({\theta})
  = 
  \sum_{i=1}^{N_{\rm dat}} \left(\frac{D_i 
    + 
    \sum_{j=1}^{n_{\rm corr}} r_j \sigma_{j,i}^{\rm corr}
    - T_i(\theta)}{\sigma_i^{\rm unc}}\right)^2
  + 
  \sum_{j=1}^{n_{\rm corr}} r_j^2
  \,,
  \label{eq:chi2penalty}
\end{equation}
which becomes equivalent to Eq.~\eqref{eq:chi2cov} after minimisation with
respect to the shift parameters, $r_j$. This also allows one to study the 
behaviour of the shift parameters at the minimum, where
their distribution is expected to be a univariate Gaussian with mean zero.

When correlated multiplicative uncertainties are present ({\it e.g.}~a
luminosity normalisation uncertainty), 
they must be treated carefully in order to avoid the so-called 
d'Agostini bias~\citep{DAgostini:1993arp}, which biases the fitted theoretical
prediction downwards. 
To avoid this, multiplicative uncertainties in Eq.~\eqref{eq:covariance}
can be treated iteratively by multiplying them by central theory 
predictions from a previous fit instead of by the experimental data. This
procedure, called the $t_0$-method~\citep{Ball:2009qv}, has been shown to
converge rapidly. Alternatively, one can fit an overall normalisation parameter 
and allow it to fluctuate within the multiplicative uncertainty range. 
This can be included naturally in Eq.~\eqref{eq:chi2penalty} as one of the
$r_j$  parameters. 

The optimisation of Eq.~\eqref{eq:chi2cov} is performed numerically, 
which means DGLAP evolution and convolutions between PDFs and 
partonic cross sections must be evaluated both rapidly and accurately to make
global fits viable. Several publicly available codes solve
DGLAP evolution equations efficiently~\citep{Vogt:2004ns,Salam:2008qg,
  Botje:2010ay,Bertone:2013vaa,Bertone:2015cwa,Candido:2022tld}.
These programs, as well as most private evolution codes used in PDF fits, have
been benchmarked against standard tables~\citep{Dittmar:2005ed,Hekhorn:2024tqm}.
The calculation of hadronic cross sections is performed using lookup
tables, where partonic cross sections convolved with evolution kernels are
precomputed and stored for each data set on a suitable interpolation grid.
Hadronic cross sections then reduce to the scalar product between such
interpolation tables and the PDFs parametrised at the scale $\mu_0^2$.
This method is usually implemented within private optimisation codes
(see, for example, Ref.~\citep{Stratmann:2001pb}). The {\sc PineAPPL}
program~\citep{Carrazza:2020gss,Jezo:2026adf,christopher_schwan_2025_15635174}
provides a public implementation. PineAPPL has been interfaced to
{\sc Yadism}~\citep{Candido:2024rkr} for the computation of polarised DIS
structure functions and to codes based on~\citep{Jager:2002xm,Jager:2004jh,
  Boughezal:2021wjw}
for single-inclusive-hadron, jet/dijet, and gauge-boson production spin
asymmetries in $pp$ collisions.

The choice of an efficient optimisation strategy is crucial for an accurate
exploration of the parameter space, whose effectiveness depends strongly on
its dimension. Numerical gradient-based algorithms such as Newton's method, the
Levenberg--Marquardt method (which supplements Newton's method with a
steepest-descent component), or variable metric methods
(which only rely on gradient information) are used for spaces of 
moderate dimension (roughly on the order of 40 free parameters). If the 
dimension of the PDF parameter space is larger, these methods start to become
unsuitable due to inefficient numerical inversions of large matrices 
and the increased possibility of ending in a local minimum. Therefore, they
are appropriate only if polarised PDFs are parametrised in terms of relatively
simple functional forms. On the other hand, stochastic genetic
algorithms~\citep{Ball:2013lla,NNPDF:2014otw} or deterministic 
gradient descent methods~\citep{Carrazza:2019mzf,Bertone:2024taw,
  Cruz-Martinez:2025ahf} can be used to efficiently explore the parameter
space if polarised PDFs are parametrised using neural networks.
To avoid fitting the noise in the data, it is important to devise a suitable
stopping criterion. In this respect, the standard method in the literature 
is cross-validation, where the data points are randomly divided into two sets:
training and validation.  The $\chi^2$ is then computed for both sets
separately, but optimised only on the training set. The fit terminates when
the $\chi^2$ of the validation set starts to increase (while the $\chi^2$ of
the training set continues to decrease). To avoid information loss, the
procedure should be repeated a sufficiently large number of times starting from
different data partitions.

When neural networks are used, parametrisation and optimisation entail
non-trivial choices that may bias the posterior
distribution of the parameters $\theta$. One can therefore scan a range of
sensible choices, each of which represents a different methodology, and
select those that lead to the best optimisation performance. The optimum is
often not unique, because several choices may lead to equivalent
performance. This process is called hyperparameter
optimisation~\citep{Carrazza:2021yrg,Cruz-Martinez:2024wiu,
  Cruz-Martinez:2026aqe,Cruz-Martinez:2025ahf} and consists of an automated
scan of
the space of models, which are appraised according to a properly defined figure
of merit (see~\citep{Cruz-Martinez:2025ahf} for details). To ensure that the
optimal model does not lead to overfitted PDFs, a $k$-fold partition of the
data set is used to assess each model on subsets of data excluded in turn from
the fit. The hyperparameters, folds, and figure of merit must be selected case
by case.

\subsection{Uncertainty representation}
\label{subsec:uncertainty_representation}

The posterior probability distribution $p(\theta\mid D)$ contains the
uncertainty on the fitted parameters and, through the PDF parametrisation, on
the PDFs and any observable computed from them. This uncertainty may be
summarised by posterior moments or by credible regions. Credible
regions depend on the quantification of uncertainties, which are grouped below
into three categories: experimental, procedural, and theoretical uncertainties.

\begin{description}

\item[Experimental uncertainties.]
  Modern determinations of polarised PDFs propagate experimental
  uncertainties by means of a Monte Carlo replica method. Starting from the
  measured data vector $D$ and its experimental covariance matrix
  $\operatorname{cov}^{\rm exp}$, one generates $N_{\rm rep}$ artificial data
  replicas, $D^{(k)}\sim\mathcal{N}(D,\operatorname{cov}^{\rm exp})$,
  $k=1,\ldots,N_{\rm rep}$. Each replica is fitted independently,
  yielding a set of best-fit or MAP parameters $\widehat{\theta}^{(k)}$. The
  resulting ensemble defines the empirical distribution
  \begin{equation}
    \widehat p_{\rm rep}(\theta\mid D)
    =
    \frac{1}{N_{\rm rep}}
    \sum_{k=1}^{N_{\rm rep}}
    \delta^{(n)}
    (\theta-\widehat{\theta}^{(k)})\,.
    \label{eq:replica_distribution}
  \end{equation}
  This construction propagates fluctuations of the experimental data through
  the generally nonlinear fitting procedure and into the fitted PDFs. The
  replica distribution $\widehat p_{\rm rep}(\theta\mid D)$ should not, in
  general, be identified automatically with the Bayesian posterior
  $p(\theta\mid D)$. It is more directly interpreted as an empirical
  distribution of fitted estimators induced by repeated fluctuations of the
  data, together with any additional randomness introduced by the
  parametrisation and optimisation procedure. Its interpretation as an
  approximation to the posterior requires additional assumptions concerning the
  likelihood, the prior, the generation of the replicas, and the fitting
  prescription. By contrast, an ensemble generated by drawing $\theta^{(k)}$
  directly from $p(\theta\mid D)$, for example with a posterior-sampling
  algorithm, is by construction a sample from the posterior.
  The central value and uncertainty of an observable $\mathcal{O}$ can be
  estimated from the replica ensemble as
  \begin{equation}
    {\mathbb E}_{\rm rep}[\mathcal{O}]
    =
    \frac{1}{N_{\rm rep}}
    \sum_{k=1}^{N_{\rm rep}}
    \mathcal{O}
    \left(
    \widehat{\theta}^{(k)}
    \right)\,,
    \qquad
    {\mathbb V}_{\rm rep}[\mathcal{O}]
    =
    \frac{1}{N_{\rm rep}-1}
    \sum_{k=1}^{N_{\rm rep}}
    \left[
      \mathcal{O}
      \left(
      \widehat{\theta}^{(k)}
      \right)
      -
      {\mathbb E}_{\rm rep}[\mathcal{O}]
      \right]^2\,.
    \label{eq:replica_moments}
  \end{equation}
  If the ensemble can be regarded as a
  sample from the posterior, these estimators approximate the posterior moments
  in Eq.~\eqref{eq:posterior_moments}. More generally, they characterise the
  empirical distribution generated by the replica fitting procedure.

  Although computationally expensive, the Monte Carlo method is flexible: data
  can be sampled from non-Gaussian distributions when needed; no linear
  approximation of the observables is required; the method does not rely on a
  low-dimensional parameter space; and, when the PDF ensemble admits a
  probabilistic interpretation, it can be updated after the inclusion of new
  data without repeating the full fit. This is achieved through Bayesian
  reweighting~\citep{Ball:2010gb,Ball:2011gg}, in which each replica is assigned
  a weight proportional to the likelihood of the new data.
  The averages in Eq.~\eqref{eq:replica_moments} are then replaced by weighted
  averages. If only a small number of replicas acquire appreciable weight, the
  effective statistical size of the ensemble is reduced and a new fit may be
  required. The Bayesian interpretation of the update relies on the ensemble
  being an adequate representation of the probability distribution that is
  being reweighted. Since replicas with small weights become
  immaterial, the PDF ensemble loses part of its statistical power, a drawback
  of the method when such a loss is excessive.
  
\item[Procedural uncertainties.] There are three classes of procedural
  uncertainties, all connected to optimisation and naturally incorporated in
  $p({\theta} | D)$. The first is the methodological uncertainty associated
  with choices made in a fit, namely the basis functions, the functional form,
  the number of parameters, and the optimisation strategy. The second is the
  extrapolation uncertainty arising because the data, even if
  infinitely precise, do not cover the entire phase space. The third is the
  functional uncertainty arising because a set of functions, which are
  infinite-dimensional objects, is inferred from a finite amount of information.
  Methodological uncertainty can be reduced through hyperparameter optimisation,
  while interpolation and extrapolation uncertainties can be reduced by
  increasing the precision and kinematic coverage of the measurements.
  Functional uncertainty is instead intrinsic to the inverse problem. 

\item[Theoretical uncertainties.] Theoretical uncertainties arise from
  assumptions made in the computation of hadronic observables and are more
  challenging to propagate into PDF uncertainties.
  These include missing higher-order uncertainties (MHOU) due to the truncation
  of the perturbative expansion in the strong coupling to a given order,
  uncertainties in the input values of the physical parameters such as the
  strong coupling and the heavy-quark masses, uncertainties due to the neglect
  of power-suppressed corrections when they are omitted from the factorisation
  formul\ae, and uncertainties due to nuclear corrections when data on nuclear
  targets are used in fits of proton PDFs. The first is usually estimated via
  scale variations. The second is accounted for by performing different fits
  with varied values of the input physical parameters and combining the results.
  Power-suppressed effects can usually be kept under control by removing data
  points from the fit that are particularly sensitive to power-suppressed
  corrections and by studying the stability of the fit under variations of this
  cut-off~\citep{Simolo:2006iw}. Alternatively, power
  corrections can be modelled and fitted along with PDFs~\citep{Sato:2016tuz},
  as can nuclear corrections~\citep{Sato:2016tuz}; their uncertainties can then
  be estimated from the fitted variations.

  A general procedure for representing theoretical uncertainties in PDFs has
  recently been proposed in the framework of the Monte Carlo
  method~\citep{Ball:2018lag}. Assuming a Gaussian model for these
  uncertainties, it follows from
  Bayes' theorem that they can be included by redefining the covariance matrix
  in Eq.~\eqref{eq:chi2cov} as the sum of an experimental and a theoretical
  part, ${\rm cov}_{ij}={\rm cov}_{ij}^{\rm exp}+{\rm cov}_{ij}^{\rm th}$.
  The resulting $\chi^2$ is then used both in the sampling of the data replicas
  and in the optimisation of the fit. Propagating theory uncertainties into PDFs
  is therefore reduced to estimating the theoretical
  covariance matrix ${\rm cov}_{ij}^{\rm th}$ through a model of correlated theory
  shifts.
  For MHOU, estimates have been formulated by defining the matrix elements of 
  ${\rm cov}_{ij}^{\rm th}$ from differences between theoretical predictions
  obtained with central and
  varied factorisation and renormalisation scales according to various
  prescriptions~\citep{NNPDF:2019vjt,NNPDF:2024dpb,
    Cruz-Martinez:2025ahf}. In these studies, correlations across data points
  induced by the structure of higher-order corrections in the partonic cross
  sections and splitting functions were treated consistently.
  At NLO, MHOU were validated against the exact NNLO result, to test
  the validity of the Gaussian hypothesis. The inclusion of such theoretical
  uncertainties improves the description of the data, shifts the central value
  of the PDFs towards the higher-order result, and slightly increases their
  uncertainties. A similar strategy can be used to account for other sources of
  theoretical uncertainties, {\it e.g.}~nuclear and higher-twist uncertainties.
  Such an exercise was performed for unpolarised PDFs~\citep{Ball:2018twp,
    Ball:2020xqw,Ball:2025xtj}, though not yet for polarised PDFs. Extending
  this strategy to the polarised case is conceptually straightforward,
  although its numerical impact may be smaller because many of these
  uncertainties partially cancel when measuring spin asymmetries.

\end{description}

\subsection{Uncertainty validation}
\label{subsec:unc_validation}

The Monte Carlo method allows for a rigorous statistical validation of PDF
uncertainties. This validation can be performed by means of closure
tests~\citep{DelDebbio:2021whr}, where one assumes that the underlying PDFs
are known by using a specific (previously determined) PDF set to generate
artificial data. Fits are then performed to check whether the output PDFs
provide consistent and unbiased estimators of the underlying truth.
Closure tests have been extensively performed for unpolarised
PDFs~\citep{NNPDF:2014otw,NNPDF:2021njg,Harland-Lang:2024kvt}, though not yet
for polarised PDFs. Nevertheless, it is useful to describe the three types of
closure tests, denoted as levels, because they may find application to
polarised PDFs in the future. They are summarised below.

\begin{description}

\item[Level-0.] A level-0 closure test consists of generating unfluctuated
  artificial data with the same covariance as the real measurements. Their
  central value is therefore perfectly described by the
  underlying PDF. In this case, there are infinitely many PDF replicas
  (differing only in the initialisation of the parameters $\theta$) that fit
  the data equally well. If the methodology is sufficiently flexible and the
  optimisation is successful, by construction all
  replicas should fit the data perfectly, with vanishing $\chi^2$ in the limit
  of infinite training. The residual uncertainty is an interpolation and
  extrapolation uncertainty resulting from the fact that the available
  measurements cover a restricted kinematic region. A level-0 closure test
  therefore serves two purposes: the first is to validate the methodology, by
  effectively ensuring that it leads to PDF replicas with
  vanishing $\chi^2$; the second is to characterise the size of interpolation
  and extrapolation uncertainty.

\item[Level-1.] A level-1 closure test is equivalent to a level-0 test, except
  that the central values of the artificial data are fluctuated according to the
  experimental covariance matrix. Infinitely many PDF replicas can again fit
  the data equally well, and the expected $\chi^2$ per data point is of order
  unity. The residual uncertainty is irreducible and represents the functional
  uncertainty intrinsic to the inverse problem. A level-1 closure test therefore
  characterises this uncertainty.

\item[Level-2.] A level-2 closure test is equivalent to a level-1 test, except
  that each replica is fitted to a different fluctuation of the level-1
  artificial data. The Monte Carlo propagation of data uncertainties into PDF
  uncertainties is thereby incorporated into the closure test. One expects an
  ensemble of PDF replicas that describe the data equally well, with a $\chi^2$
  per data point of order unity. A level-2 closure test first allows the
  experimental component of the uncertainty
  to be assessed relative to the interpolation, extrapolation, and functional
  components, by comparison with level-0 and level-1 tests. On the other hand,
  one may want to perform a so-called multi-closure test, in which a level-2
  closure test is run a large number of times, each time starting from a
  different level-1 fluctuation of the artificial data. In this case, one can
  reconstruct the distribution of the bias (the distance of the central value
  of each level-2 closure test from the truth) and the variance (the
  dispersion of all level-2 closure tests around their average). Two properties
  can then be tested quantitatively. First, statistical faithfulness requires
  the underlying truth to lie within the nominal 68\% uncertainty interval of
  the fitted ensemble in 68\% of repeated closure tests. This property can be
  assessed directly from the empirical coverage over the ensemble of level-2
  tests. Second, the complexity of the methodology should be chosen so that
  uncertainties are neither underestimated nor overestimated. This
  fact can be checked by computing the ratio of the squared bias to the
  variance, which should be close to unity.

\end{description}
  
\section{State of the art}
\label{sec:state_of_the_art}

As is clear from the discussion in
Sects.~\ref{sec:theoretical_input}--\ref{sec:methodological_input}, determining
polarised PDFs is a particularly involved problem. Several collaborations have
produced and updated global analyses over more than two decades. This section
presents the most recent polarised PDF sets and assesses current knowledge of
the quark and gluon distributions and their moments. It also discusses the
implications for the proton-spin decomposition in Eq.~\eqref{eq:total_orbital}
and the expected impact of future EIC measurements.

\subsection{Parton distributions}
\label{subsec:parton_distributions}

The first determinations of polarised PDFs based on a global QCD analysis of
experimental data were performed in the mid-1990s at LO and NLO
accuracy~\citep{Gehrmann:1994rb,Gehrmann:1995ag,Gluck:1995yq,Gluck:1995yr}.
At that time, only the available measurements of polarised NC DIS
asymmetries and structure functions were analysed; therefore, only valence and
sea quark PDFs were determined. Because of the weak sensitivity of the data to
the polarised gluon PDF, different forms of the fitted distributions were
explored; their spread revealed that the gluon PDF was essentially
unconstrained. For almost two decades, determinations of helicity PDFs were
based predominantly on inclusive NC DIS data, with updates in the 
included data sets as the SLAC and CERN polarised DIS programmes progressed,
together with methodological improvements in the analyses~\citep{Leader:1998qv,
  Leader:2001kh,Leader:2005ci,Leader:2006xc,
  AsymmetryAnalysis:1999gsr,Hirai:2003pm,Blumlein:2002qeu,Blumlein:2010rn}.
More recent DIS-only determinations introduced neural-network Monte Carlo
methods~\citep{Ball:2013lla}, a simultaneous treatment of leading- and
higher-twist contributions~\citep{Sato:2016tuz}, and, eventually, NNLO
perturbative accuracy~\citep{Taghavi-Shahri:2016idw}. They improved the
determination of PDF uncertainties but confirmed that the
polarised gluon PDF remained essentially unconstrained.

The extension from inclusive-DIS-only determinations to genuinely
multi-process analyses occurred progressively. A significant change was
brought about by the first DSSV
determination~\citep{deFlorian:2008mr,deFlorian:2009vb}, which combined, for
the first time within a consistent NLO framework, inclusive DIS, SIDIS, and
longitudinal-spin asymmetries for hadron and jet production in polarised
$pp$ collisions. The inclusion of SIDIS data enabled a separation
of quark and antiquark helicity distributions, albeit through the additional
nonperturbative input provided by FFs, while the RHIC measurements supplied
direct constraints on the gluon helicity distribution.
Contemporaneously, the AAC08 analysis~\citep{Hirai:2008aj} combined inclusive
DIS measurements with the PHENIX neutral-pion production asymmetry,
concentrating primarily on the impact of collider data on $\Delta f_g$ and
considering both positive and sign-changing gluon parametrisations. The
LSS10 determination~\citep{Leader:2010rb} followed a different route,
combining DIS and SIDIS data while retaining target-mass
and dynamical higher-twist corrections in the description of the inclusive
structure function $g_1$. The subsequent development of global analyses was
driven largely by the increasing precision and variety of the RHIC
measurements. The DSSV14 update~\citep{deFlorian:2014yva}, incorporating
high-statistics jet and neutral-pion data, provided the first clear evidence
for a positive gluon polarisation over the range of momentum fractions directly
probed by RHIC, although the contribution from the unmeasured small-$x$ region
remained poorly constrained. At the same time, the NNPDFpol1.1
analysis~\citep{Nocera:2014gqa} provided an
independent global determination based on inclusive DIS, COMPASS open-charm
production, and RHIC jet and $W^\pm$ asymmetries. It introduced a
neural-network parametrisation and a Monte Carlo representation of
uncertainties into a multi-process helicity-PDF analysis, and used
$W^\pm$ production rather than SIDIS for light-quark and antiquark flavour
separation, thereby avoiding dependence on light-hadron FFs.

The most recent broad global determinations have enlarged the data sets further
and improved the theoretical and statistical treatment. The BDSSV24
analysis~\citep{Borsa:2024mss} combined inclusive DIS, SIDIS, and
$W$-boson, jet, and identified-hadron production in polarised
$pp$ collisions at NNLO accuracy, using approximate higher-order
corrections where complete NNLO partonic cross sections are not available.
The NNPDFpol2.0 analysis~\citep{Cruz-Martinez:2025ahf} provided an
independent NNLO determination from inclusive DIS and RHIC $W$-boson,
single-inclusive-jet, and dijet measurements. It incorporated heavy-quark mass
corrections in DIS, missing-higher-order uncertainties through a theory
covariance matrix, and a machine-learning methodology based on neural networks,
stochastic-gradient optimisation, and hyperparameter selection.
The MAPPDFpol1.0 analysis~\citep{Bertone:2024taw} revisited the combined
analysis of DIS and SIDIS within a neural-network and Monte Carlo framework,
providing the first determination in which both processes were consistently
treated at NNLO accuracy. Finally, JAMpol25~\citep{Cocuzza:2025qvf}
brought together the different components of the JAM programme in a
comprehensive NLO analysis of inclusive and semi-inclusive DIS,
$W$-boson production, and jet production in
polarised $pp$ collisions, simultaneously determining
helicity-dependent PDFs, spin-averaged PDFs, and fragmentation
functions (see also the earlier JAM17 analysis~\citep{Ethier:2017zbq}).
This analysis incorporated recent high-$x$ JLab measurements and
treated target-mass and higher-twist corrections explicitly, thereby
extending the fitted region towards larger $x$. It also investigated the
impact of lattice-QCD information on gluonic pseudo-Ioffe-time distributions,
thereby providing a complementary constraint on the gluon helicity distribution.

\begin{figure}[!t]
  \centering
  \includegraphics[width=0.49\textwidth]{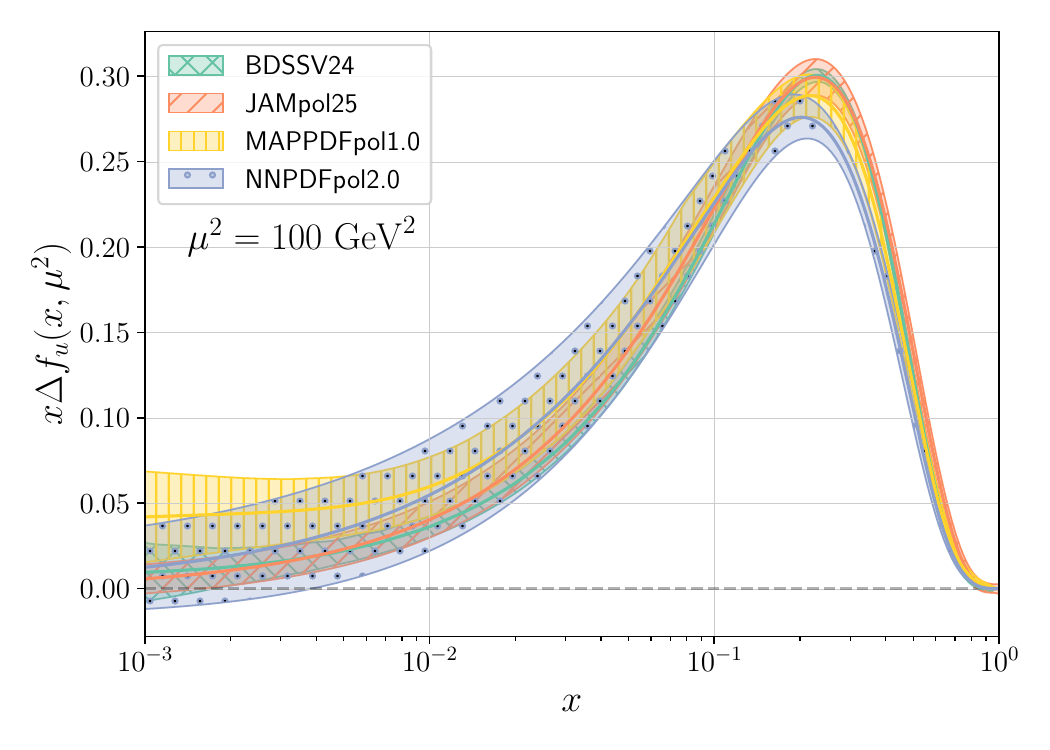}
  \includegraphics[width=0.49\textwidth]{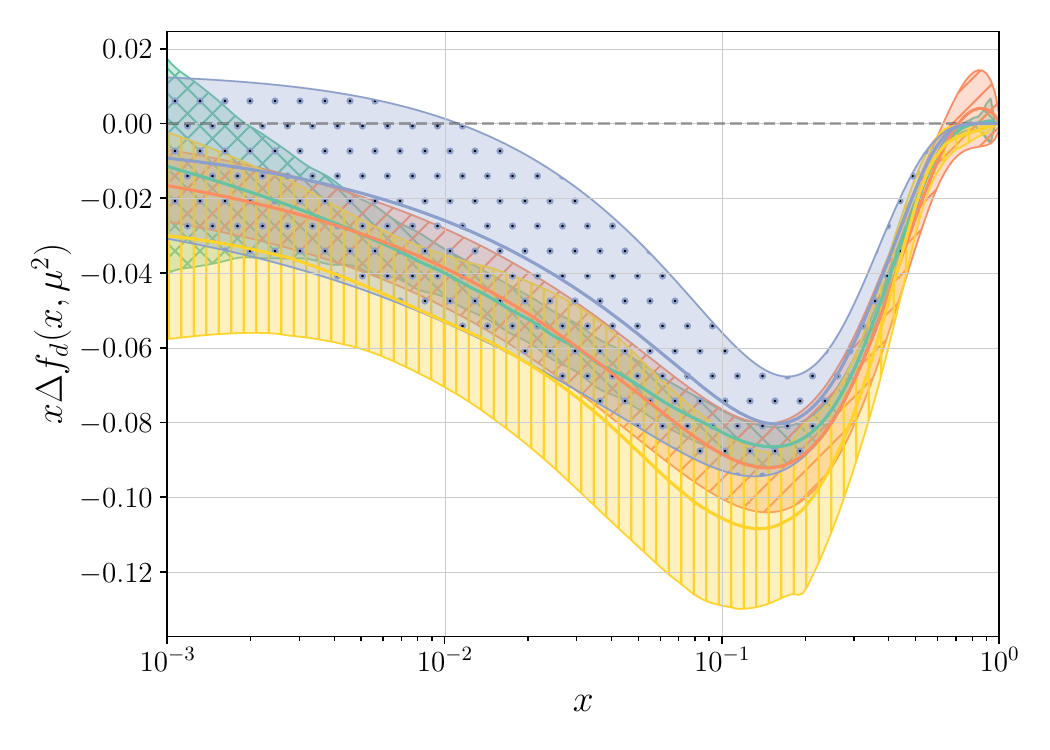}\\
  \includegraphics[width=0.49\textwidth]{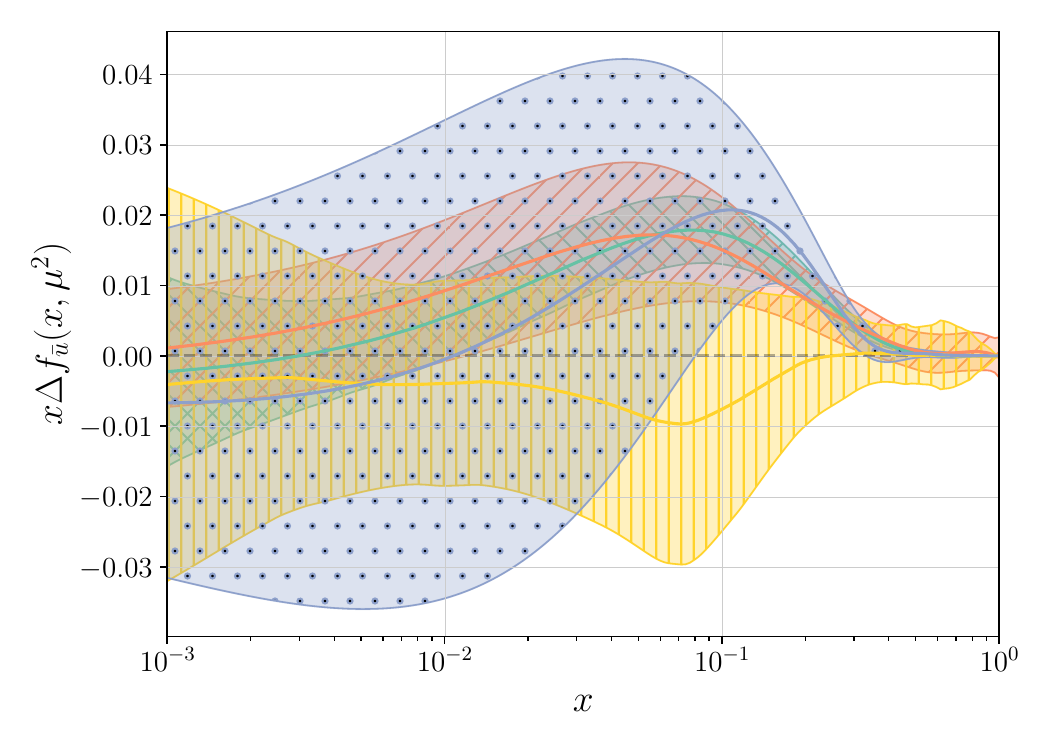}
  \includegraphics[width=0.49\textwidth]{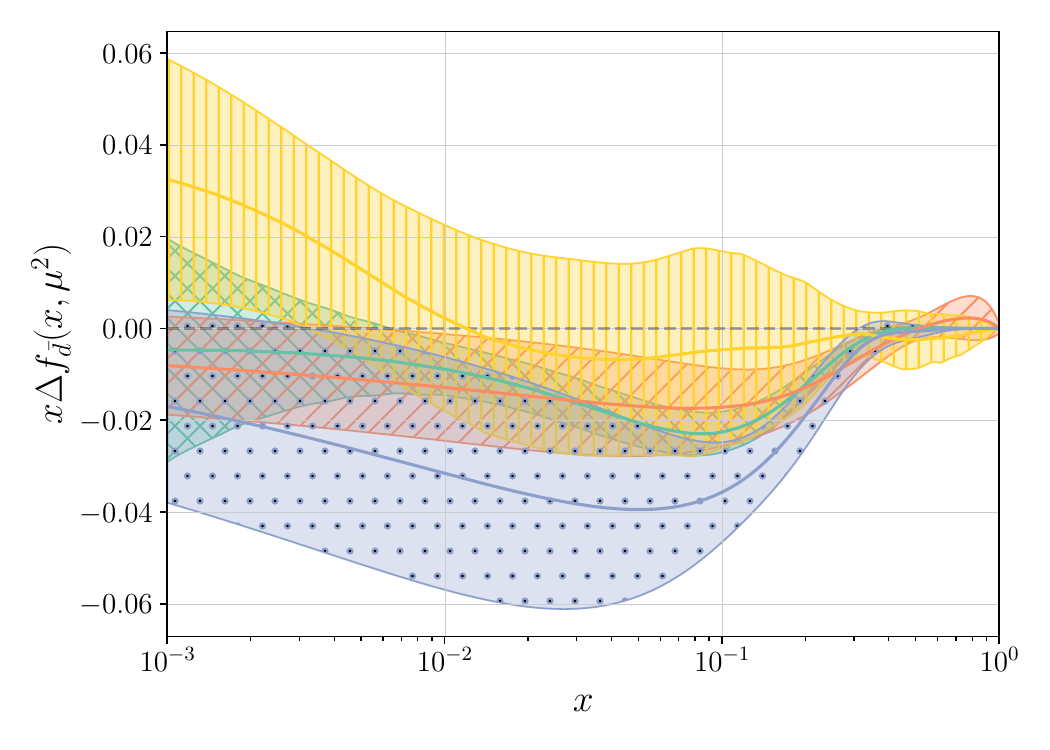}\\
  \includegraphics[width=0.49\textwidth]{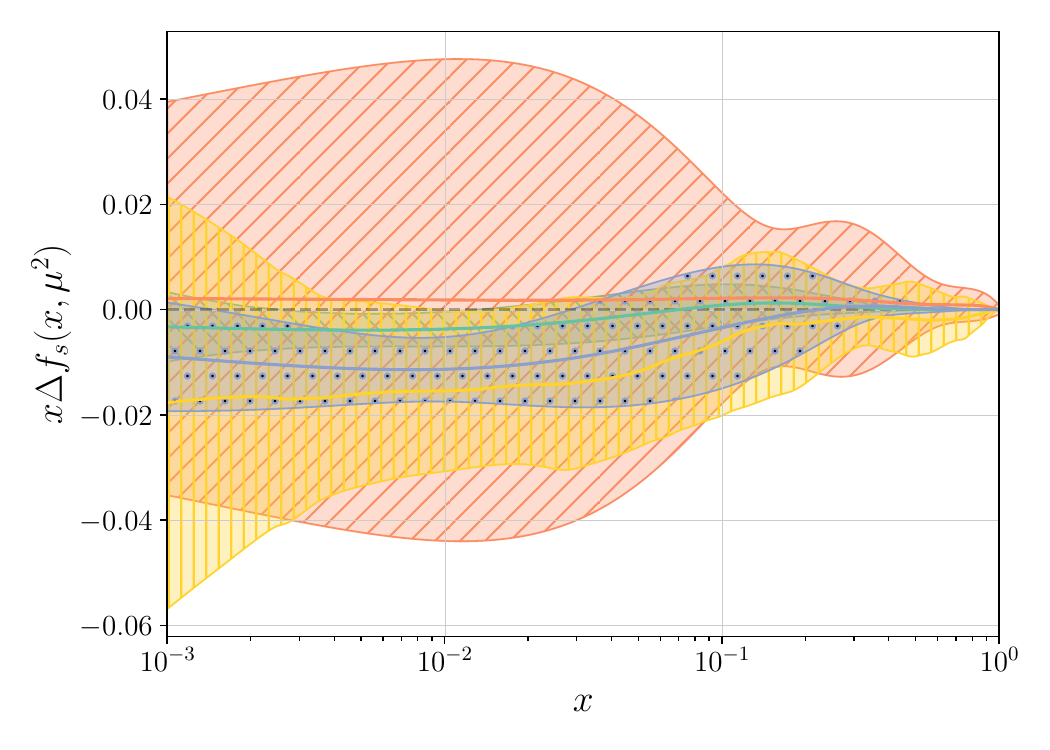}
  \includegraphics[width=0.49\textwidth]{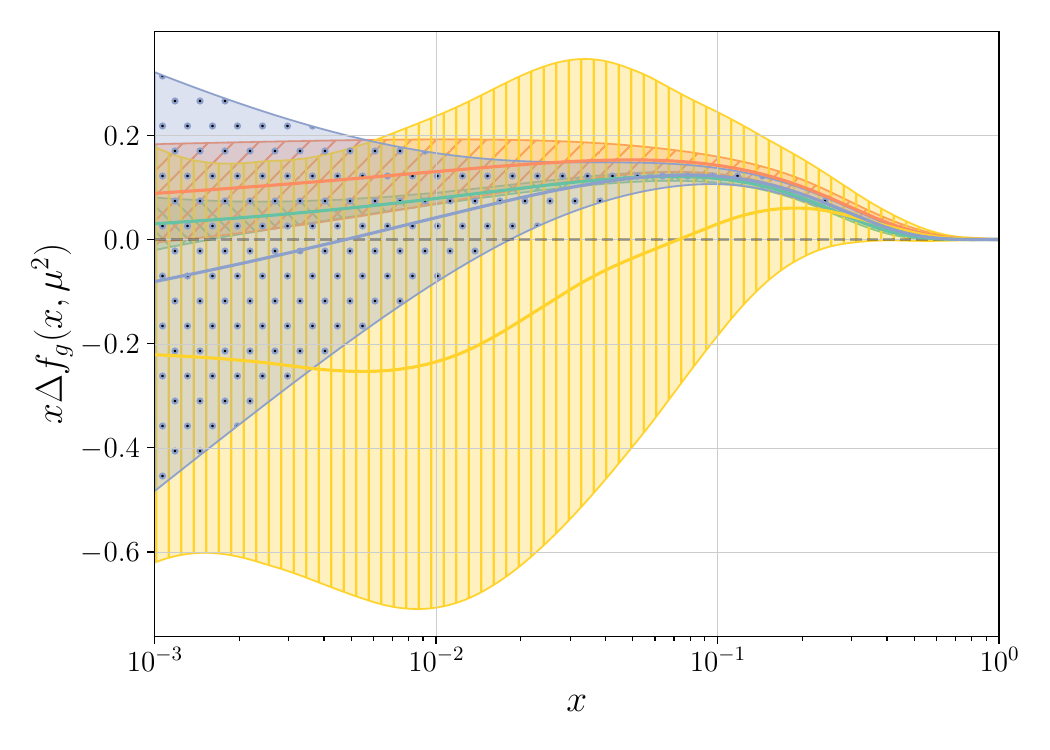}\\  
  \caption{The polarised PDFs $\Delta f_u$, $\Delta f_d$,  $\Delta f_{\bar u}$,
    $\Delta f_{\bar d}$, $\Delta f_{s}$, and $\Delta f_g$ from the
    BDSSV24, JAMpol25, MAPPDFpol1.0, and NNPDFpol2.0 sets as a function of $x$
    in logarithmic scale at $\mu^2=100$~GeV$^2$. Error bands correspond to
    one-standard-deviation PDF uncertainties. The BDSSV24, MAPPDFpol1.0, and
    NNPDFpol2.0 sets are accurate to NNLO; the JAMpol25 set is accurate to NLO.}
  \label{fig:polPDFs}
\end{figure}

Taken together, these developments illustrate how helicity-PDF phenomenology
has progressed not only through increasingly complementary experimental
information, but also through higher perturbative accuracy, more flexible
parametrisations, and a more systematic treatment of correlations among PDFs,
fragmentation functions, and theoretical uncertainties. The most recent
sets of polarised PDFs are summarised in Table~\ref{tab:polPDF_sets}
with their theoretical, experimental, and methodological features,
according to the description given in
Sects.~\ref{sec:theoretical_input}--\ref{sec:methodological_input}. The
corresponding polarised PDFs $\Delta f_u$, $\Delta f_d$, $\Delta f_{\bar u}$,
$\Delta f_{\bar d}$, $\Delta f_{s}$, and $\Delta f_g$ are
compared in Fig.~\ref{fig:polPDFs} as a function of $x$ in logarithmic scale at
$\mu^2=100$~GeV$^2$. Error bands correspond to one-standard-deviation PDF
uncertainties.
The BDSSV24, MAPPDFpol1.0, and NNPDFpol2.0 sets are accurate to NNLO; the
JAMpol25 set is accurate to NLO. 

%-------------------------------------------------------------------------------
\begin{table}[!t]
  \renewcommand{\arraystretch}{1.4}
  \TBL{\caption{The most recent sets of polarised PDFs:
      BDSSV24~\citep{Borsa:2024mss}, JAMpol25~\citep{Cocuzza:2025qvf},
      MAPPDFpol1.0~\citep{Bertone:2024taw}, and
      NNPDFpol2.0~\citep{Cruz-Martinez:2025ahf}. Their theoretical,
      experimental, and methodological features are summarised,
      following the discussion in
      Sects.~\ref{sec:theoretical_input}--\ref{sec:methodological_input}.}
    \label{tab:polPDF_sets}}
  \begin{tabular*}{\textwidth}{@{\extracolsep{\fill}}@{}lcccc@{}}
\toprule
\multicolumn{1}{@{}l}{\TCH{}} &
\multicolumn{1}{c}{\TCH{BDSSV}} &
\multicolumn{1}{c}{\TCH{JAM}} &
\multicolumn{1}{c}{\TCH{MAP}} &
\multicolumn{1}{c}{\TCH{NNPDF}} \\
\colrule
perturbative order
& NNLO
& NLO
& NLO, NNLO
& LO, NLO, NNLO\\
MHOU included
& no
& no
& no
& yes \\
heavy quark scheme
& ZM-VFN
& ZM-VFN
& ZM-VFN
& GM-VFN (FONLL)\\
value of $\alpha_s(m_Z)$
& 0.118
& 0.118
& 0.118
& 0.118\\
input scale $\mu_0$
& 1.00~GeV
& 1.28~GeV
& 1.00~GeV
& 1.00~GeV\\
\colrule
DIS
& $\checkmark$
& $\checkmark$
& $\checkmark$
& $\checkmark$\\
SIDIS
& $\checkmark$
& $\checkmark$
& $\checkmark$
& \\
$pp$ ($W^\pm$)
& $\checkmark$
& $\checkmark$
&
& $\checkmark$\\
$pp$ (jets)
& $\checkmark$
& $\checkmark$
&
& $\checkmark$\\
$pp$ (dijets)
&
&
&
& $\checkmark$\\
$pp$ ($\pi^0$)
& $\checkmark$
&
&
&\\
\colrule
independent PDFs
& 6
& 6
& 7
& 6\\
fitted PDFs
& $\Delta f_u^+, \Delta f_d^+, \Delta f_{\bar u},$
& $\Delta f_u^-, \Delta f_d^-, \Delta f_{\bar u},$
& $\Delta f_u, \Delta f_{\bar u}, \Delta f_d,$
& $\Delta f_\Sigma, \Delta f_{T_3}, \Delta f_{T_8},$\\
& $\Delta f_{\bar d}, \Delta f_{\bar s}, \Delta f_g$
& $\Delta f_{\bar d}, \Delta f_{\bar s}, \Delta f_g$
& $\Delta f_{\bar d}, \Delta f_s, \Delta f_{\bar s}, \Delta f_g$
& $\Delta f_V, \Delta f_{V_3}, \Delta f_g$\\
parametrisation
& polynomial
& polynomial
& shallow neural network
& deep neural network\\
free parameters
& 22
& 24
& 97
& 467\\
hyperoptimisation
& no
& no
& no
& yes \\
statistical treatment
& Monte Carlo
& Monte Carlo
& Monte Carlo
& Monte Carlo\\
\colrule
latest update
& BDSSV24
& JAMpol25
& MAPPDFpol1.0
& NNPDFpol2.0\\
reference
& \citep{Borsa:2024mss}
& \citep{Cocuzza:2025qvf}
& \citep{Bertone:2024taw}
& \citep{Cruz-Martinez:2025ahf}\\
\botrule
\end{tabular*}

\end{table}
%-------------------------------------------------------------------------------

%-------------------------------------------------------------------------------
%-------------------------------------------------------------------------------

The comparison in Fig.~\ref{fig:polPDFs} reveals a clear hierarchy in the
present knowledge of polarised PDFs. The up- and down-quark distributions,
which are dominated by their valence components over most of the experimentally
accessible region, are comparatively well determined. The light antiquark
distributions are smaller and less precise, while the strange-quark and gluon
distributions retain substantial uncertainties. The agreement among the
different determinations is generally best in the region directly covered by
data. Towards small $x$, where no measurements are available below
$x \simeq 0.004$, the uncertainty bands rapidly broaden and differences among
the fits increasingly reflect extrapolation and parametrisation assumptions
rather than tensions among the fitted data. A detailed survey of quark and
gluon polarised PDFs is presented next.

\begin{description}

\item[Up and down quark PDFs.]
  The up-quark polarised distribution is the best determined of all polarised
  PDFs. As shown in Fig.~\ref{fig:polPDFs}, $x\Delta f_u$ is positive throughout
  the region in which it is appreciable, increases from small $x$ to a broad
  maximum at $x \simeq 0.2$, and then decreases towards the kinematic endpoint.
  Conversely, $x\Delta f_d$ is negative over almost the entire displayed range,
  with its largest magnitude at intermediate $x$. These opposite signs express
  the preferential alignment of up-quark helicities and anti-alignment of
  down-quark helicities with the proton helicity, respectively.

  The four analyses agree rather well for both flavours in the approximate range
  $10^{-2} \lesssim x \lesssim 0.5$. The agreement is particularly striking for
  $\Delta f_u$, whereas the spread is somewhat larger for $\Delta f_d$.
  NNPDFpol2.0, for instance, favours a less negative down-quark distribution
  than the other sets at intermediate $x$, while MAPPDFpol1.0 generally
  displays a broader uncertainty band. These differences nevertheless remain
  compatible within the quoted one-standard-deviation uncertainties.
  The greater precision of
  $\Delta f_u$ originates primarily from the extensive measurements of the
  proton structure function $g_1^\gamma$, in which the up-quark contribution is
  enhanced by its electric-charge weight. The determination of $\Delta f_d$
  relies more strongly on neutron and deuteron measurements and is consequently
  affected by their lower statistical precision and by uncertainties associated
  with nuclear corrections.

  The large-$x$ region deserves particular attention. Fixed-target measurements
  at large $x$, especially those performed at JLab, necessarily lie at
  comparatively low values of $Q^2$ and of the final-state invariant mass $W^2$.
  Target-mass corrections and dynamical power corrections can therefore no
  longer be neglected if these data are to be retained in a global analysis.
  JAMpol25 incorporates recent high-$x$ JLab measurements and treats target-mass
  and higher-twist corrections explicitly. This allows the fitted region to be
  extended towards the endpoint without forcing power-suppressed
  contributions into the leading-twist PDFs. The resulting JAMpol25 up- and
  down-quark distributions remain consistent with the
  other global determinations, while being more directly informed by the
  available high-$x$ data.

  This treatment forms part of a broader JAM programme devoted to separating
  leading- and subleading-twist contributions. Dedicated JAM studies have found
  evidence for non-vanishing twist-three quark distributions, while the fitted
  twist-four quark contributions were compatible with zero~\citep{Sato:2016tuz}.
  The distinction is important: even when a twist-four correction is
  statistically consistent with zero, allowing for it in the fit prevents its
  possible contribution from being absorbed into the extracted leading-twist up-
  and down-quark PDFs. The conclusion is therefore not that power corrections
  are generically irrelevant, but rather that their size and twist structure
  can be tested only in an analysis that retains the low-$Q^2$, high-$x$
  information and treats such effects explicitly.

  The up- and down-quark sector is also the one in which first-principles
  lattice QCD information is currently most mature. JAMpol25 and an earlier JAM
  analysis~\citep{Bringewatt:2020ixn} have explored the inclusion
  of coordinate-space nucleon isovector matrix elements, which constrain
  predominantly the non-singlet combination of up- and down-quark polarised
  PDFs. The lattice information was found to be compatible with the experimental
  data and can reduce uncertainties in directions that are poorly constrained by
  the measurements, particularly at large $x$. Nevertheless, because the up- and
  down-quark polarised distributions are already accurately determined
  phenomenologically, estimates indicate that lattice uncertainties on their
  lowest moments would need to reach approximately the $1$--$2\%$ level in order
  to improve substantially upon present experimental constraints. At the current
  level of precision, lattice results therefore provide primarily a valuable
  consistency test and a complementary constraint on the large-$x$ behaviour.

\item[Sea quark PDFs.]
  The polarisation of the light-quark sea is considerably smaller than that of
  the up and down quarks and is correspondingly more difficult to determine.
  Figure~\ref{fig:polPDFs} nevertheless displays a reasonably coherent
  qualitative pattern: the central values of $x\Delta f_{\bar u}$ tend to be
  positive at intermediate $x$, whereas those of $x\Delta f_{\bar d}$ tend to be
  negative. This points towards a positive polarised light-sea asymmetry,
  $\Delta f_{\bar u}(x,\mu^2) - \Delta f_{\bar d}(x,\mu^2)> 0$,
  over the region $x \sim 10^{-2}$--$10^{-1}$. Its magnitude and detailed
  $x$ dependence, however, remain less firmly established than the corresponding
  unpolarised sea asymmetry~\citep{Nocera:2014rea}.
  
  The separation of quarks from antiquarks relies on the complementarity of
  SIDIS and weak-boson production in polarised $pp$ collisions. Identified-pion
  and identified-kaon asymmetries in SIDIS provide flavour sensitivity through
  FFs, whereas single-spin asymmetries for $W^\pm$ production
  at RHIC provide a fragmentation-independent constraint on particular
  quark--antiquark combinations, see Table~\ref{tab:factorisation}. Earlier
  global analyses based predominantly on one or the other class of observables
  found noticeable differences, most visibly in the sign and magnitude of
  $\Delta f_{\bar u}$. Modern fits show substantially greater consistency
  between the two sources of information.

  The JAM analyses have helped clarify this issue by
  determining helicity PDFs and FFs simultaneously. In this
  framework, correlations between the initial-state flavour decomposition and
  the final-state hadronisation process are propagated directly into the PDF
  uncertainties instead of being hidden by the use of a fixed external
  FF set. The JAM analysis found good consistency between
  the constraints supplied by $W$ production and those supplied by pion and
  kaon SIDIS. It also showed that the residual dependence on FFs is relatively
  mild when spin asymmetries, rather than absolute cross sections, are fitted,
  because part of the FF dependence cancels between the polarised and
  unpolarised cross sections.

  The strange-quark helicity distribution remains substantially less well
  determined. In Fig.~\ref{fig:polPDFs} all four analyses are compatible with a
  relatively small strange polarisation. Their central values are generally
  negative over part of the intermediate-$x$ region, but the distributions are
  statistically compatible with zero over a broad interval. The particularly
  large JAMpol25 uncertainty illustrates the limited direct information
  available once restrictive flavour assumptions are relaxed and the kaon FFs
  are determined simultaneously with the PDFs.
  
  Historically, the strange helicity distribution has been strongly affected by
  the use of the octet axial charge extracted from hyperon decays and by the
  assumption of exact SU$_f$(3) flavour symmetry. The simultaneous JAM analysis
  found the isospin SU$_f$(2) relation to be respected at approximately the
  $2\%$ level, but indicated that a breaking of SU$_f$(3) symmetry at the level
  of about $20\%$ may be required. This helps explain why analyses imposing
  exact SU$_f$(3) symmetry can obtain a rather definite negative strange moment
  even when the available SIDIS data do not determine the local shape of
  $\Delta f_s$ with comparable precision~\citep{Sato:2016tuz}. More accurate
  charged-kaon spin asymmetries, together with improved FFs, are therefore
  necessary to establish the sign and shape of the strange distribution and
  eventually to separate the strange- and antistrange-quark polarisations.

  The sea sector is also where lattice QCD may have the greatest near-term
  phenomenological impact. For the accurately known up- and down-quark
  combinations, percent-level lattice precision would be required to compete
  with experiment. By contrast, estimates indicate that a lattice determination
  of the strange helicity moment with an uncertainty of order $15\%$ could
  already provide a useful constraint~\citep{Lin:2017snn}. Such information
  would not by itself determine the full $x$ dependence of $\Delta f_s$, but it
  could restrict its normalisation and reduce the dependence on flavour-symmetry
  assumptions. Lattice constraints on isovector combinations can also sharpen
  the separation of valence and sea contributions indirectly, although a
  complete flavour decomposition ultimately requires experimental information
  from SIDIS and weak-boson production.

\item[Gluon PDF.]
  The gluon helicity distribution has evolved from being almost unconstrained
  in inclusive-DIS-only analyses to being directly probed by jet,
  dijet, and high-transverse-momentum hadron production at RHIC. In the region
  to which these measurements are most sensitive, approximately
  $0.02 \lesssim x \lesssim 0.3$--$0.4$, the modern global analyses favour a
  positive gluon polarisation. In Fig.~\ref{fig:polPDFs}, BDSSV24, JAMpol25 and
  NNPDFpol2.0 all display positive central values throughout most of this
  interval, with a broad maximum around $x \sim 0.05$--$0.1$. The positive
  solution in the RHIC region is therefore a robust phenomenological conclusion
  shared by analyses with rather different data sets and fitting methodologies.

  The sign of $\Delta f_g$ has nevertheless prompted a detailed discussion
  within the JAM programme. A JAM analysis in which the usual
  leading-order positivity condition $|\Delta f_i(x,\mu^2)|\leq f_i(x,\mu^2)$
  was relaxed found that a negative gluon solution could describe the fitted
  jet data~\citep{Zhou:2022wzm}. The argument was that an important contribution
  to inclusive-jet and dijet double-spin asymmetries arises from gluon-gluon
  scattering and is therefore approximately quadratic in $\Delta f_g$, making
  these observables alone comparatively insensitive to its sign.
  This observation correctly identifies a potential sign ambiguity in a
  restricted set of gluon-dominated observables, but it does not survive the
  full set of phenomenological and theoretical constraints. Negative
  moderate-$x$ solutions were not reproduced in the more flexible
  MAPPDFpol1.0 and NNPDFpol2.0 analyses. Moreover, a negative gluon helicity
  distribution in the relevant region can lead to a violation of the
  positivity of the polarised Higgs-production cross section, as shown
  in~\citep{deFlorian:2024utd}. Subsequent JAM
  work also showed that the negative branch is ruled out phenomenologically
  once large-$x$ inclusive-DIS asymmetries and high-transverse-momentum
  hadron-production asymmetries in SIDIS are taken into
  account~\citep{Hunt-Smith:2024khs}. The combined evidence therefore supports a
  positive $\Delta f_g$ at moderate and large $x$, even though jet and dijet
  asymmetries considered in isolation may not determine its sign uniquely.

  This conclusion cannot be extrapolated uncritically to the small-$x$
  region. Below the reach of present RHIC measurements, the uncertainty bands
  in Fig.~\ref{fig:polPDFs} broaden rapidly and become strongly dependent on the
  functional form and extrapolation assumptions. MAPPDFpol1.0, in particular,
  allows broad variations, including sign changes, at small $x$, while remaining
  compatible within uncertainties with the positive solutions obtained by the
  other analyses. The sign of the gluon helicity distribution at moderate $x$
  is thus much better established than either its sign or its magnitude at
  genuinely small $x$. Consequently, the full first moment of $\Delta f_g$
  remains dominated by the uncertainty associated with the unmeasured region.

  A complementary approach to the uncertainty in the unmeasured small-$x$
  region was developed in the JAMsmallx analysis~\citep{Adamiak:2021ppq}, which
  performed the first fit of the world data on polarised DIS at $x<0.1$ using
  the small-$x$ helicity evolution equations developed
  in a series of papers~\citep{Kovchegov:2015pbl,Kovchegov:2016weo,
    Kovchegov:2017lsr,Kovchegov:2017jxc,Kovchegov:2020hgb,Kovchegov:2021lvz,
    Cougoulic:2022gbk,Borden:2024bxa}. In this formalism, powers of
  $\alpha_s\ln^2(1/x)$ are resummed in the double-logarithmic approximation,
  so that the small-$x$ behaviour is generated dynamically by evolution in
  $x$, rather than being inherited from an assumed input parametrisation and
  subsequently evolved in $\mu^2$ through the standard DGLAP equations. The
  analysis showed that the existing proton, deuteron, and ${}^3{\rm He}$
  double-spin asymmetries can be described successfully within this framework
  and that the resulting evolution provides a substantially more controlled
  extrapolation of the $g_1$ structure function into the unmeasured region.
  Current data, however, do not yet determine precisely the value $x_0$ at
  which small-$x$ evolution becomes applicable: acceptable descriptions were
  obtained for values as large as $x_0\simeq 0.2$, while the fit quality
  deteriorated when data at $x\gtrsim 0.25$ were included. This matching
  ambiguity constitutes an additional theoretical uncertainty that is not
  captured by the experimental error bands alone. Moreover, at the accuracy
  considered in that study (LO in the double-logarithmic
  approximation and in the large-$N_c$ limit) the analysis constrained the
  quark-helicity combinations entering $g_1$, but did not extract the gluon
  helicity distribution directly. Extensions incorporating subleading
  $\alpha_s\ln(1/x)$ contributions and evolution beyond the large-$N_c$
  limit are required for a direct determination of $\Delta f_g$ and its
  small-$x$ contribution. The main significance of the JAM study for
  the gluon sector is therefore methodological: it demonstrated that
  QCD-based small-$x$ helicity evolution can replace largely
  parametrisation-driven extrapolations and may eventually provide the
  theoretical control needed to determine the contribution of the unmeasured
  region to the full gluon moment, especially once precise EIC data become
  available.
  
  Lattice QCD provides an important complementary perspective on the polarised
  gluon PDF, although the calculation of gluonic observables is considerably
  more difficult than that of quark observables. Polarised gluon matrix elements
  suffer from contamination terms in their Euclidean correlation functions,
  and gluonic observables generally require much larger statistical samples.
  Recent methods based on simultaneous analyses at several nucleon momenta allow
  the leading contamination to be isolated and removed. Existing lattice
  calculations tend to disfavour a negative gluon polarisation at moderate and
  large $x$, consistently with the phenomenological and cross-section
  positivity arguments (for details, see
  {\it e.g.}~\citep{Cichy:2018mum,Constantinou:2020hdm,Constantinou:2020pek}).
  A related JAM analysis has incorporated lattice results for gluonic
  pseudo-Ioffe-time distributions into a global fit~\citep{Karpie:2023nyg}.
  The lattice and experimental information were found to be mutually consistent,
  but the present pseudo-distribution data produced only a limited reduction of
  the gluon PDF uncertainty. This contrasts with the more visible impact
  obtained from isovector quark matrix elements in the quark sector. The
  difference reflects both the larger systematic and statistical uncertainties
  of gluonic lattice observables and the indirect relation between a finite set
  of Euclidean matrix elements and the complete $x$ dependence of the PDF.
  At present, lattice input should therefore be regarded as a complementary
  constraint and consistency test rather than as a replacement for collider
  data. Fits with and without lattice information should be reported
  separately so that the impact of the lattice input and its systematic
  uncertainties remain transparent.
  
\item[Perturbative convergence.]
  A final observation concerns the perturbative convergence of polarised PDFs.
  A systematic comparison of the LO, NLO and NNLO
  NNPDFpol2.0 determinations, including correlated uncertainties from missing
  higher orders, shows that the main change occurs between LO and NLO, whereas
  the differences between NLO and NNLO are generally much smaller than the PDF
  uncertainties and have little impact on the global fit
  quality~\citep{Cruz-Martinez:2025ahf}. Missing-higher-order
  uncertainties are moderate at LO and NLO and become very limited at NNLO; when
  included, they comfortably encompass the shifts produced by the subsequent
  known perturbative order without substantially enlarging the PDF
  uncertainties. This unusually good convergence, stronger than is commonly
  observed for unpolarised PDFs, can be traced in part to the fact that much of
  the polarised data consist of spin asymmetries, in which perturbative
  corrections to the polarised numerator and unpolarised denominator partially
  cancel. This conclusion should nevertheless be understood within the accuracy
  of present data and theory: the NNLO determination remains approximate for
  single-inclusive jet and dijet production, whose polarised hard-scattering
  matrix elements are currently known only to NLO. Complete NNLO calculations
  for these processes, together with the substantially greater precision and
  kinematic reach expected at the EIC, will therefore be required to establish
  whether the same perturbative stability persists in the next generation of
  polarised PDF determinations.
  
\end{description}

\subsection{Polarised PDF moments}
\label{subsec:moments}

The lowest Mellin moments of polarised PDFs quantify the net helicity
carried by each partonic species. For a generic polarised distribution
$\Delta f_i$, it is useful to define its truncated lowest moment as
\begin{equation}
  \langle \Delta f_i(\mu^2)\rangle_{[x_{\min},x_{\max}]}
  =
  \int_{x_{\min}}^{x_{\max}}
  dx\,\Delta f_i(x,\mu^2)\,.
  \label{eq:truncated_moment}
\end{equation}
The full lowest moment is recovered by taking $x_{\min}=0$ and
$x_{\max}=1$, provided that the integral converges, see Eq.~\eqref{eq:moments}.
The distributions of interest are $f_i=\Sigma,g,u^+,d^+,s^+$,
where $q^+=q+\bar q$ and $q=u,d,s$. The quantities entering the canonical
proton-spin decomposition, Eq.~\eqref{eq:Jaffe_and_Manohar}, are then
one half of the full singlet moment,
$\frac12\langle\Delta f_\Sigma\rangle_{[0,1]}$, and the full gluon moment,
usually denoted by $\Delta G(\mu^2)=\langle\Delta f_g(\mu^2)\rangle_{[0,1]}$.
The distinction between full and truncated moments is essential in the
polarised case. As discussed in Sect.~\ref{sec:experimental_input}, current
measurements do not extend below $x\simeq 0.004$, and the behaviour of
polarised PDFs in the unmeasured small-$x$ region is only weakly constrained.
A numerical evaluation of the full moments would therefore necessarily rely on
an extrapolation whose uncertainty depends on the flexibility of the PDF
parametrisation and on the assumptions imposed on its small-$x$ behaviour.
Truncated moments isolate the contribution from a specified range of
momentum fractions and make it possible to study explicitly how the result
changes as the lower integration boundary is moved into the unmeasured
region. They remain dependent on the factorisation scale and, in the
singlet sector, on the factorisation scheme; all results presented below
are understood in the common $\overline{\rm MS}$ scheme.

\begin{figure}[!t]
  \centering
  \includegraphics[width=0.49\textwidth]{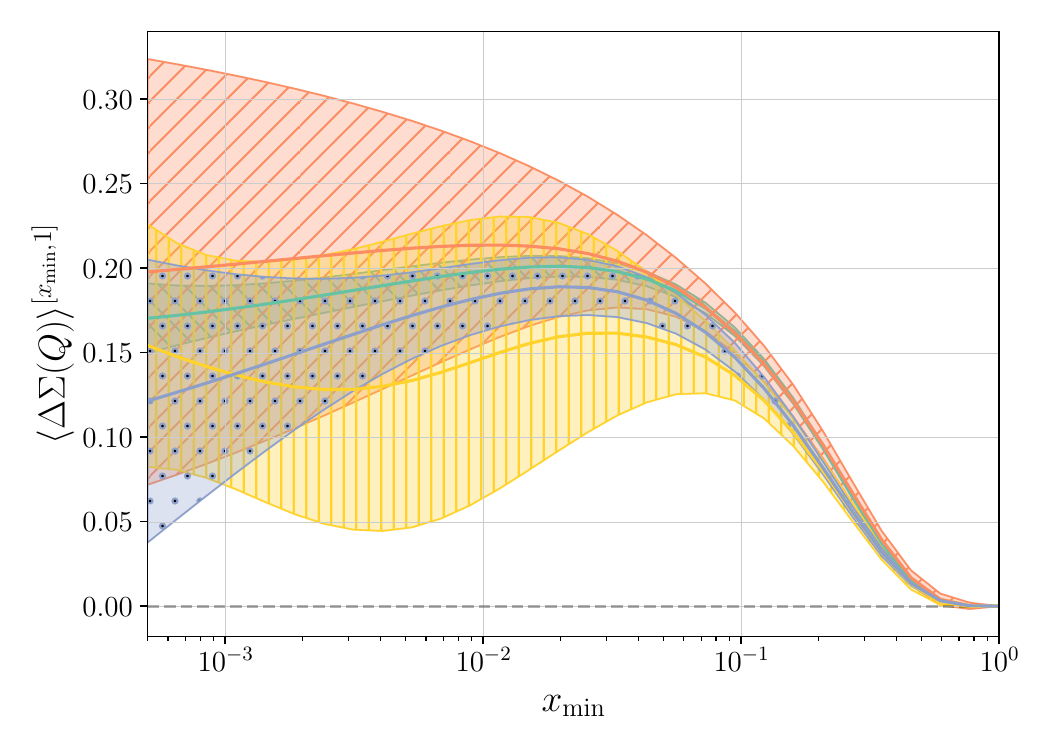}
  \includegraphics[width=0.49\textwidth]{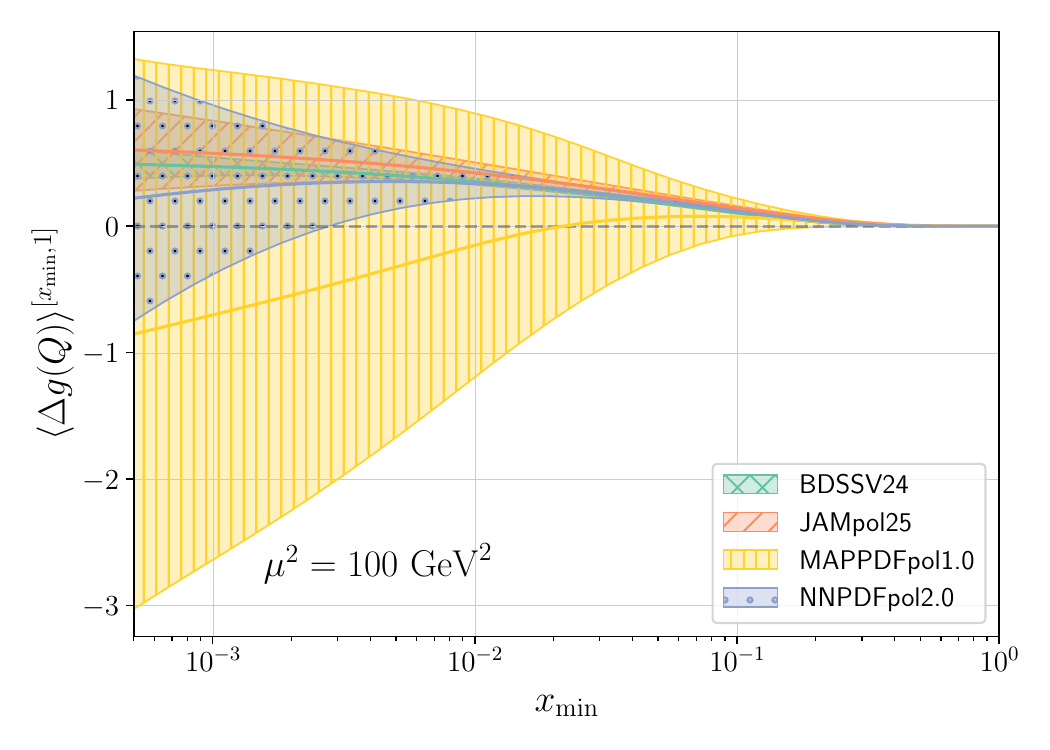}\\ 
  \caption{The singlet (left) and gluon (right) lowest truncated moments,
    Eq.~\eqref{eq:truncated_moment}, as a function of $x_{\min}$ at
    $\mu^2=100$~GeV$^2$, computed with the BDSSV24, JAMpol25, MAPPDFpol1.0, and
    NNPDFpol2.0 polarised PDF sets. Error bands correspond to
    one-standard-deviation PDF uncertainties.}
  \label{fig:moments}
\end{figure}

Figure~\ref{fig:moments} displays the quark and gluon helicity
contributions as functions of $x_{\min}$ at $\mu^2=100~{\rm GeV}^2$, for the
BDSSV24, JAMpol25, MAPPDFpol1.0, and NNPDFpol2.0 sets. The corresponding
one-standard-deviation PDF uncertainties are shown as bands.
Table~\ref{tab:moments} complements the figure by reporting the truncated
moments of the singlet, gluon, and individual light-flavour charge-even
distributions for the fixed interval $10^{-3}\leq x\leq1$, at
$\mu^2=4~{\rm GeV}^2$ and $\mu^2=100~{\rm GeV}^2$. The lower boundary
$x_{\min}=10^{-3}$ provides a useful estimate of the spin
contribution over an extended but still truncated range. It should not be
confused with the full moment: it lies below the region directly
covered by present measurements and therefore already includes an
extrapolated component. Figure~\ref{fig:moments} and Table~\ref{tab:moments}
support the following observations.

%-------------------------------------------------------------------------------
%-------------------------------------------------------------------------------

%-------------------------------------------------------------------------------
\begin{table}[!t]
  \renewcommand{\arraystretch}{1.4}
  \TBL{\caption{The truncated moments,
      Eq.~\eqref{eq:truncated_moment}, of the singlet, gluon, total up, down,
      and strange polarised PDFs, computed with $x_{\min}=10^{-3}$ at
      $\mu^2=4$~GeV$^2$ (top) and at $\mu^2=100$~GeV$^2$ (bottom) with the
      BDSSV24, JAMpol25, MAPPDFpol1.0, and NNPDFpol2.0 polarised PDF sets. }
    \label{tab:moments}}
  \begin{tabular*}{\textwidth}{@{\extracolsep{\fill}}@{}lccccc@{}}
\toprule
\multicolumn{1}{@{}l}{\TCH{PDF set ($\mu^2=4~{\rm GeV}^2$)}} &
\multicolumn{1}{c}{\TCH{$\langle\Delta\Sigma (\mu^2)\rangle_{[0.001,1]}$}} &
\multicolumn{1}{c}{\TCH{$\langle\Delta g (\mu^2)\rangle_{[0.001,1]}$}} &
\multicolumn{1}{c}{\TCH{$\langle\Delta u^+ (\mu^2)\rangle_{[0.001,1]}$}} &
\multicolumn{1}{c}{\TCH{$\langle\Delta d^+ (\mu^2)\rangle_{[0.001,1]}$}} &
\multicolumn{1}{c}{\TCH{$\langle\Delta s^+ (\mu^2)\rangle_{[0.001,1]}$}}\\
\colrule
BDSSV24
& $+0.34 \pm 0.03$
& $+0.24 \pm 0.09$
& $+0.76 \pm 0.02$
& $-0.39 \pm 0.02$
& $-0.03 \pm 0.04$
\\
JAMpol25
& $+0.41 \pm 0.24$
& $+0.44 \pm 0.25$
& $+0.80 \pm 0.09$
& $-0.42 \pm 0.08$
& $+0.02 \pm 0.40$
\\
MAPPDFpol1.0
& $+0.37 \pm 0.08$
& $-0.75 \pm 1.56$
& $+0.83 \pm 0.03$
& $-0.36 \pm 0.04$
& $-0.10 \pm 0.09$
\\
NNPDFpol2.0
& $+0.27 \pm 0.08$
& $+0.03 \pm 0.79$
& $+0.79 \pm 0.04$
& $-0.42 \pm 0.04$
& $-0.09 \pm 0.06$
\\
\colrule
\multicolumn{1}{@{}l}{\TCH{PDF set ($\mu^2=100~{\rm GeV}^2$)}} &
\multicolumn{1}{c}{\TCH{$\langle\Delta\Sigma (\mu^2)\rangle_{[0.001,1]}$}} &
\multicolumn{1}{c}{\TCH{$\langle\Delta g (\mu^2)\rangle_{[0.001,1]}$}} &
\multicolumn{1}{c}{\TCH{$\langle\Delta u^+ (\mu^2)\rangle_{[0.001,1]}$}} &
\multicolumn{1}{c}{\TCH{$\langle\Delta d^+ (\mu^2)\rangle_{[0.001,1]}$}} &
\multicolumn{1}{c}{\TCH{$\langle\Delta s^+ (\mu^2)\rangle_{[0.001,1]}$}}\\
\colrule
BDSSV24
& $+0.35 \pm 0.03$
& $+0.47 \pm 0.07$
& $+0.75 \pm 0.01$
& $-0.38 \pm 0.02$
& $-0.02 \pm 0.04$
\\
JAMpol25
& $+0.43 \pm 0.23$
& $+0.73 \pm 0.26$
& $+0.80 \pm 0.08$
& $-0.40 \pm 0.08$
& $+0.03 \pm 0.37$
\\
MAPPDFpol1.0
& $+0.28 \pm 0.13$
& $-0.70 \pm 1.95$
& $+0.79 \pm 0.05$
& $-0.38 \pm 0.06$
& $-0.11 \pm 0.07$
\\
NNPDFpol2.0
& $+0.27 \pm 0.12$
& $+0.29 \pm 0.68$
& $+0.78 \pm 0.05$
& $-0.42 \pm 0.05$
& $-0.09 \pm 0.06$
\\
\botrule
\end{tabular*}

\end{table}
%-------------------------------------------------------------------------------

The singlet moment is considerably better determined than the gluon
moment. In the region $x_{\min}\gtrsim10^{-2}$, the four analyses display
a similar accumulation of quark helicity as the lower integration boundary
is decreased. Differences and uncertainties grow once $x_{\min}$ enters
the region below the experimental coverage, although the deterioration is
much less dramatic than for the gluon. At $x_{\min}=10^{-3}$ and
$\mu^2=100~{\rm GeV}^2$, the quark-spin contribution is approximately
$27$--$43\%$ of the proton spin, within the specified
integration region. The central values are therefore reasonably consistent,
although the precision varies substantially among the
analyses. BDSSV24 yields the smallest uncertainty, while JAMpol25 gives
the largest. MAPPDFpol1.0 and NNPDFpol2.0 lie between
these two cases, with uncertainties of order $0.1$.

The flavour decomposition in Table~\ref{tab:moments} clarifies the
origin of this pattern. The total up-quark moment is positive and remarkably
stable among the four determinations, whereas the total down-quark moment is
negative. Their partial cancellation is the principal reason why the singlet
moment is much smaller than either of the two individual contributions. The
spread in $\Delta\Sigma$ is then largely associated with the strange
sector. BDSSV24 and JAMpol25 find strange central values compatible with
zero, while MAPPDFpol1.0 and NNPDFpol2.0 favour a modestly negative
moment. The uncertainty is especially large in JAMpol25, reflecting
its simultaneous determination of PDFs and FFs. Consequently, the
comparatively large JAMpol25 uncertainty on the singlet moment originates
primarily from the strange contribution, rather than from the well constrained
up- and down-quark sectors.

The interpretation of the gluon moment is qualitatively different.
For lower limits within the region probed by RHIC, approximately
$x_{\min}\gtrsim0.02$, the analyses that include polarised jet or hadron
production generally favour a positive accumulated gluon helicity and are
mutually compatible within uncertainties. As $x_{\min}$ is lowered below
this region, however, the uncertainty bands expand rapidly. The integral
becomes sensitive to a gluon distribution whose sign and magnitude are not
directly constrained by data, and different admissible small-$x$ shapes
can generate either a substantial enhancement or a cancellation of the
moderate-$x$ contribution.

The positive BDSSV24 and JAMpol25 results are consistent with the
positive gluon distribution found in the RHIC region. The MAPPDFpol1.0
central value is negative, but its uncertainty is so large that it is
compatible with both positive and negative gluon contributions. This is
expected because MAPPDFpol1.0 does not include the RHIC jet and dijet
measurements that provide the most direct constraint on $\Delta f_g$.
NNPDFpol2.0 includes an extensive set of such measurements and determines
a positive gluon distribution at moderate $x$, but its flexible
parametrisation propagates the lack of information at small $x$ into a
large uncertainty on the integral down to $10^{-3}$. The differences among
the quoted uncertainties should consequently not be interpreted solely as
differences in the constraining power of the data: they also reflect the
parametrisation, theoretical constraints, and uncertainty methodology
adopted in each analysis.

The comparison of the two scales in Table~\ref{tab:moments} also
illustrates the different evolution of the quark and gluon moments. The
up-, down-, and strange-quark moments vary only mildly between
$\mu^2=4~{\rm GeV}^2$ and $\mu^2=100~{\rm GeV}^2$, and the singlet moment
is correspondingly rather stable. Its residual scale dependence in the
$\overline{\rm MS}$ scheme follows from the anomalous dimension of the
flavour-singlet axial current. The gluon moment displays a more pronounced
scale dependence. This behaviour is consistent with singlet DGLAP
evolution and with the asymptotic tendency of the gluon moment to grow as
the strong coupling decreases. The MAPPDFpol1.0 moment remains poorly
determined at both scales, so that little significance should be attached
to the evolution of its central value.

Finally, one should refrain from using the numbers in
Table~\ref{tab:moments} to infer the orbital-angular-momentum
contribution by a direct subtraction from the proton-spin sum rule.
The moments are truncated rather than complete, and the omitted
$x<10^{-3}$ region can contribute significantly, particularly in the
gluon sector. Moreover, the separate quark, gluon, and orbital terms are
scale- and scheme-dependent. The robust conclusion from the present
comparison is instead that the quark helicity contribution over the
measured and moderately extrapolated region is relatively stable, while
the integrated gluon contribution remains limited by the unknown
small-$x$ behaviour. Extending measurements to lower $x$, together with
the consistent implementation of small-$x$ helicity evolution, is
therefore indispensable for converting the present determination of
truncated moments into a controlled determination of the complete quark
and gluon spin contributions.

\section{Summary and outlook}
\label{sec:conclusions}

Over almost five decades, the experimental study of the longitudinal spin
structure of the proton has evolved from inclusive fixed-target DIS
measurements to a genuinely global programme involving inclusive and
semi-inclusive lepton scattering and a variety of processes in polarised
$pp$ collisions. In parallel, the theoretical description has progressed from a
leading-order parton-model interpretation to QCD analyses based on collinear
factorisation, DGLAP evolution, and perturbative predictions that have reached
NNLO accuracy for the relevant DIS and SIDIS observables. Modern fitting
methodologies now
provide increasingly flexible PDF parametrisations, statistically meaningful
representations of experimental uncertainties, and systematic treatments of
theoretical and procedural effects.

This progress has led to a robust qualitative picture of the helicity structure
of the proton in the region covered by present data. The up-quark polarised
distribution is positive, whereas the down-quark distribution is negative.
These features are consistently reproduced by all modern analyses and are
comparatively stable against variations in the fitted data and methodology.
The polarised antiquark distributions are smaller and less precisely known,
although present determinations generally favour a positive light-sea asymmetry
at intermediate $x$. The strange distribution remains considerably more
uncertain and is especially sensitive to assumptions concerning flavour
symmetry, to the treatment of SIDIS measurements, and to the FFs used in their
interpretation. Substantial progress has also been made in determining the
gluon helicity distribution. Measurements of jet, dijet, and hadron production
at RHIC provide evidence for a positive gluon polarisation in the moderate-$x$
region to which they are sensitive. Nevertheless, the comparison of present PDF
sets shows that neither the shape nor even the sign of $\Delta f_g$ is
determined over the full range of momentum fractions. In particular, its
behaviour below the experimentally probed region remains weakly constrained and
strongly dependent on the flexibility of the PDF parametrisation and on the
assumptions adopted in the fit.

The corresponding truncated moments make this hierarchy especially transparent.
For $x\gtrsim10^{-3}$, the quark-singlet moment is consistently found to be
positive and of order $0.3$--$0.4$, although its quoted uncertainty varies
among analyses. The up- and down-quark moments are mutually consistent across
modern PDF sets, whereas the strange contribution remains compatible with a
broad range of values. The gluon moment is qualitatively different: it is
reasonably constrained only when the lower integration boundary lies within the
region directly probed by RHIC measurements, and its uncertainty increases
rapidly when the integral is extended towards small $x$. A controlled
determination of the complete moment of the polarised gluon distribution is
therefore not yet available. For the same reason, present truncated moments
should not be inserted directly into the proton-spin sum rule to infer an
orbital-angular-momentum contribution.

The perturbative behaviour of present determinations is encouraging.
Comparisons between NLO and NNLO fits generally show shifts that are smaller
than the current PDF uncertainties, while estimates of missing-higher-order
effects become small at NNLO. Part of this stability may be attributed to the
widespread use of spin asymmetries, in which perturbative corrections to
polarised and unpolarised cross sections partially cancel. This conclusion
remains provisional, however, because some of the hadronic processes that
provide the strongest constraints on the gluon distribution are not yet known
at complete NNLO accuracy. More precise measurements will therefore provide not
only tighter PDF constraints, but also a more stringent test of the
perturbative stability of the theoretical framework.

The principal limitations of current polarised PDF determinations can thus be
traced to a restricted kinematic reach, incomplete flavour separation, and the
extrapolation into the unmeasured small-$x$ region. Addressing these limitations
is a central objective of the EIC. The potential impact of polarised
lepton-hadron collisions was recognised from the early formulation of the EIC
science case and has since been developed in the reference community white
papers and design reports~\citep{Accardi:2012qut,AbdulKhalek:2021gbh}. With
longitudinally polarised lepton and proton beams, high luminosity, variable
centre-of-mass energies, and extensive coverage in both $x$ and $Q^2$, the EIC
will enlarge the kinematic domain explored by present measurements by several
orders of magnitude, reaching values of $x$ as small as approximately
$10^{-4}$ while retaining sufficiently large perturbative scales.

Numerous impact studies have investigated the consequences of this
extended coverage for helicity PDFs~\citep{Aschenauer:2012ve,Aschenauer:2013iia,
  Ball:2013tyh,Aschenauer:2015ata,Aschenauer:2019kzf,Borsa:2020lsz,Zhou:2021llj,
  Khanpour:2026erj}. Despite differences in the assumed beam configurations,
luminosities, observables, and baseline PDF determinations, these studies lead
to a consistent qualitative picture. Precise inclusive DIS measurements over a
wide range of scales will substantially improve the determination of the
quark-singlet distribution and, through scaling violations, of the gluon
helicity distribution. The availability of several centre-of-mass energies will
be especially important in this respect, because the determination of
$\Delta f_g$ requires not only access to small $x$, but also a sufficiently
broad lever arm in $Q^2$ at fixed $x$. The resulting reduction of the
extrapolation uncertainty is expected to translate into a significantly more
precise determination of the lowest moments $\Delta\Sigma$ and $\Delta G$.
In particular, the region that currently dominates the uncertainty on the gluon
moment will become, at least in part, directly accessible to experiment.

Inclusive measurements with proton beams will be complemented by
semi-inclusive, charged-current, and parity-violating neutral-current
observables. Identified-pion and identified-kaon production in SIDIS will
provide additional flavour separation and improve the determination of the
polarised light-antiquark and strange distributions. Charged-current
measurements, especially if both electron and positron beams become available,
will probe different quark and antiquark combinations without requiring
identified hadrons in the final state. If available, polarised light-ion beams
would provide complementary neutron information. Taken together, these
observables should make it possible
to replace several assumptions that remain necessary in present analyses
(concerning small-$x$ behaviour, flavour symmetry, or the relative
polarisation of sea quarks) with direct experimental constraints.

Exploiting this experimental precision will require a corresponding development
of the theoretical description. QED corrections and QCD$\otimes$QED evolution
may become relevant at the accuracy envisaged for EIC measurements. In this
context, the recent construction of a polarised photon PDF provides the
necessary starting point for incorporating a helicity-dependent photon
component consistently into the partonic description of a polarised
proton~\citep{deFlorian:2024hsu}. Jet observables in polarised DIS offer
another important opportunity. Single-inclusive jet and dijet production
provide sensitivity to the gluon helicity distribution through partonic channels
that differ from those entering inclusive DIS and can therefore complement the
information obtained from scaling violations. The development of increasingly
accurate calculations for these processes~\citep{Boughezal:2018azh,
  Borsa:2020yxh,Borsa:2021afb} opens the possibility of including them
systematically in future global analyses. More generally, the precision of EIC
measurements will require perturbative predictions, electroweak corrections,
heavy-quark treatments, resummation, and estimates of missing-higher-order
uncertainties to be developed and implemented at a comparable level of accuracy.

The arrival of EIC data will not, by itself, guarantee a correspondingly
precise and reliable determination of helicity PDFs. As uncertainties decrease,
methodological and computational differences that are presently concealed by
broad error bands will become increasingly important. A coordinated programme
of PDF benchmarking should therefore be undertaken before the first precision
EIC measurements are included in global analyses. Different fitting groups
should analyse a common experimental data set, with a standardised treatment of
correlated uncertainties, normalisations, nuclear corrections, and kinematic
cuts, using common theory predictions and physical settings. Controlled
variations of one ingredient at a time would then make it possible to determine
whether differences among PDF sets arise from the data, the perturbative theory,
the parametrisation, the optimisation procedure, or the uncertainty
prescription. The experience accumulated in benchmarking unpolarised
PDFs~\citep{PDF4LHCWorkingGroup:2022cjn} provides a useful model, although the
specific problems of polarised observables will require dedicated solutions.

This effort must be supported by an adequate computational infrastructure.
Global analyses increasingly rely on a chain of specialised components for PDF
evolution, the computation of DIS and SIDIS observables, fast interpolation of
hadronic cross sections, the delivery of PDF ensembles, statistical fitting,
and validation. These components should be publicly available whenever possible,
thoroughly documented, independently benchmarked, versioned, and maintained
over timescales comparable to those of the experimental programme. Public code
is not only a matter of reproducibility: it allows theoretical improvements to
be incorporated rapidly by different groups, makes independent comparisons
possible, and lowers the barrier for new researchers to enter the field.
The development and long-term maintenance of such software should therefore be
treated as a scientific contribution in its own right and supported and
recognised accordingly.

A similarly coordinated strategy will be needed for the interplay between
polarised PDFs and FFs. SIDIS is indispensable for separating quark and
antiquark flavours, but its interpretation necessarily introduces FFs. In many
present measurements their impact is mitigated by the use of spin asymmetries,
in which some of the fragmentation dependence cancels. This cancellation will
not necessarily remain sufficient once EIC measurements reach percent-level
precision, particularly if differential cross sections are analysed in addition
to asymmetries. Uncertainties in FFs, their flavour decomposition, and their
correlations with helicity PDFs may then become a limiting systematic effect.
Future analyses should therefore move towards simultaneous PDF and FF
determinations, or at least towards coordinated iterative fits in which both
sets of nonperturbative functions and their correlations are propagated
consistently. Measurements from electron--positron annihilation, unpolarised
SIDIS, hadron collisions, and the EIC itself will all be required to achieve
this goal.

Finally, the relation between phenomenological determinations and lattice QCD
calculations should continue to be strengthened. Lattice calculations of
moments, coordinate-space matrix elements, and pseudo- or quasi-distributions
provide information that is complementary to experimental measurements and may
be particularly useful for poorly constrained flavour combinations. At present,
their impact depends strongly on the observable considered and on the control
of lattice systematic uncertainties. Lattice input should therefore be
incorporated transparently, with its correlations and theoretical uncertainties
propagated as completely as possible. Fits performed with and without lattice
information should be made available separately, so that its effect can be
identified unambiguously. Common benchmarks, conventions, and interfaces
between the lattice and global-analysis communities will be essential if
lattice information is to become a quantitative component of precision
polarised PDF determinations rather than only a qualitative consistency test.

The EIC thus represents both an exceptional experimental opportunity and a
challenge for the organisation of the field. It will directly address the
small-$x$, flavour-separation, and scale-evolution limitations that currently
prevent a complete determination of the quark and gluon helicity contributions
to the proton spin. Realising this potential will require more than new data:
it will demand precision theoretical calculations, common benchmarks,
sustainable public software, a consistent treatment of fragmentation, and a
controlled integration of lattice information. With these developments in
place, the next generation of global analyses should be able to transform the
present qualitative understanding of proton spin into a quantitatively precise,
reproducible, and systematically improvable one.

\begin{ack}[Acknowledgments]{}
  I would like to thank Stefano Forte and Richard D.~Ball for their continuous
  support over many years of collaboration.
\end{ack}

%\seealso{article title article title}

\bibliographystyle{Harvard}
\bibliography{30025_Nocera}

\begin{thebibliography*}{237}
\providecommand{\bibtype}[1]{}
\providecommand{\natexlab}[1]{#1}
{\catcode`\|=0\catcode`\#=12\catcode`\@=11\catcode`\\=12
|immediate|write|@auxout{\expandafter\ifx\csname
  natexlab\endcsname\relax\gdef\natexlab#1{#1}\fi}}
\renewcommand{\url}[1]{{\tt #1}}
\providecommand{\urlprefix}{URL }
\expandafter\ifx\csname urlstyle\endcsname\relax
  \providecommand{\doi}[1]{doi:\discretionary{}{}{}#1}\else
  \providecommand{\doi}{doi:\discretionary{}{}{}\begingroup
  \urlstyle{rm}\Url}\fi
\providecommand{\bibinfo}[2]{#2}
\providecommand{\eprint}[2][]{\url{#2}}

\bibtype{Article}%
\bibitem[Abdallah et al.(2021)]{STAR:2021mfd}
\bibinfo{author}{Abdallah MS} and  et al. (\bibinfo{collaboration}{STAR})
  (\bibinfo{year}{2021}).
\bibinfo{title}{{Longitudinal double-spin asymmetry for inclusive jet and dijet
  production in polarized proton collisions at $\sqrt{s}=200$ GeV}}.
\bibinfo{journal}{{\em Phys. Rev. D}} \bibinfo{volume}{103}
  (\bibinfo{number}{9}): \bibinfo{pages}{L091103}.
  \bibinfo{doi}{\doi{10.1103/PhysRevD.103.L091103}}.
\eprint{2103.05571}.

\bibtype{Article}%
\bibitem[Abdallah et al.(2022)]{STAR:2021mqa}
\bibinfo{author}{Abdallah MS} and  et al. (\bibinfo{collaboration}{STAR})
  (\bibinfo{year}{2022}).
\bibinfo{title}{{Longitudinal double-spin asymmetry for inclusive jet and dijet
  production in polarized proton collisions at $\sqrt{s}=510$ GeV}}.
\bibinfo{journal}{{\em Phys. Rev. D}} \bibinfo{volume}{105}
  (\bibinfo{number}{9}): \bibinfo{pages}{092011}.
  \bibinfo{doi}{\doi{10.1103/PhysRevD.105.092011}}.
\eprint{2110.11020}.

\bibtype{Article}%
\bibitem[Abdul~Khalek et al.(2019{\natexlab{a}})]{NNPDF:2019vjt}
\bibinfo{author}{Abdul~Khalek R} and  et al. (\bibinfo{collaboration}{NNPDF})
  (\bibinfo{year}{2019}{\natexlab{a}}).
\bibinfo{title}{{A first determination of parton distributions with theoretical
  uncertainties}}.
\bibinfo{journal}{{\em Eur. Phys. J.}} \bibinfo{volume}{C}:
  \bibinfo{pages}{79:838}. \bibinfo{doi}{\doi{10.1140/epjc/s10052-019-7364-5}}.
\eprint{1905.04311}.

\bibtype{Article}%
\bibitem[Abdul~Khalek et al.(2019{\natexlab{b}})]{NNPDF:2019ubu}
\bibinfo{author}{Abdul~Khalek R} and  et al. (\bibinfo{collaboration}{NNPDF})
  (\bibinfo{year}{2019}{\natexlab{b}}).
\bibinfo{title}{{Parton Distributions with Theory Uncertainties: General
  Formalism and First Phenomenological Studies}}.
\bibinfo{journal}{{\em Eur. Phys. J. C}} \bibinfo{volume}{79}
  (\bibinfo{number}{11}): \bibinfo{pages}{931}.
  \bibinfo{doi}{\doi{10.1140/epjc/s10052-019-7401-4}}.
\eprint{1906.10698}.

\bibtype{Article}%
\bibitem[Abdul~Khalek et al.(2022)]{AbdulKhalek:2021gbh}
\bibinfo{author}{Abdul~Khalek R} and  et al. (\bibinfo{year}{2022}).
\bibinfo{title}{{Science Requirements and Detector Concepts for the
  Electron-Ion Collider}: {EIC Yellow Report}}.
\bibinfo{journal}{{\em Nucl. Phys. A}} \bibinfo{volume}{1026}:
  \bibinfo{pages}{122447}.
  \bibinfo{doi}{\doi{10.1016/j.nuclphysa.2022.122447}}.
\eprint{2103.05419}.

\bibtype{Article}%
\bibitem[Abe et al.(1997)]{E154:1997xfa}
\bibinfo{author}{Abe K} and  et al. (\bibinfo{collaboration}{E154})
  (\bibinfo{year}{1997}).
\bibinfo{title}{{Precision determination of the neutron spin structure function
  g1(n)}}.
\bibinfo{journal}{{\em Phys. Rev. Lett.}} \bibinfo{volume}{79}:
  \bibinfo{pages}{26--30}. \bibinfo{doi}{\doi{10.1103/PhysRevLett.79.26}}.
\eprint{hep-ex/9705012}.

\bibtype{Article}%
\bibitem[Abe et al.(1998)]{E143:1998hbs}
\bibinfo{author}{Abe K} and  et al. (\bibinfo{collaboration}{E143})
  (\bibinfo{year}{1998}).
\bibinfo{title}{{Measurements of the proton and deuteron spin structure
  functions g(1) and g(2)}}.
\bibinfo{journal}{{\em Phys. Rev. D}} \bibinfo{volume}{58}:
  \bibinfo{pages}{112003}. \bibinfo{doi}{\doi{10.1103/PhysRevD.58.112003}}.
\eprint{hep-ph/9802357}.

\bibtype{Article}%
\bibitem[Abele et al.(2021)]{Abele:2021nyo}
\bibinfo{author}{Abele M}, \bibinfo{author}{de~Florian D} and
  \bibinfo{author}{Vogelsang W} (\bibinfo{year}{2021}).
\bibinfo{title}{{Approximate NNLO QCD corrections to semi-inclusive DIS}}.
\bibinfo{journal}{{\em Phys. Rev. D}} \bibinfo{volume}{104}
  (\bibinfo{number}{9}): \bibinfo{pages}{094046}.
  \bibinfo{doi}{\doi{10.1103/PhysRevD.104.094046}}.
\eprint{2109.00847}.

\bibtype{Article}%
\bibitem[Ablinger et al.(2020)]{Ablinger:2019etw}
\bibinfo{author}{Ablinger J}, \bibinfo{author}{Behring A},
  \bibinfo{author}{Bl{\"u}mlein J}, \bibinfo{author}{De~Freitas A},
  \bibinfo{author}{von Manteuffel A}, \bibinfo{author}{Schneider C} and
  \bibinfo{author}{Sch{\"o}nwald K} (\bibinfo{year}{2020}).
\bibinfo{title}{{The three-loop single mass polarized pure singlet operator
  matrix element}}.
\bibinfo{journal}{{\em Nucl. Phys. B}} \bibinfo{volume}{953}:
  \bibinfo{pages}{114945}.
  \bibinfo{doi}{\doi{10.1016/j.nuclphysb.2020.114945}}.
\eprint{1912.02536}.

\bibtype{Article}%
\bibitem[Ablinger et al.(2022)]{Ablinger:2022wbb}
\bibinfo{author}{Ablinger J}, \bibinfo{author}{Behring A},
  \bibinfo{author}{Bl{\"u}mlein J}, \bibinfo{author}{De~Freitas A},
  \bibinfo{author}{Goedicke A}, \bibinfo{author}{von Manteuffel A},
  \bibinfo{author}{Schneider C} and  \bibinfo{author}{Sch{\"o}nwald K}
  (\bibinfo{year}{2022}).
\bibinfo{title}{{The unpolarized and polarized single-mass three-loop heavy
  flavor operator matrix elements A$_{gg,Q}$ and
  {\ensuremath{\Delta}}A$_{gg,Q}$}}.
\bibinfo{journal}{{\em JHEP}} \bibinfo{volume}{12}: \bibinfo{pages}{134}.
  \bibinfo{doi}{\doi{10.1007/JHEP12(2022)134}}.
\eprint{2211.05462}.

\bibtype{Article}%
\bibitem[Ablinger et al.(2024{\natexlab{a}})]{Ablinger:2023ahe}
\bibinfo{author}{Ablinger J}, \bibinfo{author}{Behring A},
  \bibinfo{author}{Bl{\"u}mlein J}, \bibinfo{author}{De~Freitas A},
  \bibinfo{author}{von Manteuffel A}, \bibinfo{author}{Schneider C} and
  \bibinfo{author}{Sch{\"o}nwald K} (\bibinfo{year}{2024}{\natexlab{a}}).
\bibinfo{title}{{The first{\textendash}order factorizable contributions to the
  three{\textendash}loop massive operator matrix elements AQg(3) and
  {\ensuremath{\Delta}}AQg(3)}}.
\bibinfo{journal}{{\em Nucl. Phys. B}} \bibinfo{volume}{999}:
  \bibinfo{pages}{116427}.
  \bibinfo{doi}{\doi{10.1016/j.nuclphysb.2023.116427}}.
\eprint{2311.00644}.

\bibtype{Article}%
\bibitem[Ablinger et al.(2024{\natexlab{b}})]{Ablinger:2024xtt}
\bibinfo{author}{Ablinger J}, \bibinfo{author}{Behring A},
  \bibinfo{author}{Bl{\"u}mlein J}, \bibinfo{author}{De~Freitas A},
  \bibinfo{author}{von Manteuffel A}, \bibinfo{author}{Schneider C} and
  \bibinfo{author}{Sch{\"o}nwald K} (\bibinfo{year}{2024}{\natexlab{b}}).
\bibinfo{title}{{The non-first-order-factorizable contributions to the
  three-loop single-mass operator matrix elements AQg(3) and
  {\ensuremath{\Delta}}AQg(3)}}.
\bibinfo{journal}{{\em Phys. Lett. B}} \bibinfo{volume}{854}:
  \bibinfo{pages}{138713}. \bibinfo{doi}{\doi{10.1016/j.physletb.2024.138713}}.
\eprint{2403.00513}.

\bibtype{Article}%
\bibitem[Accardi et al.(2016)]{Accardi:2012qut}
\bibinfo{author}{Accardi A} and  et al. (\bibinfo{year}{2016}).
\bibinfo{title}{{Electron Ion Collider: The Next QCD Frontier}: {Understanding
  the glue that binds us all}}.
\bibinfo{journal}{{\em Eur. Phys. J. A}} \bibinfo{volume}{52}
  (\bibinfo{number}{9}): \bibinfo{pages}{268}.
  \bibinfo{doi}{\doi{10.1140/epja/i2016-16268-9}}.
\eprint{1212.1701}.

\bibtype{Article}%
\bibitem[Ackerstaff et al.(1997)]{HERMES:1997hjr}
\bibinfo{author}{Ackerstaff K} and  et al. (\bibinfo{collaboration}{HERMES})
  (\bibinfo{year}{1997}).
\bibinfo{title}{{Measurement of the neutron spin structure function g1(n) with
  a polarized He-3 internal target}}.
\bibinfo{journal}{{\em Phys. Lett. B}} \bibinfo{volume}{404}:
  \bibinfo{pages}{383--389}.
  \bibinfo{doi}{\doi{10.1016/S0370-2693(97)00611-4}}.
\eprint{hep-ex/9703005}.

\bibtype{Article}%
\bibitem[Adam et al.(2018)]{STAR:2018iyz}
\bibinfo{author}{Adam J} and  et al. (\bibinfo{collaboration}{STAR})
  (\bibinfo{year}{2018}).
\bibinfo{title}{{Longitudinal Double-Spin Asymmetries for $\pi^{0}$s in the
  Forward Direction for 510 GeV Polarized $pp$ Collisions}}.
\bibinfo{journal}{{\em Phys. Rev. D}} \bibinfo{volume}{98}
  (\bibinfo{number}{3}): \bibinfo{pages}{032013}.
  \bibinfo{doi}{\doi{10.1103/PhysRevD.98.032013}}.
\eprint{1805.09745}.

\bibtype{Article}%
\bibitem[Adam et al.(2019{\natexlab{a}})]{STAR:2019yqm}
\bibinfo{author}{Adam J} and  et al. (\bibinfo{collaboration}{STAR})
  (\bibinfo{year}{2019}{\natexlab{a}}).
\bibinfo{title}{{Longitudinal double-spin asymmetry for inclusive jet and dijet
  production in pp collisions at $\sqrt{s} = 510$ GeV}}.
\bibinfo{journal}{{\em Phys. Rev. D}} \bibinfo{volume}{100}
  (\bibinfo{number}{5}): \bibinfo{pages}{052005}.
  \bibinfo{doi}{\doi{10.1103/PhysRevD.100.052005}}.
\eprint{1906.02740}.

\bibtype{Article}%
\bibitem[Adam et al.(2019{\natexlab{b}})]{STAR:2018fty}
\bibinfo{author}{Adam J} and  et al. (\bibinfo{collaboration}{STAR})
  (\bibinfo{year}{2019}{\natexlab{b}}).
\bibinfo{title}{{Measurement of the longitudinal spin asymmetries for weak
  boson production in proton-proton collisions at $\sqrt{s}$ = 510 GeV}}.
\bibinfo{journal}{{\em Phys. Rev. D}} \bibinfo{volume}{99}
  (\bibinfo{number}{5}): \bibinfo{pages}{051102}.
  \bibinfo{doi}{\doi{10.1103/PhysRevD.99.051102}}.
\eprint{1812.04817}.

\bibtype{Article}%
\bibitem[Adamczyk et al.(2012)]{STAR:2012hth}
\bibinfo{author}{Adamczyk L} and  et al. (\bibinfo{collaboration}{STAR})
  (\bibinfo{year}{2012}).
\bibinfo{title}{{Longitudinal and transverse spin asymmetries for inclusive jet
  production at mid-rapidity in polarized $p+p$ collisions at $\sqrt{s}=200$
  GeV}}.
\bibinfo{journal}{{\em Phys. Rev. D}} \bibinfo{volume}{86}:
  \bibinfo{pages}{032006}. \bibinfo{doi}{\doi{10.1103/PhysRevD.86.032006}}.
\eprint{1205.2735}.

\bibtype{Article}%
\bibitem[Adamczyk et al.(2014)]{STAR:2014afm}
\bibinfo{author}{Adamczyk L} and  et al. (\bibinfo{collaboration}{STAR})
  (\bibinfo{year}{2014}).
\bibinfo{title}{{Measurement of longitudinal spin asymmetries for weak boson
  production in polarized proton-proton collisions at RHIC}}.
\bibinfo{journal}{{\em Phys. Rev. Lett.}} \bibinfo{volume}{113}:
  \bibinfo{pages}{072301}. \bibinfo{doi}{\doi{10.1103/PhysRevLett.113.072301}}.
\eprint{1404.6880}.

\bibtype{Article}%
\bibitem[Adamczyk et al.(2015)]{STAR:2014wox}
\bibinfo{author}{Adamczyk L} and  et al. (\bibinfo{collaboration}{STAR})
  (\bibinfo{year}{2015}).
\bibinfo{title}{{Precision Measurement of the Longitudinal Double-spin
  Asymmetry for Inclusive Jet Production in Polarized Proton Collisions at
  $\sqrt{s}=200$ GeV}}.
\bibinfo{journal}{{\em Phys. Rev. Lett.}} \bibinfo{volume}{115}
  (\bibinfo{number}{9}): \bibinfo{pages}{092002}.
  \bibinfo{doi}{\doi{10.1103/PhysRevLett.115.092002}}.
\eprint{1405.5134}.

\bibtype{Article}%
\bibitem[Adamczyk et al.(2017)]{STAR:2016kpm}
\bibinfo{author}{Adamczyk L} and  et al. (\bibinfo{collaboration}{STAR})
  (\bibinfo{year}{2017}).
\bibinfo{title}{{Measurement of the cross section and longitudinal double-spin
  asymmetry for di-jet production in polarized $pp$ collisions at $\sqrt{s}$ =
  200 GeV}}.
\bibinfo{journal}{{\em Phys. Rev. D}} \bibinfo{volume}{95}
  (\bibinfo{number}{7}): \bibinfo{pages}{071103}.
  \bibinfo{doi}{\doi{10.1103/PhysRevD.95.071103}}.
\eprint{1610.06616}.

\bibtype{Article}%
\bibitem[Adamiak et al.(2021)]{Adamiak:2021ppq}
\bibinfo{author}{Adamiak D}, \bibinfo{author}{Kovchegov YV},
  \bibinfo{author}{Melnitchouk W}, \bibinfo{author}{Pitonyak D},
  \bibinfo{author}{Sato N} and  \bibinfo{author}{Sievert MD}
  (\bibinfo{collaboration}{Jefferson Lab Angular Momentum})
  (\bibinfo{year}{2021}).
\bibinfo{title}{{First analysis of world polarized DIS data with small-x
  helicity evolution}}.
\bibinfo{journal}{{\em Phys. Rev. D}} \bibinfo{volume}{104}
  (\bibinfo{number}{3}): \bibinfo{pages}{L031501}.
  \bibinfo{doi}{\doi{10.1103/PhysRevD.104.L031501}}.
\eprint{2102.06159}.

\bibtype{Article}%
\bibitem[Adamiak et al.(2025)]{JAMCollaborationSmall-xAnalysisGroup:2025tfa}
\bibinfo{author}{Adamiak D}, \bibinfo{author}{Baldonado N},
  \bibinfo{author}{Kovchegov YV}, \bibinfo{author}{Li M},
  \bibinfo{author}{Melnitchouk W}, \bibinfo{author}{Pitonyak D},
  \bibinfo{author}{Sato N}, \bibinfo{author}{Sievert MD},
  \bibinfo{author}{Tarasov A} and  \bibinfo{author}{Tawabutr Y}
  (\bibinfo{collaboration}{JAM Collaboration (Small-x Analysis Group)})
  (\bibinfo{year}{2025}).
\bibinfo{title}{{First study of polarized proton-proton scattering with small-x
  helicity evolution}}.
\bibinfo{journal}{{\em Phys. Rev. D}} \bibinfo{volume}{112}
  (\bibinfo{number}{9}): \bibinfo{pages}{094032}.
  \bibinfo{doi}{\doi{10.1103/9gnx-ycs4}}.
\eprint{2503.21006}.

\bibtype{Article}%
\bibitem[Adare et al.(2007)]{PHENIX:2007kqm}
\bibinfo{author}{Adare A} and  et al. (\bibinfo{collaboration}{PHENIX})
  (\bibinfo{year}{2007}).
\bibinfo{title}{{Inclusive cross-section and double helicity asymmetry for
  $\pi^0$ production in $p + p$ collisions at $\sqrt{s} =$ 200 GeV:
  Implications for the polarized gluon distribution in the proton}}.
\bibinfo{journal}{{\em Phys. Rev. D}} \bibinfo{volume}{76}:
  \bibinfo{pages}{051106}. \bibinfo{doi}{\doi{10.1103/PhysRevD.76.051106}}.
\eprint{0704.3599}.

\bibtype{Article}%
\bibitem[Adare et al.(2009{\natexlab{a}})]{PHENIX:2008sgl}
\bibinfo{author}{Adare A} and  et al. (\bibinfo{collaboration}{PHENIX})
  (\bibinfo{year}{2009}{\natexlab{a}}).
\bibinfo{title}{{Inclusive cross section and double helicity asymmetry for
  pi{\textasciicircum}0 production in $p^+ p$ collisions at $\sqrt{s}=62.4$
  GeV}}.
\bibinfo{journal}{{\em Phys. Rev. D}} \bibinfo{volume}{79}:
  \bibinfo{pages}{012003}. \bibinfo{doi}{\doi{10.1103/PhysRevD.79.012003}}.
\eprint{0810.0701}.

\bibtype{Article}%
\bibitem[Adare et al.(2009{\natexlab{b}})]{PHENIX:2008swq}
\bibinfo{author}{Adare A} and  et al. (\bibinfo{collaboration}{PHENIX})
  (\bibinfo{year}{2009}{\natexlab{b}}).
\bibinfo{title}{{The Polarized gluon contribution to the proton spin from the
  double helicity asymmetry in inclusive pi0 production in polarized p + p
  collisions at s**(1/2) = 200-GeV}}.
\bibinfo{journal}{{\em Phys. Rev. Lett.}} \bibinfo{volume}{103}:
  \bibinfo{pages}{012003}. \bibinfo{doi}{\doi{10.1103/PhysRevLett.103.012003}}.
\eprint{0810.0694}.

\bibtype{Article}%
\bibitem[Adare et al.(2011)]{PHENIX:2010aru}
\bibinfo{author}{Adare A} and  et al. (\bibinfo{collaboration}{PHENIX})
  (\bibinfo{year}{2011}).
\bibinfo{title}{{Event Structure and Double Helicity Asymmetry in Jet
  Production from Polarized $p+p$ Collisions at $\sqrt{s} =
  200${\textasciitilde}GeV}}.
\bibinfo{journal}{{\em Phys. Rev. D}} \bibinfo{volume}{84}:
  \bibinfo{pages}{012006}. \bibinfo{doi}{\doi{10.1103/PhysRevD.84.012006}}.
\eprint{1009.4921}.

\bibtype{Article}%
\bibitem[Adare et al.(2014)]{PHENIX:2014gbf}
\bibinfo{author}{Adare A} and  et al. (\bibinfo{collaboration}{PHENIX})
  (\bibinfo{year}{2014}).
\bibinfo{title}{{Inclusive double-helicity asymmetries in neutral-pion and
  eta-meson production in $\vec{p}+\vec{p}$ collisions at $\sqrt{s}=200$ GeV}}.
\bibinfo{journal}{{\em Phys. Rev. D}} \bibinfo{volume}{90}
  (\bibinfo{number}{1}): \bibinfo{pages}{012007}.
  \bibinfo{doi}{\doi{10.1103/PhysRevD.90.012007}}.
\eprint{1402.6296}.

\bibtype{Article}%
\bibitem[Adare et al.(2016{\natexlab{a}})]{PHENIX:2015fxo}
\bibinfo{author}{Adare A} and  et al. (\bibinfo{collaboration}{PHENIX})
  (\bibinfo{year}{2016}{\natexlab{a}}).
\bibinfo{title}{{Inclusive cross section and double-helicity asymmetry for
  $\pi^{0}$ production at midrapidity in $p+p$ collisions at $\sqrt{s}=510$
  GeV}}.
\bibinfo{journal}{{\em Phys. Rev. D}} \bibinfo{volume}{93}
  (\bibinfo{number}{1}): \bibinfo{pages}{011501}.
  \bibinfo{doi}{\doi{10.1103/PhysRevD.93.011501}}.
\eprint{1510.02317}.

\bibtype{Article}%
\bibitem[Adare et al.(2016{\natexlab{b}})]{PHENIX:2015ade}
\bibinfo{author}{Adare A} and  et al. (\bibinfo{collaboration}{PHENIX})
  (\bibinfo{year}{2016}{\natexlab{b}}).
\bibinfo{title}{{Measurement of parity-violating spin asymmetries in W$^{\pm}$
  production at midrapidity in longitudinally polarized $p+p$ collisions}}.
\bibinfo{journal}{{\em Phys. Rev. D}} \bibinfo{volume}{93}
  (\bibinfo{number}{5}): \bibinfo{pages}{051103}.
  \bibinfo{doi}{\doi{10.1103/PhysRevD.93.051103}}.
\eprint{1504.07451}.

\bibtype{Article}%
\bibitem[Adeva et al.(1998)]{SpinMuon:1998eqa}
\bibinfo{author}{Adeva B} and  et al. (\bibinfo{collaboration}{Spin Muon})
  (\bibinfo{year}{1998}).
\bibinfo{title}{{Spin asymmetries A(1) and structure functions g1 of the proton
  and the deuteron from polarized high-energy muon scattering}}.
\bibinfo{journal}{{\em Phys. Rev. D}} \bibinfo{volume}{58}:
  \bibinfo{pages}{112001}. \bibinfo{doi}{\doi{10.1103/PhysRevD.58.112001}}.

\bibtype{Article}%
\bibitem[Adeva et al.(1999)]{SpinMuon:1999udj}
\bibinfo{author}{Adeva B} and  et al. (\bibinfo{collaboration}{Spin Muon})
  (\bibinfo{year}{1999}).
\bibinfo{title}{{Spin asymmetries A(1) of the proton and the deuteron in the
  low x and low Q**2 region from polarized high-energy muon scattering}}.
\bibinfo{journal}{{\em Phys. Rev. D}} \bibinfo{volume}{60}:
  \bibinfo{pages}{072004}. \bibinfo{doi}{\doi{10.1103/PhysRevD.60.072004}}.
\bibinfo{note}{[Erratum: Phys.Rev.D 62, 079902 (2000)]}.

\bibtype{Article}%
\bibitem[Adolph et al.(2016)]{COMPASS:2015mhb}
\bibinfo{author}{Adolph C} and  et al. (\bibinfo{collaboration}{COMPASS})
  (\bibinfo{year}{2016}).
\bibinfo{title}{{The spin structure function $g_1^{\rm p}$ of the proton and a
  test of the Bjorken sum rule}}.
\bibinfo{journal}{{\em Phys. Lett. B}} \bibinfo{volume}{753}:
  \bibinfo{pages}{18--28}. \bibinfo{doi}{\doi{10.1016/j.physletb.2015.11.064}}.
\eprint{1503.08935}.

\bibtype{Article}%
\bibitem[Adolph et al.(2017)]{COMPASS:2016jwv}
\bibinfo{author}{Adolph C} and  et al. (\bibinfo{collaboration}{COMPASS})
  (\bibinfo{year}{2017}).
\bibinfo{title}{{Final COMPASS results on the deuteron spin-dependent structure
  function $g_1^{\rm d}$ and the Bjorken sum rule}}.
\bibinfo{journal}{{\em Phys. Lett. B}} \bibinfo{volume}{769}:
  \bibinfo{pages}{34--41}. \bibinfo{doi}{\doi{10.1016/j.physletb.2017.03.018}}.
\eprint{1612.00620}.

\bibtype{Article}%
\bibitem[Aidala et al.(2013)]{Aidala:2012mv}
\bibinfo{author}{Aidala CA}, \bibinfo{author}{Bass SD}, \bibinfo{author}{Hasch
  D} and  \bibinfo{author}{Mallot GK} (\bibinfo{year}{2013}).
\bibinfo{title}{{The Spin Structure of the Nucleon}}.
\bibinfo{journal}{{\em Rev. Mod. Phys.}} \bibinfo{volume}{85}:
  \bibinfo{pages}{655--691}. \bibinfo{doi}{\doi{10.1103/RevModPhys.85.655}}.
\eprint{1209.2803}.

\bibtype{Article}%
\bibitem[Airapetian et al.(2007)]{HERMES:2006jyl}
\bibinfo{author}{Airapetian A} and  et al. (\bibinfo{collaboration}{HERMES})
  (\bibinfo{year}{2007}).
\bibinfo{title}{{Precise determination of the spin structure function g(1) of
  the proton, deuteron and neutron}}.
\bibinfo{journal}{{\em Phys. Rev. D}} \bibinfo{volume}{75}:
  \bibinfo{pages}{012007}. \bibinfo{doi}{\doi{10.1103/PhysRevD.75.012007}}.
\eprint{hep-ex/0609039}.

\bibtype{Article}%
\bibitem[Airapetian et al.(2019)]{HERMES:2018awh}
\bibinfo{author}{Airapetian A} and  et al. (\bibinfo{collaboration}{HERMES})
  (\bibinfo{year}{2019}).
\bibinfo{title}{{Longitudinal double-spin asymmetries in semi-inclusive
  deep-inelastic scattering of electrons and positrons by protons and
  deuterons}}.
\bibinfo{journal}{{\em Phys. Rev. D}} \bibinfo{volume}{99}
  (\bibinfo{number}{11}): \bibinfo{pages}{112001}.
  \bibinfo{doi}{\doi{10.1103/PhysRevD.99.112001}}.
\eprint{1810.07054}.

\bibtype{Article}%
\bibitem[Alekseev et al.(2010)]{COMPASS:2010hwr}
\bibinfo{author}{Alekseev MG} and  et al. (\bibinfo{collaboration}{COMPASS})
  (\bibinfo{year}{2010}).
\bibinfo{title}{{Quark helicity distributions from longitudinal spin
  asymmetries in muon-proton and muon-deuteron scattering}}.
\bibinfo{journal}{{\em Phys. Lett. B}} \bibinfo{volume}{693}:
  \bibinfo{pages}{227--235}.
  \bibinfo{doi}{\doi{10.1016/j.physletb.2010.08.034}}.
\eprint{1007.4061}.

\bibtype{Article}%
\bibitem[Alguard et al.(1976)]{Alguard:1976bm}
\bibinfo{author}{Alguard MJ} and  et al. (\bibinfo{year}{1976}).
\bibinfo{title}{{Deep Inelastic Scattering of Polarized Electrons by Polarized
  Protons}}.
\bibinfo{journal}{{\em Phys. Rev. Lett.}} \bibinfo{volume}{37}:
  \bibinfo{pages}{1261}. \bibinfo{doi}{\doi{10.1103/PhysRevLett.37.1261}}.

\bibtype{Article}%
\bibitem[Almasy et al.(2012)]{Almasy:2011eq}
\bibinfo{author}{Almasy AA}, \bibinfo{author}{Moch S} and
  \bibinfo{author}{Vogt A} (\bibinfo{year}{2012}).
\bibinfo{title}{{On the Next-to-Next-to-Leading Order Evolution of
  Flavour-Singlet Fragmentation Functions}}.
\bibinfo{journal}{{\em Nucl. Phys. B}} \bibinfo{volume}{854}:
  \bibinfo{pages}{133--152}.
  \bibinfo{doi}{\doi{10.1016/j.nuclphysb.2011.08.028}}.
\eprint{1107.2263}.

\bibtype{Article}%
\bibitem[Altarelli and Parisi(1977)]{Altarelli:1977zs}
\bibinfo{author}{Altarelli G} and  \bibinfo{author}{Parisi G}
  (\bibinfo{year}{1977}).
\bibinfo{title}{{Asymptotic Freedom in Parton Language}}.
\bibinfo{journal}{{\em Nucl. Phys. B}} \bibinfo{volume}{126}:
  \bibinfo{pages}{298--318}. \bibinfo{doi}{\doi{10.1016/0550-3213(77)90384-4}}.

\bibtype{Article}%
\bibitem[Altarelli and Ross(1988)]{Altarelli:1988nr}
\bibinfo{author}{Altarelli G} and  \bibinfo{author}{Ross GG}
  (\bibinfo{year}{1988}).
\bibinfo{title}{{The Anomalous Gluon Contribution to Polarized
  Leptoproduction}}.
\bibinfo{journal}{{\em Phys. Lett. B}} \bibinfo{volume}{212}:
  \bibinfo{pages}{391--396}. \bibinfo{doi}{\doi{10.1016/0370-2693(88)91335-4}}.

\bibtype{Article}%
\bibitem[Altarelli et al.(1998)]{Altarelli:1998gn}
\bibinfo{author}{Altarelli G}, \bibinfo{author}{Forte S} and
  \bibinfo{author}{Ridolfi G} (\bibinfo{year}{1998}).
\bibinfo{title}{{On positivity of parton distributions}}.
\bibinfo{journal}{{\em Nucl. Phys. B}} \bibinfo{volume}{534}:
  \bibinfo{pages}{277--296}.
  \bibinfo{doi}{\doi{10.1016/S0550-3213(98)00661-0}}.
\eprint{hep-ph/9806345}.

\bibtype{Article}%
\bibitem[Anastasiou et al.(2003)]{Anastasiou:2003yy}
\bibinfo{author}{Anastasiou C}, \bibinfo{author}{Dixon LJ},
  \bibinfo{author}{Melnikov K} and  \bibinfo{author}{Petriello F}
  (\bibinfo{year}{2003}).
\bibinfo{title}{{Dilepton rapidity distribution in the Drell-Yan process at
  NNLO in QCD}}.
\bibinfo{journal}{{\em Phys. Rev. Lett.}} \bibinfo{volume}{91}:
  \bibinfo{pages}{182002}. \bibinfo{doi}{\doi{10.1103/PhysRevLett.91.182002}}.
\eprint{hep-ph/0306192}.

\bibtype{Article}%
\bibitem[Anastasiou et al.(2004)]{Anastasiou:2003ds}
\bibinfo{author}{Anastasiou C}, \bibinfo{author}{Dixon LJ},
  \bibinfo{author}{Melnikov K} and  \bibinfo{author}{Petriello F}
  (\bibinfo{year}{2004}).
\bibinfo{title}{{High precision QCD at hadron colliders: Electroweak gauge
  boson rapidity distributions at NNLO}}.
\bibinfo{journal}{{\em Phys. Rev. D}} \bibinfo{volume}{69}:
  \bibinfo{pages}{094008}. \bibinfo{doi}{\doi{10.1103/PhysRevD.69.094008}}.
\eprint{hep-ph/0312266}.

\bibtype{Article}%
\bibitem[Anderle et al.(2013)]{Anderle:2013lka}
\bibinfo{author}{Anderle DP}, \bibinfo{author}{Ringer F} and
  \bibinfo{author}{Vogelsang W} (\bibinfo{year}{2013}).
\bibinfo{title}{{Threshold resummation for polarized (semi-)inclusive deep
  inelastic scattering}}.
\bibinfo{journal}{{\em Phys. Rev. D}} \bibinfo{volume}{87}:
  \bibinfo{pages}{094021}. \bibinfo{doi}{\doi{10.1103/PhysRevD.87.094021}}.
\eprint{1304.1373}.

\bibtype{Article}%
\bibitem[Anselmino et al.(1994)]{Anselmino:1993tc}
\bibinfo{author}{Anselmino M}, \bibinfo{author}{Gambino P} and
  \bibinfo{author}{Kalinowski J} (\bibinfo{year}{1994}).
\bibinfo{title}{{Polarized deep inelastic scattering at high-energies and
  parity violating structure functions}}.
\bibinfo{journal}{{\em Z. Phys. C}} \bibinfo{volume}{64}:
  \bibinfo{pages}{267--274}. \bibinfo{doi}{\doi{10.1007/BF01557397}}.
\eprint{hep-ph/9401264}.

\bibtype{Article}%
\bibitem[Anthony et al.(1996)]{E142:1996thl}
\bibinfo{author}{Anthony PL} and  et al. (\bibinfo{collaboration}{E142})
  (\bibinfo{year}{1996}).
\bibinfo{title}{{Deep inelastic scattering of polarized electrons by polarized
  He-3 and the study of the neutron spin structure}}.
\bibinfo{journal}{{\em Phys. Rev. D}} \bibinfo{volume}{54}:
  \bibinfo{pages}{6620--6650}. \bibinfo{doi}{\doi{10.1103/PhysRevD.54.6620}}.
\eprint{hep-ex/9610007}.

\bibtype{Article}%
\bibitem[Anthony et al.(2000)]{E155:2000qdr}
\bibinfo{author}{Anthony PL} and  et al. (\bibinfo{collaboration}{E155})
  (\bibinfo{year}{2000}).
\bibinfo{title}{{Measurements of the Q**2 dependence of the proton and neutron
  spin structure functions g(1)**p and g(1)**n}}.
\bibinfo{journal}{{\em Phys. Lett. B}} \bibinfo{volume}{493}:
  \bibinfo{pages}{19--28}. \bibinfo{doi}{\doi{10.1016/S0370-2693(00)01014-5}}.
\eprint{hep-ph/0007248}.

\bibtype{Article}%
\bibitem[Appelquist and Carazzone(1975)]{Appelquist:1974tg}
\bibinfo{author}{Appelquist T} and  \bibinfo{author}{Carazzone J}
  (\bibinfo{year}{1975}).
\bibinfo{title}{{Infrared Singularities and Massive Fields}}.
\bibinfo{journal}{{\em Phys. Rev. D}} \bibinfo{volume}{11}:
  \bibinfo{pages}{2856}. \bibinfo{doi}{\doi{10.1103/PhysRevD.11.2856}}.

\bibtype{Article}%
\bibitem[Aschenauer et al.(2012)]{Aschenauer:2012ve}
\bibinfo{author}{Aschenauer EC}, \bibinfo{author}{Sassot R} and
  \bibinfo{author}{Stratmann M} (\bibinfo{year}{2012}).
\bibinfo{title}{{Helicity Parton Distributions at a Future Electron-Ion
  Collider: A Quantitative Appraisal}}.
\bibinfo{journal}{{\em Phys. Rev. D}} \bibinfo{volume}{86}:
  \bibinfo{pages}{054020}. \bibinfo{doi}{\doi{10.1103/PhysRevD.86.054020}}.
\eprint{1206.6014}.

\bibtype{Article}%
\bibitem[Aschenauer et al.(2013)]{Aschenauer:2013iia}
\bibinfo{author}{Aschenauer EC}, \bibinfo{author}{Burton T},
  \bibinfo{author}{Martini T}, \bibinfo{author}{Spiesberger H} and
  \bibinfo{author}{Stratmann M} (\bibinfo{year}{2013}).
\bibinfo{title}{{Prospects for Charged Current Deep-Inelastic Scattering off
  Polarized Nucleons at a Future Electron-Ion Collider}}.
\bibinfo{journal}{{\em Phys. Rev. D}} \bibinfo{volume}{88}:
  \bibinfo{pages}{114025}. \bibinfo{doi}{\doi{10.1103/PhysRevD.88.114025}}.
\eprint{1309.5327}.

\bibtype{Article}%
\bibitem[Aschenauer et al.(2015)]{Aschenauer:2015ata}
\bibinfo{author}{Aschenauer EC}, \bibinfo{author}{Sassot R} and
  \bibinfo{author}{Stratmann M} (\bibinfo{year}{2015}).
\bibinfo{title}{{Unveiling the Proton Spin Decomposition at a Future
  Electron-Ion Collider}}.
\bibinfo{journal}{{\em Phys. Rev. D}} \bibinfo{volume}{92}
  (\bibinfo{number}{9}): \bibinfo{pages}{094030}.
  \bibinfo{doi}{\doi{10.1103/PhysRevD.92.094030}}.
\eprint{1509.06489}.

\bibtype{Article}%
\bibitem[Aschenauer et al.(2019)]{Aschenauer:2019kzf}
\bibinfo{author}{Aschenauer EC}, \bibinfo{author}{Borsa I},
  \bibinfo{author}{Sassot R} and  \bibinfo{author}{Van~Hulse C}
  (\bibinfo{year}{2019}).
\bibinfo{title}{{Semi-inclusive Deep-Inelastic Scattering, Parton Distributions
  and Fragmentation Functions at a Future Electron-Ion Collider}}.
\bibinfo{journal}{{\em Phys. Rev. D}} \bibinfo{volume}{99}
  (\bibinfo{number}{9}): \bibinfo{pages}{094004}.
  \bibinfo{doi}{\doi{10.1103/PhysRevD.99.094004}}.
\eprint{1902.10663}.

\bibtype{Article}%
\bibitem[Ashman et al.(1989)]{EuropeanMuon:1989yki}
\bibinfo{author}{Ashman J} and  et al. (\bibinfo{collaboration}{European Muon})
  (\bibinfo{year}{1989}).
\bibinfo{title}{{An Investigation of the Spin Structure of the Proton in Deep
  Inelastic Scattering of Polarized Muons on Polarized Protons}}.
\bibinfo{journal}{{\em Nucl. Phys. B}} \bibinfo{volume}{328}:
  \bibinfo{pages}{1}. \bibinfo{doi}{\doi{10.1016/0550-3213(89)90089-8}}.

\bibtype{inbook}%
\bibitem[Ball and Deshpande(2019)]{Ball:2018lag}
\bibinfo{author}{Ball RD} and  \bibinfo{author}{Deshpande A}
  (\bibinfo{year}{2019}).
\bibinfo{title}{{The proton spin, semi-inclusive processes, and measurements at
  a future Electron Ion Collider}}.
 \bibinfo{pages}{205--226}.
\bibinfo{doi}{\doi{10.1142/9789813238053_0011}}.
\eprint{1801.04842}.

\bibtype{Article}%
\bibitem[Ball et al.(2010)]{Ball:2009qv}
\bibinfo{author}{Ball RD}, \bibinfo{author}{Del~Debbio L},
  \bibinfo{author}{Forte S}, \bibinfo{author}{Guffanti A},
  \bibinfo{author}{Latorre JI}, \bibinfo{author}{Rojo J} and
  \bibinfo{author}{Ubiali M} (\bibinfo{collaboration}{NNPDF})
  (\bibinfo{year}{2010}).
\bibinfo{title}{{Fitting Parton Distribution Data with Multiplicative
  Normalization Uncertainties}}.
\bibinfo{journal}{{\em JHEP}} \bibinfo{volume}{05}: \bibinfo{pages}{075}.
  \bibinfo{doi}{\doi{10.1007/JHEP05(2010)075}}.
\eprint{0912.2276}.

\bibtype{Article}%
\bibitem[Ball et al.(2011)]{Ball:2010gb}
\bibinfo{author}{Ball RD}, \bibinfo{author}{Bertone V},
  \bibinfo{author}{Cerutti F}, \bibinfo{author}{Del~Debbio L},
  \bibinfo{author}{Forte S}, \bibinfo{author}{Guffanti A},
  \bibinfo{author}{Latorre JI}, \bibinfo{author}{Rojo J} and
  \bibinfo{author}{Ubiali M} (\bibinfo{collaboration}{NNPDF})
  (\bibinfo{year}{2011}).
\bibinfo{title}{{Reweighting NNPDFs: the W lepton asymmetry}}.
\bibinfo{journal}{{\em Nucl. Phys. B}} \bibinfo{volume}{849}:
  \bibinfo{pages}{112--143}.
  \bibinfo{doi}{\doi{10.1016/j.nuclphysb.2011.03.017}}.
\bibinfo{note}{[Erratum: Nucl.Phys.B 854, 926--927 (2012), Erratum: Nucl.Phys.B
  855, 927--928 (2012)]}, \eprint{1012.0836}.

\bibtype{Article}%
\bibitem[Ball et al.(2012)]{Ball:2011gg}
\bibinfo{author}{Ball RD}, \bibinfo{author}{Bertone V},
  \bibinfo{author}{Cerutti F}, \bibinfo{author}{Del~Debbio L},
  \bibinfo{author}{Forte S}, \bibinfo{author}{Guffanti A},
  \bibinfo{author}{Hartland NP}, \bibinfo{author}{Latorre JI},
  \bibinfo{author}{Rojo J} and  \bibinfo{author}{Ubiali M}
  (\bibinfo{year}{2012}).
\bibinfo{title}{{Reweighting and Unweighting of Parton Distributions and the
  LHC W lepton asymmetry data}}.
\bibinfo{journal}{{\em Nucl. Phys. B}} \bibinfo{volume}{855}:
  \bibinfo{pages}{608--638}.
  \bibinfo{doi}{\doi{10.1016/j.nuclphysb.2011.10.018}}.
\eprint{1108.1758}.

\bibtype{Article}%
\bibitem[Ball et al.(2013)]{Ball:2013lla}
\bibinfo{author}{Ball RD}, \bibinfo{author}{Forte S}, \bibinfo{author}{Guffanti
  A}, \bibinfo{author}{Nocera ER}, \bibinfo{author}{Ridolfi G} and
  \bibinfo{author}{Rojo J} (\bibinfo{collaboration}{NNPDF})
  (\bibinfo{year}{2013}).
\bibinfo{title}{{Unbiased determination of polarized parton distributions and
  their uncertainties}}.
\bibinfo{journal}{{\em Nucl. Phys. B}} \bibinfo{volume}{874}:
  \bibinfo{pages}{36--84}.
  \bibinfo{doi}{\doi{10.1016/j.nuclphysb.2013.05.007}}.
\eprint{1303.7236}.

\bibtype{Article}%
\bibitem[Ball et al.(2014)]{Ball:2013tyh}
\bibinfo{author}{Ball RD}, \bibinfo{author}{Forte S}, \bibinfo{author}{Guffanti
  A}, \bibinfo{author}{Nocera ER}, \bibinfo{author}{Ridolfi G} and
  \bibinfo{author}{Rojo J} (\bibinfo{collaboration}{NNPDF})
  (\bibinfo{year}{2014}).
\bibinfo{title}{{Polarized Parton Distributions at an Electron-Ion Collider}}.
\bibinfo{journal}{{\em Phys. Lett. B}} \bibinfo{volume}{728}:
  \bibinfo{pages}{524--531}.
  \bibinfo{doi}{\doi{10.1016/j.physletb.2013.12.023}}.
\eprint{1310.0461}.

\bibtype{Article}%
\bibitem[Ball et al.(2015)]{NNPDF:2014otw}
\bibinfo{author}{Ball RD} and  et al. (\bibinfo{collaboration}{NNPDF})
  (\bibinfo{year}{2015}).
\bibinfo{title}{{Parton distributions for the LHC Run II}}.
\bibinfo{journal}{{\em JHEP}} \bibinfo{volume}{04}: \bibinfo{pages}{040}.
  \bibinfo{doi}{\doi{10.1007/JHEP04(2015)040}}.
\eprint{1410.8849}.

\bibtype{Article}%
\bibitem[Ball et al.(2019)]{Ball:2018twp}
\bibinfo{author}{Ball RD}, \bibinfo{author}{Nocera ER} and
  \bibinfo{author}{Pearson RL} (\bibinfo{collaboration}{NNPDF})
  (\bibinfo{year}{2019}).
\bibinfo{title}{{Nuclear Uncertainties in the Determination of Proton PDFs}}.
\bibinfo{journal}{{\em Eur. Phys. J. C}} \bibinfo{volume}{79}
  (\bibinfo{number}{3}): \bibinfo{pages}{282}.
  \bibinfo{doi}{\doi{10.1140/epjc/s10052-019-6793-5}}.
\eprint{1812.09074}.

\bibtype{Article}%
\bibitem[Ball et al.(2021)]{Ball:2020xqw}
\bibinfo{author}{Ball RD}, \bibinfo{author}{Nocera ER} and
  \bibinfo{author}{Pearson RL} (\bibinfo{year}{2021}).
\bibinfo{title}{{Deuteron Uncertainties in the Determination of Proton PDFs}}.
\bibinfo{journal}{{\em Eur. Phys. J. C}} \bibinfo{volume}{81}
  (\bibinfo{number}{1}): \bibinfo{pages}{37}.
  \bibinfo{doi}{\doi{10.1140/epjc/s10052-020-08826-7}}.
\eprint{2011.00009}.

\bibtype{Article}%
\bibitem[Ball et al.(2022{\natexlab{a}})]{NNPDF:2021njg}
\bibinfo{author}{Ball RD} and  et al. (\bibinfo{collaboration}{NNPDF})
  (\bibinfo{year}{2022}{\natexlab{a}}).
\bibinfo{title}{{The path to proton structure at 1{\%} accuracy}}.
\bibinfo{journal}{{\em Eur. Phys. J. C}} \bibinfo{volume}{82}
  (\bibinfo{number}{5}): \bibinfo{pages}{428}.
  \bibinfo{doi}{\doi{10.1140/epjc/s10052-022-10328-7}}.
\eprint{2109.02653}.

\bibtype{Article}%
\bibitem[Ball et al.(2022{\natexlab{b}})]{PDF4LHCWorkingGroup:2022cjn}
\bibinfo{author}{Ball RD} and  et al. (\bibinfo{collaboration}{PDF4LHC Working
  Group}) (\bibinfo{year}{2022}{\natexlab{b}}).
\bibinfo{title}{{The PDF4LHC21 combination of global PDF fits for the LHC Run
  III}}.
\bibinfo{journal}{{\em J. Phys. G}} \bibinfo{volume}{49} (\bibinfo{number}{8}):
  \bibinfo{pages}{080501}. \bibinfo{doi}{\doi{10.1088/1361-6471/ac7216}}.
\eprint{2203.05506}.

\bibtype{Article}%
\bibitem[Ball et al.(2024)]{NNPDF:2024dpb}
\bibinfo{author}{Ball RD} and  et al. (\bibinfo{collaboration}{NNPDF})
  (\bibinfo{year}{2024}).
\bibinfo{title}{{Determination of the theory uncertainties from missing higher
  orders on NNLO parton distributions with percent accuracy}}.
\bibinfo{journal}{{\em Eur. Phys. J. C}} \bibinfo{volume}{84}
  (\bibinfo{number}{5}): \bibinfo{pages}{517}.
  \bibinfo{doi}{\doi{10.1140/epjc/s10052-024-12772-z}}.
\eprint{2401.10319}.

\bibtype{Article}%
\bibitem[Ball et al.(2026)]{Ball:2025xtj}
\bibinfo{author}{Ball RD}, \bibinfo{author}{Chiefa A} and
  \bibinfo{author}{Stegeman R} (\bibinfo{year}{2026}).
\bibinfo{title}{{Parton distributions with higher twist and jet power
  corrections}}.
\bibinfo{journal}{{\em Eur. Phys. J. C}} \bibinfo{volume}{86}
  (\bibinfo{number}{3}): \bibinfo{pages}{281}.
  \bibinfo{doi}{\doi{10.1140/epjc/s10052-026-15485-7}}.
\eprint{2511.14387}.

\bibtype{Article}%
\bibitem[Bardeen et al.(1978)]{Bardeen:1978yd}
\bibinfo{author}{Bardeen WA}, \bibinfo{author}{Buras AJ}, \bibinfo{author}{Duke
  DW} and  \bibinfo{author}{Muta T} (\bibinfo{year}{1978}).
\bibinfo{title}{{Deep Inelastic Scattering Beyond the Leading Order in
  Asymptotically Free Gauge Theories}}.
\bibinfo{journal}{{\em Phys. Rev. D}} \bibinfo{volume}{18}:
  \bibinfo{pages}{3998}. \bibinfo{doi}{\doi{10.1103/PhysRevD.18.3998}}.

\bibtype{Article}%
\bibitem[Bass and Thomas(2010)]{Bass:2009ed}
\bibinfo{author}{Bass SD} and  \bibinfo{author}{Thomas AW}
  (\bibinfo{year}{2010}).
\bibinfo{title}{{The nucleon's octet axial-charge $g_A^{(8)}$ with chiral
  corrections}}.
\bibinfo{journal}{{\em Phys. Lett. B}} \bibinfo{volume}{684}:
  \bibinfo{pages}{216--220}.
  \bibinfo{doi}{\doi{10.1016/j.physletb.2010.01.008}}.
\eprint{0912.1765}.

\bibtype{Article}%
\bibitem[Baum et al.(1983)]{Baum:1983ha}
\bibinfo{author}{Baum G} and  et al. (\bibinfo{year}{1983}).
\bibinfo{title}{{A New Measurement of Deep Inelastic e p Asymmetries}}.
\bibinfo{journal}{{\em Phys. Rev. Lett.}} \bibinfo{volume}{51}:
  \bibinfo{pages}{1135}. \bibinfo{doi}{\doi{10.1103/PhysRevLett.51.1135}}.

\bibtype{Article}%
\bibitem[Behring et al.(2015)]{Behring:2015zaa}
\bibinfo{author}{Behring A}, \bibinfo{author}{Bl{\"u}mlein J},
  \bibinfo{author}{De~Freitas A}, \bibinfo{author}{von Manteuffel A} and
  \bibinfo{author}{Schneider C} (\bibinfo{year}{2015}).
\bibinfo{title}{{The 3-Loop Non-Singlet Heavy Flavor Contributions to the
  Structure Function $g_{1}(x,Q^{2})$ at Large Momentum Transfer}}.
\bibinfo{journal}{{\em Nucl. Phys. B}} \bibinfo{volume}{897}:
  \bibinfo{pages}{612--644}.
  \bibinfo{doi}{\doi{10.1016/j.nuclphysb.2015.06.007}}.
\eprint{1504.08217}.

\bibtype{Article}%
\bibitem[Behring et al.(2021)]{Behring:2021asx}
\bibinfo{author}{Behring A}, \bibinfo{author}{Bl{\"u}mlein J},
  \bibinfo{author}{De~Freitas A}, \bibinfo{author}{von Manteuffel A},
  \bibinfo{author}{Sch{\"o}nwald K} and  \bibinfo{author}{Schneider C}
  (\bibinfo{year}{2021}).
\bibinfo{title}{{The polarized transition matrix element $A_{gq}(N)$ of the
  variable flavor number scheme at $O(\alpha^3_s)$}}.
\bibinfo{journal}{{\em Nucl. Phys. B}} \bibinfo{volume}{964}:
  \bibinfo{pages}{115331}.
  \bibinfo{doi}{\doi{10.1016/j.nuclphysb.2021.115331}}.
\eprint{2101.05733}.

\bibtype{Article}%
\bibitem[Bertone et al.(2014)]{Bertone:2013vaa}
\bibinfo{author}{Bertone V}, \bibinfo{author}{Carrazza S} and
  \bibinfo{author}{Rojo J} (\bibinfo{year}{2014}).
\bibinfo{title}{{APFEL: A PDF Evolution Library with QED corrections}}.
\bibinfo{journal}{{\em Comput. Phys. Commun.}} \bibinfo{volume}{185}:
  \bibinfo{pages}{1647--1668}. \bibinfo{doi}{\doi{10.1016/j.cpc.2014.03.007}}.
\eprint{1310.1394}.

\bibtype{Article}%
\bibitem[Bertone et al.(2015)]{Bertone:2015cwa}
\bibinfo{author}{Bertone V}, \bibinfo{author}{Carrazza S} and
  \bibinfo{author}{Nocera ER} (\bibinfo{year}{2015}).
\bibinfo{title}{{Reference results for time-like evolution up to $
  \mathcal{O}\left({\alpha}_s^3\right) $}}.
\bibinfo{journal}{{\em JHEP}} \bibinfo{volume}{03}: \bibinfo{pages}{046}.
  \bibinfo{doi}{\doi{10.1007/JHEP03(2015)046}}.
\eprint{1501.00494}.

\bibtype{Article}%
\bibitem[Bertone et al.(2025)]{Bertone:2024taw}
\bibinfo{author}{Bertone V}, \bibinfo{author}{Chiefa A} and
  \bibinfo{author}{Nocera ER} (\bibinfo{collaboration}{MAP (Multi-dimensional
  Analyses of Partonic distributions)}) (\bibinfo{year}{2025}).
\bibinfo{title}{{Helicity-dependent parton distribution functions at
  next-to-next-to-leading order accuracy from inclusive and semi-inclusive
  deep-inelastic scattering data}}.
\bibinfo{journal}{{\em Phys. Lett. B}} \bibinfo{volume}{865}:
  \bibinfo{pages}{139497}. \bibinfo{doi}{\doi{10.1016/j.physletb.2025.139497}}.
\eprint{2404.04712}.

\bibtype{Article}%
\bibitem[Bierenbaum et al.(2009)]{Bierenbaum:2009mv}
\bibinfo{author}{Bierenbaum I}, \bibinfo{author}{Blumlein J} and
  \bibinfo{author}{Klein S} (\bibinfo{year}{2009}).
\bibinfo{title}{{Mellin Moments of the O(alpha**3(s)) Heavy Flavor
  Contributions to unpolarized Deep-Inelastic Scattering at Q**2
  {\ensuremath{>}}{\ensuremath{>}} m**2 and Anomalous Dimensions}}.
\bibinfo{journal}{{\em Nucl. Phys. B}} \bibinfo{volume}{820}:
  \bibinfo{pages}{417--482}.
  \bibinfo{doi}{\doi{10.1016/j.nuclphysb.2009.06.005}}.
\eprint{0904.3563}.

\bibtype{Article}%
\bibitem[Bierenbaum et al.(2023)]{Bierenbaum:2022biv}
\bibinfo{author}{Bierenbaum I}, \bibinfo{author}{Bl{\"u}mlein J},
  \bibinfo{author}{De~Freitas A}, \bibinfo{author}{Goedicke A},
  \bibinfo{author}{Klein S} and  \bibinfo{author}{Sch{\"o}nwald K}
  (\bibinfo{year}{2023}).
\bibinfo{title}{{$O(\alpha_s^2$) polarized heavy flavor corrections to
  deep-inelastic scattering at $Q^2 \gg m^2$}}.
\bibinfo{journal}{{\em Nucl. Phys. B}} \bibinfo{volume}{988}:
  \bibinfo{pages}{116114}.
  \bibinfo{doi}{\doi{10.1016/j.nuclphysb.2023.116114}}.
\eprint{2211.15337}.

\bibtype{Inproceedings}%
\bibitem[Binoth et al.(2010)]{SM:2010nsa}
\bibinfo{author}{Binoth T} and  et al. (\bibinfo{collaboration}{SM, NLO
  Multileg Working Group}) (\bibinfo{year}{2010}), \bibinfo{month}{3},
  \bibinfo{title}{{The SM and NLO Multileg Working Group: Summary report}},
  \bibinfo{booktitle}{{6th Les Houches Workshop on Physics at TeV Colliders}},
  \bibinfo{pages}{21--189}, \eprint{1003.1241}.

\bibtype{Article}%
\bibitem[Bjorken(1966)]{Bjorken:1966jh}
\bibinfo{author}{Bjorken JD} (\bibinfo{year}{1966}).
\bibinfo{title}{{Applications of the Chiral U(6) x (6) Algebra of Current
  Densities}}.
\bibinfo{journal}{{\em Phys. Rev.}} \bibinfo{volume}{148}:
  \bibinfo{pages}{1467--1478}. \bibinfo{doi}{\doi{10.1103/PhysRev.148.1467}}.

\bibtype{Article}%
\bibitem[Bjorken(1969)]{Bjorken:1968dy}
\bibinfo{author}{Bjorken JD} (\bibinfo{year}{1969}).
\bibinfo{title}{{Asymptotic Sum Rules at Infinite Momentum}}.
\bibinfo{journal}{{\em Phys. Rev.}} \bibinfo{volume}{179}:
  \bibinfo{pages}{1547--1553}. \bibinfo{doi}{\doi{10.1103/PhysRev.179.1547}}.

\bibtype{Article}%
\bibitem[Bjorken(1970)]{Bjorken:1969mm}
\bibinfo{author}{Bjorken JD} (\bibinfo{year}{1970}).
\bibinfo{title}{{Inelastic Scattering of Polarized Leptons from Polarized
  Nucleons}}.
\bibinfo{journal}{{\em Phys. Rev. D}} \bibinfo{volume}{1}:
  \bibinfo{pages}{1376--1379}. \bibinfo{doi}{\doi{10.1103/PhysRevD.1.1376}}.

\bibtype{Article}%
\bibitem[Blumlein and Bottcher(2002)]{Blumlein:2002qeu}
\bibinfo{author}{Blumlein J} and  \bibinfo{author}{Bottcher H}
  (\bibinfo{year}{2002}).
\bibinfo{title}{{QCD analysis of polarized deep inelastic data and parton
  distributions}}.
\bibinfo{journal}{{\em Nucl. Phys. B}} \bibinfo{volume}{636}:
  \bibinfo{pages}{225--263}.
  \bibinfo{doi}{\doi{10.1016/S0550-3213(02)00342-5}}.
\eprint{hep-ph/0203155}.

\bibtype{Article}%
\bibitem[Blumlein and Bottcher(2010)]{Blumlein:2010rn}
\bibinfo{author}{Blumlein J} and  \bibinfo{author}{Bottcher H}
  (\bibinfo{year}{2010}).
\bibinfo{title}{{QCD Analysis of Polarized Deep Inelastic Scattering Data}}.
\bibinfo{journal}{{\em Nucl. Phys. B}} \bibinfo{volume}{841}:
  \bibinfo{pages}{205--230}.
  \bibinfo{doi}{\doi{10.1016/j.nuclphysb.2010.08.005}}.
\eprint{1005.3113}.

\bibtype{Article}%
\bibitem[Bl{\"u}mlein et al.(2021{\natexlab{a}})]{Blumlein:2021xlc}
\bibinfo{author}{Bl{\"u}mlein J}, \bibinfo{author}{De~Freitas A},
  \bibinfo{author}{Saragnese M}, \bibinfo{author}{Schneider C} and
  \bibinfo{author}{Sch{\"o}nwald K} (\bibinfo{year}{2021}{\natexlab{a}}).
\bibinfo{title}{{Logarithmic contributions to the polarized
  O({\ensuremath{\alpha}}s3) asymptotic massive Wilson coefficients and
  operator matrix elements in deeply inelastic scattering}}.
\bibinfo{journal}{{\em Phys. Rev. D}} \bibinfo{volume}{104}
  (\bibinfo{number}{3}): \bibinfo{pages}{034030}.
  \bibinfo{doi}{\doi{10.1103/PhysRevD.104.034030}}.
\eprint{2105.09572}.

\bibtype{Article}%
\bibitem[Bl{\"u}mlein et al.(2021{\natexlab{b}})]{Blumlein:2021enk}
\bibinfo{author}{Bl{\"u}mlein J}, \bibinfo{author}{Marquard P},
  \bibinfo{author}{Schneider C} and  \bibinfo{author}{Sch{\"o}nwald K}
  (\bibinfo{year}{2021}{\natexlab{b}}).
\bibinfo{title}{{The three-loop unpolarized and polarized non-singlet anomalous
  dimensions from off shell operator matrix elements}}.
\bibinfo{journal}{{\em Nucl. Phys. B}} \bibinfo{volume}{971}:
  \bibinfo{pages}{115542}.
  \bibinfo{doi}{\doi{10.1016/j.nuclphysb.2021.115542}}.
\eprint{2107.06267}.

\bibtype{Article}%
\bibitem[Bl{\"u}mlein et al.(2022{\natexlab{a}})]{Blumlein:2022gpp}
\bibinfo{author}{Bl{\"u}mlein J}, \bibinfo{author}{Marquard P},
  \bibinfo{author}{Schneider C} and  \bibinfo{author}{Sch{\"o}nwald K}
  (\bibinfo{year}{2022}{\natexlab{a}}).
\bibinfo{title}{{The massless three-loop Wilson coefficients for the
  deep-inelastic structure functions F$_{2}$, F$_{L}$, xF$_{3}$ and g$_{1}$}}.
\bibinfo{journal}{{\em JHEP}} \bibinfo{volume}{11}: \bibinfo{pages}{156}.
  \bibinfo{doi}{\doi{10.1007/JHEP11(2022)156}}.
\eprint{2208.14325}.

\bibtype{Article}%
\bibitem[Bl{\"u}mlein et al.(2022{\natexlab{b}})]{Blumlein:2021ryt}
\bibinfo{author}{Bl{\"u}mlein J}, \bibinfo{author}{Marquard P},
  \bibinfo{author}{Schneider C} and  \bibinfo{author}{Sch{\"o}nwald K}
  (\bibinfo{year}{2022}{\natexlab{b}}).
\bibinfo{title}{{The three-loop polarized singlet anomalous dimensions from
  off-shell operator matrix elements}}.
\bibinfo{journal}{{\em JHEP}} \bibinfo{volume}{01}: \bibinfo{pages}{193}.
  \bibinfo{doi}{\doi{10.1007/JHEP01(2022)193}}.
\bibinfo{note}{[Erratum: JHEP 02, 049 (2026)]}, \eprint{2111.12401}.

\bibtype{Article}%
\bibitem[Bonino et al.(2024{\natexlab{a}})]{Bonino:2024wgg}
\bibinfo{author}{Bonino L}, \bibinfo{author}{Gehrmann T},
  \bibinfo{author}{L{\"o}chner M}, \bibinfo{author}{Sch{\"o}nwald K} and
  \bibinfo{author}{Stagnitto G} (\bibinfo{year}{2024}{\natexlab{a}}).
\bibinfo{title}{{Polarized Semi-Inclusive Deep-Inelastic Scattering at
  Next-to-Next-to-Leading Order in QCD}}.
\bibinfo{journal}{{\em Phys. Rev. Lett.}} \bibinfo{volume}{133}
  (\bibinfo{number}{21}): \bibinfo{pages}{211904}.
  \bibinfo{doi}{\doi{10.1103/PhysRevLett.133.211904}}.
\eprint{2404.08597}.

\bibtype{Article}%
\bibitem[Bonino et al.(2024{\natexlab{b}})]{Bonino:2024qbh}
\bibinfo{author}{Bonino L}, \bibinfo{author}{Gehrmann T} and
  \bibinfo{author}{Stagnitto G} (\bibinfo{year}{2024}{\natexlab{b}}).
\bibinfo{title}{{Semi-Inclusive Deep-Inelastic Scattering at
  Next-to-Next-to-Leading Order in QCD}}.
\bibinfo{journal}{{\em Phys. Rev. Lett.}} \bibinfo{volume}{132}
  (\bibinfo{number}{25}): \bibinfo{pages}{251901}.
  \bibinfo{doi}{\doi{10.1103/PhysRevLett.132.251901}}.
\eprint{2401.16281}.

\bibtype{Article}%
\bibitem[Bonino et al.(2025{\natexlab{a}})]{Bonino:2025tnf}
\bibinfo{author}{Bonino L}, \bibinfo{author}{Gehrmann T},
  \bibinfo{author}{L{\"o}chner M}, \bibinfo{author}{Sch{\"o}nwald K} and
  \bibinfo{author}{Stagnitto G} (\bibinfo{year}{2025}{\natexlab{a}}).
\bibinfo{title}{{Identified Hadron Production in Deeply Inelastic
  Neutrino-Nucleon Scattering}}.
\bibinfo{journal}{{\em Phys. Rev. Lett.}} \bibinfo{volume}{135}
  (\bibinfo{number}{21}): \bibinfo{pages}{211902}.
  \bibinfo{doi}{\doi{10.1103/rr84-wlt3}}.
\eprint{2504.05376}.

\bibtype{Article}%
\bibitem[Bonino et al.(2025{\natexlab{b}})]{Bonino:2025qta}
\bibinfo{author}{Bonino L}, \bibinfo{author}{Gehrmann T},
  \bibinfo{author}{L{\"o}chner M}, \bibinfo{author}{Sch{\"o}nwald K} and
  \bibinfo{author}{Stagnitto G} (\bibinfo{year}{2025}{\natexlab{b}}).
\bibinfo{title}{{Neutral and charged current semi-inclusive deep-inelastic
  scattering at NNLO QCD}}.
\bibinfo{journal}{{\em JHEP}} \bibinfo{volume}{10}: \bibinfo{pages}{016}.
  \bibinfo{doi}{\doi{10.1007/JHEP10(2025)016}}.
\eprint{2506.19926}.

\bibtype{Article}%
\bibitem[Bonino et al.(2026)]{Bonino:2025bqa}
\bibinfo{author}{Bonino L}, \bibinfo{author}{Gehrmann T},
  \bibinfo{author}{L{\"o}chner M}, \bibinfo{author}{Sch{\"o}nwald K} and
  \bibinfo{author}{Stagnitto G} (\bibinfo{year}{2026}).
\bibinfo{title}{{Polarized neutral and charged current semi-inclusive
  deep-inelastic scattering at NNLO in QCD}}.
\bibinfo{journal}{{\em JHEP}} \bibinfo{volume}{03}: \bibinfo{pages}{109}.
  \bibinfo{doi}{\doi{10.1007/JHEP03(2026)109}}.
\eprint{2510.00100}.

\bibtype{Article}%
\bibitem[Borden et al.(2024)]{Borden:2024bxa}
\bibinfo{author}{Borden J}, \bibinfo{author}{Kovchegov YV} and
  \bibinfo{author}{Li M} (\bibinfo{year}{2024}).
\bibinfo{title}{{Helicity evolution at small x: quark to gluon and gluon to
  quark transition operators}}.
\bibinfo{journal}{{\em JHEP}} \bibinfo{volume}{09}: \bibinfo{pages}{037}.
  \bibinfo{doi}{\doi{10.1007/JHEP09(2024)037}}.
\eprint{2406.11647}.

\bibtype{Article}%
\bibitem[Borsa et al.(2020)]{Borsa:2020lsz}
\bibinfo{author}{Borsa I}, \bibinfo{author}{Lucero G}, \bibinfo{author}{Sassot
  R}, \bibinfo{author}{Aschenauer EC} and  \bibinfo{author}{Nunes AS}
  (\bibinfo{year}{2020}).
\bibinfo{title}{{Revisiting helicity parton distributions at a future
  electron-ion collider}}.
\bibinfo{journal}{{\em Phys. Rev. D}} \bibinfo{volume}{102}
  (\bibinfo{number}{9}): \bibinfo{pages}{094018}.
  \bibinfo{doi}{\doi{10.1103/PhysRevD.102.094018}}.
\eprint{2007.08300}.

\bibtype{Article}%
\bibitem[Borsa et al.(2021)]{Borsa:2020yxh}
\bibinfo{author}{Borsa I}, \bibinfo{author}{de~Florian D} and
  \bibinfo{author}{Pedron I} (\bibinfo{year}{2021}).
\bibinfo{title}{{Inclusive-jet and dijet production in polarized deep inelastic
  scattering}}.
\bibinfo{journal}{{\em Phys. Rev. D}} \bibinfo{volume}{103}
  (\bibinfo{number}{1}): \bibinfo{pages}{014008}.
  \bibinfo{doi}{\doi{10.1103/PhysRevD.103.014008}}.
\eprint{2010.07354}.

\bibtype{Article}%
\bibitem[Borsa et al.(2022{\natexlab{a}})]{Borsa:2021afb}
\bibinfo{author}{Borsa I}, \bibinfo{author}{de~Florian D} and
  \bibinfo{author}{Pedron I} (\bibinfo{year}{2022}{\natexlab{a}}).
\bibinfo{title}{{Dijet production in neutral current and charged current
  polarized deep inelastic scattering}}.
\bibinfo{journal}{{\em Phys. Rev. D}} \bibinfo{volume}{105}
  (\bibinfo{number}{7}): \bibinfo{pages}{074025}.
  \bibinfo{doi}{\doi{10.1103/PhysRevD.105.074025}}.
\eprint{2112.08223}.

\bibtype{Article}%
\bibitem[Borsa et al.(2022{\natexlab{b}})]{Borsa:2022irn}
\bibinfo{author}{Borsa I}, \bibinfo{author}{de~Florian D} and
  \bibinfo{author}{Pedron I} (\bibinfo{year}{2022}{\natexlab{b}}).
\bibinfo{title}{{The full set of polarized deep inelastic scattering structure
  functions at NNLO accuracy}}.
\bibinfo{journal}{{\em Eur. Phys. J. C}} \bibinfo{volume}{82}
  (\bibinfo{number}{12}): \bibinfo{pages}{1167}.
  \bibinfo{doi}{\doi{10.1140/epjc/s10052-022-11140-z}}.
\eprint{2210.12014}.

\bibtype{Article}%
\bibitem[Borsa et al.(2024)]{Borsa:2024mss}
\bibinfo{author}{Borsa I}, \bibinfo{author}{Stratmann M},
  \bibinfo{author}{Vogelsang W}, \bibinfo{author}{de~Florian D} and
  \bibinfo{author}{Sassot R} (\bibinfo{year}{2024}).
\bibinfo{title}{{Next-to-Next-to-Leading Order Global Analysis of Polarized
  Parton Distribution Functions}}.
\bibinfo{journal}{{\em Phys. Rev. Lett.}} \bibinfo{volume}{133}
  (\bibinfo{number}{15}): \bibinfo{pages}{151901}.
  \bibinfo{doi}{\doi{10.1103/PhysRevLett.133.151901}}.
\eprint{2407.11635}.

\bibtype{Article}%
\bibitem[Botje(2011)]{Botje:2010ay}
\bibinfo{author}{Botje M} (\bibinfo{year}{2011}).
\bibinfo{title}{{QCDNUM: Fast QCD Evolution and Convolution}}.
\bibinfo{journal}{{\em Comput. Phys. Commun.}} \bibinfo{volume}{182}:
  \bibinfo{pages}{490--532}. \bibinfo{doi}{\doi{10.1016/j.cpc.2010.10.020}}.
\eprint{1005.1481}.

\bibtype{Article}%
\bibitem[Boughezal et al.(2018)]{Boughezal:2018azh}
\bibinfo{author}{Boughezal R}, \bibinfo{author}{Petriello F} and
  \bibinfo{author}{Xing H} (\bibinfo{year}{2018}).
\bibinfo{title}{{Inclusive jet production as a probe of polarized parton
  distribution functions at a future EIC}}.
\bibinfo{journal}{{\em Phys. Rev. D}} \bibinfo{volume}{98}
  (\bibinfo{number}{5}): \bibinfo{pages}{054031}.
  \bibinfo{doi}{\doi{10.1103/PhysRevD.98.054031}}.
\eprint{1806.07311}.

\bibtype{Article}%
\bibitem[Boughezal et al.(2021)]{Boughezal:2021wjw}
\bibinfo{author}{Boughezal R}, \bibinfo{author}{Li HT} and
  \bibinfo{author}{Petriello F} (\bibinfo{year}{2021}).
\bibinfo{title}{{$W$-boson production in polarized proton-proton collisions at
  RHIC through next-to-next-to-leading order in perturbative QCD}}.
\bibinfo{journal}{{\em Phys. Lett. B}} \bibinfo{volume}{817}:
  \bibinfo{pages}{136333}. \bibinfo{doi}{\doi{10.1016/j.physletb.2021.136333}}.
\eprint{2101.02214}.

\bibtype{Article}%
\bibitem[Bringewatt et al.(2021)]{Bringewatt:2020ixn}
\bibinfo{author}{Bringewatt J}, \bibinfo{author}{Sato N},
  \bibinfo{author}{Melnitchouk W}, \bibinfo{author}{Qiu JW},
  \bibinfo{author}{Steffens F} and  \bibinfo{author}{Constantinou M}
  (\bibinfo{year}{2021}).
\bibinfo{title}{{Confronting lattice parton distributions with global QCD
  analysis}}.
\bibinfo{journal}{{\em Phys. Rev. D}} \bibinfo{volume}{103}
  (\bibinfo{number}{1}): \bibinfo{pages}{016003}.
  \bibinfo{doi}{\doi{10.1103/PhysRevD.103.016003}}.
\eprint{2010.00548}.

\bibtype{Article}%
\bibitem[Brodsky and Farrar(1973)]{Brodsky:1973kr}
\bibinfo{author}{Brodsky SJ} and  \bibinfo{author}{Farrar GR}
  (\bibinfo{year}{1973}).
\bibinfo{title}{{Scaling Laws at Large Transverse Momentum}}.
\bibinfo{journal}{{\em Phys. Rev. Lett.}} \bibinfo{volume}{31}:
  \bibinfo{pages}{1153--1156}.
  \bibinfo{doi}{\doi{10.1103/PhysRevLett.31.1153}}.

\bibtype{Article}%
\bibitem[Cacciari et al.(2005)]{Cacciari:2005ry}
\bibinfo{author}{Cacciari M}, \bibinfo{author}{Nason P} and
  \bibinfo{author}{Oleari C} (\bibinfo{year}{2005}).
\bibinfo{title}{{Crossing heavy-flavor thresholds in fragmentation functions}}.
\bibinfo{journal}{{\em JHEP}} \bibinfo{volume}{10}: \bibinfo{pages}{034}.
  \bibinfo{doi}{\doi{10.1088/1126-6708/2005/10/034}}.
\eprint{hep-ph/0504192}.

\bibtype{Article}%
\bibitem[Candido et al.(2022)]{Candido:2022tld}
\bibinfo{author}{Candido A}, \bibinfo{author}{Hekhorn F} and
  \bibinfo{author}{Magni G} (\bibinfo{year}{2022}).
\bibinfo{title}{{EKO: evolution kernel operators}}.
\bibinfo{journal}{{\em Eur. Phys. J. C}} \bibinfo{volume}{82}
  (\bibinfo{number}{10}): \bibinfo{pages}{976}.
  \bibinfo{doi}{\doi{10.1140/epjc/s10052-022-10878-w}}.
\eprint{2202.02338}.

\bibtype{Article}%
\bibitem[Candido et al.(2024)]{Candido:2024rkr}
\bibinfo{author}{Candido A}, \bibinfo{author}{Hekhorn F},
  \bibinfo{author}{Magni G}, \bibinfo{author}{Rabemananjara TR} and
  \bibinfo{author}{Stegeman R} (\bibinfo{year}{2024}).
\bibinfo{title}{{Yadism: yet another deep-inelastic scattering module}}.
\bibinfo{journal}{{\em Eur. Phys. J. C}} \bibinfo{volume}{84}
  (\bibinfo{number}{7}): \bibinfo{pages}{698}.
  \bibinfo{doi}{\doi{10.1140/epjc/s10052-024-12972-7}}.
\eprint{2401.15187}.

\bibtype{Article}%
\bibitem[Carrazza and Cruz-Martinez(2019)]{Carrazza:2019mzf}
\bibinfo{author}{Carrazza S} and  \bibinfo{author}{Cruz-Martinez J}
  (\bibinfo{year}{2019}).
\bibinfo{title}{{Towards a new generation of parton densities with deep
  learning models}}.
\bibinfo{journal}{{\em Eur. Phys. J. C}} \bibinfo{volume}{79}
  (\bibinfo{number}{8}): \bibinfo{pages}{676}.
  \bibinfo{doi}{\doi{10.1140/epjc/s10052-019-7197-2}}.
\eprint{1907.05075}.

\bibtype{Article}%
\bibitem[Carrazza et al.(2020)]{Carrazza:2020gss}
\bibinfo{author}{Carrazza S}, \bibinfo{author}{Nocera ER},
  \bibinfo{author}{Schwan C} and  \bibinfo{author}{Zaro M}
  (\bibinfo{year}{2020}).
\bibinfo{title}{{PineAPPL: combining EW and QCD corrections for fast evaluation
  of LHC processes}}.
\bibinfo{journal}{{\em JHEP}} \bibinfo{volume}{12}: \bibinfo{pages}{108}.
  \bibinfo{doi}{\doi{10.1007/JHEP12(2020)108}}.
\eprint{2008.12789}.

\bibtype{Article}%
\bibitem[Carrazza et al.(2022)]{Carrazza:2021yrg}
\bibinfo{author}{Carrazza S}, \bibinfo{author}{Cruz-Martinez JM} and
  \bibinfo{author}{Stegeman R} (\bibinfo{year}{2022}).
\bibinfo{title}{{A data-based parametrization of parton distribution
  functions}}.
\bibinfo{journal}{{\em Eur. Phys. J. C}} \bibinfo{volume}{82}
  (\bibinfo{number}{2}): \bibinfo{pages}{163}.
  \bibinfo{doi}{\doi{10.1140/epjc/s10052-022-10136-z}}.
\eprint{2111.02954}.

\bibtype{Article}%
\bibitem[Chen et al.(2021)]{Chen:2020uvt}
\bibinfo{author}{Chen H}, \bibinfo{author}{Yang TZ}, \bibinfo{author}{Zhu HX}
  and  \bibinfo{author}{Zhu YJ} (\bibinfo{year}{2021}).
\bibinfo{title}{{Analytic Continuation and Reciprocity Relation for Collinear
  Splitting in QCD}}.
\bibinfo{journal}{{\em Chin. Phys. C}} \bibinfo{volume}{45}
  (\bibinfo{number}{4}): \bibinfo{pages}{043101}.
  \bibinfo{doi}{\doi{10.1088/1674-1137/abde2d}}.
\eprint{2006.10534}.

\bibtype{Article}%
\bibitem[Chen et al.(2022{\natexlab{a}})]{Chen:2022tpk}
\bibinfo{author}{Chen X}, \bibinfo{author}{Gehrmann T}, \bibinfo{author}{Glover
  EWN}, \bibinfo{author}{Huss A} and  \bibinfo{author}{Mo J}
  (\bibinfo{year}{2022}{\natexlab{a}}).
\bibinfo{title}{{NNLO QCD corrections in full colour for jet production
  observables at the LHC}}.
\bibinfo{journal}{{\em JHEP}} \bibinfo{volume}{09}: \bibinfo{pages}{025}.
  \bibinfo{doi}{\doi{10.1007/JHEP09(2022)025}}.
\eprint{2204.10173}.

\bibtype{Article}%
\bibitem[Chen et al.(2022{\natexlab{b}})]{Chen:2022cgv}
\bibinfo{author}{Chen X}, \bibinfo{author}{Gehrmann T}, \bibinfo{author}{Glover
  EWN}, \bibinfo{author}{Huss A}, \bibinfo{author}{Monni PF},
  \bibinfo{author}{Re E}, \bibinfo{author}{Rottoli L} and
  \bibinfo{author}{Torrielli P} (\bibinfo{year}{2022}{\natexlab{b}}).
\bibinfo{title}{{Third-Order Fiducial Predictions for Drell-Yan Production at
  the LHC}}.
\bibinfo{journal}{{\em Phys. Rev. Lett.}} \bibinfo{volume}{128}
  (\bibinfo{number}{25}): \bibinfo{pages}{252001}.
  \bibinfo{doi}{\doi{10.1103/PhysRevLett.128.252001}}.
\eprint{2203.01565}.

\bibtype{Article}%
\bibitem[Cichy and Constantinou(2019)]{Cichy:2018mum}
\bibinfo{author}{Cichy K} and  \bibinfo{author}{Constantinou M}
  (\bibinfo{year}{2019}).
\bibinfo{title}{{A guide to light-cone PDFs from Lattice QCD: an overview of
  approaches, techniques and results}}.
\bibinfo{journal}{{\em Adv. High Energy Phys.}} \bibinfo{volume}{2019}:
  \bibinfo{pages}{3036904}. \bibinfo{doi}{\doi{10.1155/2019/3036904}}.
\eprint{1811.07248}.

\bibtype{Article}%
\bibitem[Cocuzza et al.(2025)]{Cocuzza:2025qvf}
\bibinfo{author}{Cocuzza C}, \bibinfo{author}{Hunt-Smith NT},
  \bibinfo{author}{Melnitchouk W}, \bibinfo{author}{Sato N} and
  \bibinfo{author}{Thomas AW} (\bibinfo{collaboration}{JAM Collaboration (Spin
  PDF Analysis Group)}) (\bibinfo{year}{2025}).
\bibinfo{title}{{Global QCD analysis of spin PDFs in the proton with high-x and
  lattice constraints}}.
\bibinfo{journal}{{\em Phys. Rev. D}} \bibinfo{volume}{112}
  (\bibinfo{number}{11}): \bibinfo{pages}{114017}.
  \bibinfo{doi}{\doi{10.1103/6fn9-1wqb}}.
\eprint{2506.13616}.

\bibtype{Book}%
\bibitem[Collins(2011)]{Collins:2011zzd}
\bibinfo{author}{Collins J} (\bibinfo{year}{2011}).
\bibinfo{title}{{Foundations of Perturbative QCD}}, \bibinfo{volume}{32},
  \bibinfo{publisher}{Cambridge University Press}.
\bibinfo{comment}{ISBN} \bibinfo{isbn}{978-1-009-40184-5, 978-1-009-40183-8,
  978-1-009-40182-1}.
\bibinfo{doi}{\doi{10.1017/9781009401845}}.

\bibtype{Article}%
\bibitem[Collins et al.(1978)]{Collins:1978wz}
\bibinfo{author}{Collins JC}, \bibinfo{author}{Wilczek F} and
  \bibinfo{author}{Zee A} (\bibinfo{year}{1978}).
\bibinfo{title}{{Low-Energy Manifestations of Heavy Particles: Application to
  the Neutral Current}}.
\bibinfo{journal}{{\em Phys. Rev. D}} \bibinfo{volume}{18}:
  \bibinfo{pages}{242}. \bibinfo{doi}{\doi{10.1103/PhysRevD.18.242}}.

\bibtype{Article}%
\bibitem[Collins et al.(1985)]{Collins:1985ue}
\bibinfo{author}{Collins JC}, \bibinfo{author}{Soper DE} and
  \bibinfo{author}{Sterman GF} (\bibinfo{year}{1985}).
\bibinfo{title}{{Factorization for Short Distance Hadron - Hadron Scattering}}.
\bibinfo{journal}{{\em Nucl. Phys. B}} \bibinfo{volume}{261}:
  \bibinfo{pages}{104--142}. \bibinfo{doi}{\doi{10.1016/0550-3213(85)90565-6}}.

\bibtype{Article}%
\bibitem[Collins et al.(1988)]{Collins:1988ig}
\bibinfo{author}{Collins JC}, \bibinfo{author}{Soper DE} and
  \bibinfo{author}{Sterman GF} (\bibinfo{year}{1988}).
\bibinfo{title}{{Soft Gluons and Factorization}}.
\bibinfo{journal}{{\em Nucl. Phys. B}} \bibinfo{volume}{308}:
  \bibinfo{pages}{833--856}. \bibinfo{doi}{\doi{10.1016/0550-3213(88)90130-7}}.

\bibtype{Article}%
\bibitem[Collins et al.(1989)]{Collins:1989gx}
\bibinfo{author}{Collins JC}, \bibinfo{author}{Soper DE} and
  \bibinfo{author}{Sterman GF} (\bibinfo{year}{1989}).
\bibinfo{title}{{Factorization of Hard Processes in QCD}}.
\bibinfo{journal}{{\em Adv. Ser. Direct. High Energy Phys.}}
  \bibinfo{volume}{5}: \bibinfo{pages}{1--91}.
  \bibinfo{doi}{\doi{10.1142/9789814503266_0001}}.
\eprint{hep-ph/0409313}.

\bibtype{Article}%
\bibitem[Constantinou(2021)]{Constantinou:2020pek}
\bibinfo{author}{Constantinou M} (\bibinfo{year}{2021}).
\bibinfo{title}{{The x-dependence of hadronic parton distributions: A review on
  the progress of lattice QCD}}.
\bibinfo{journal}{{\em Eur. Phys. J. A}} \bibinfo{volume}{57}
  (\bibinfo{number}{2}): \bibinfo{pages}{77}.
  \bibinfo{doi}{\doi{10.1140/epja/s10050-021-00353-7}}.
\eprint{2010.02445}.

\bibtype{Article}%
\bibitem[Constantinou et al.(2021)]{Constantinou:2020hdm}
\bibinfo{author}{Constantinou M} and  et al. (\bibinfo{year}{2021}).
\bibinfo{title}{{Parton distributions and lattice-QCD calculations: Toward 3D
  structure}}.
\bibinfo{journal}{{\em Prog. Part. Nucl. Phys.}} \bibinfo{volume}{121}:
  \bibinfo{pages}{103908}. \bibinfo{doi}{\doi{10.1016/j.ppnp.2021.103908}}.
\eprint{2006.08636}.

\bibtype{Article}%
\bibitem[Cougoulic et al.(2022)]{Cougoulic:2022gbk}
\bibinfo{author}{Cougoulic F}, \bibinfo{author}{Kovchegov YV},
  \bibinfo{author}{Tarasov A} and  \bibinfo{author}{Tawabutr Y}
  (\bibinfo{year}{2022}).
\bibinfo{title}{{Quark and gluon helicity evolution at small x: revised and
  updated}}.
\bibinfo{journal}{{\em JHEP}} \bibinfo{volume}{07}: \bibinfo{pages}{095}.
  \bibinfo{doi}{\doi{10.1007/JHEP07(2022)095}}.
\bibinfo{note}{[Erratum: JHEP 09, 052 (2024)]}, \eprint{2204.11898}.

\bibtype{Article}%
\bibitem[Cridge et al.(2025)]{Cridge:2024icl}
\bibinfo{author}{Cridge T} and  et al. (\bibinfo{year}{2025}).
\bibinfo{title}{{Combination of aN$^3$LO PDFs and implications for Higgs
  production cross-sections at the LHC}}.
\bibinfo{journal}{{\em J. Phys. G}} \bibinfo{volume}{52}: \bibinfo{pages}{6}.
  \bibinfo{doi}{\doi{10.1088/1361-6471/adde78}}.
\eprint{2411.05373}.

\bibtype{Article}%
\bibitem[Cruz-Martinez et al.(2025{\natexlab{a}})]{Cruz-Martinez:2025ahf}
\bibinfo{author}{Cruz-Martinez J}, \bibinfo{author}{Hasenack T},
  \bibinfo{author}{Hekhorn F}, \bibinfo{author}{Magni G},
  \bibinfo{author}{Nocera ER}, \bibinfo{author}{Rabemananjara TR},
  \bibinfo{author}{Rojo J}, \bibinfo{author}{Sharma T} and
  \bibinfo{author}{van Seeventer G} (\bibinfo{year}{2025}{\natexlab{a}}).
\bibinfo{title}{{NNPDFpol2.0: a global determination of polarised PDFs and
  their uncertainties at next-to-next-to-leading order}}.
\bibinfo{journal}{{\em JHEP}} \bibinfo{volume}{07}: \bibinfo{pages}{168}.
  \bibinfo{doi}{\doi{10.1007/JHEP07(2025)168}}.
\eprint{2503.11814}.

\bibtype{Article}%
\bibitem[Cruz-Martinez et al.(2025{\natexlab{b}})]{Cruz-Martinez:2024wiu}
\bibinfo{author}{Cruz-Martinez J}, \bibinfo{author}{Jansen A},
  \bibinfo{author}{van Oord G}, \bibinfo{author}{Rabemananjara TR},
  \bibinfo{author}{Rocha CMR}, \bibinfo{author}{Rojo J} and
  \bibinfo{author}{Stegeman R} (\bibinfo{year}{2025}{\natexlab{b}}).
\bibinfo{title}{{Hyperparameter optimisation in deep learning from ensemble
  methods: applications to proton structure}}.
\bibinfo{journal}{{\em Mach. Learn. Sci. Tech.}} \bibinfo{volume}{6}
  (\bibinfo{number}{2}): \bibinfo{pages}{025027}.
  \bibinfo{doi}{\doi{10.1088/2632-2153/adcd39}}.
\eprint{2410.16248}.

\bibtype{Article}%
\bibitem[Cruz-Martinez et al.(2026)]{Cruz-Martinez:2026aqe}
\bibinfo{author}{Cruz-Martinez JM}, \bibinfo{author}{Giani T} and
  \bibinfo{author}{Rabemananjara TR} (\bibinfo{year}{2026}),
  \bibinfo{month}{5}.
\bibinfo{title}{{Hyperoptimisation algorithm for the next generation of PDF
  determinations: ensemble regression with an unbiased selection model}}
  \eprint{2605.31482}.

\bibtype{Article}%
\bibitem[Currie et al.(2017{\natexlab{a}})]{Currie:2017eqf}
\bibinfo{author}{Currie J}, \bibinfo{author}{Gehrmann-De~Ridder A},
  \bibinfo{author}{Gehrmann T}, \bibinfo{author}{Glover EWN},
  \bibinfo{author}{Huss A} and  \bibinfo{author}{Pires J}
  (\bibinfo{year}{2017}{\natexlab{a}}).
\bibinfo{title}{{Precise predictions for dijet production at the LHC}}.
\bibinfo{journal}{{\em Phys. Rev. Lett.}} \bibinfo{volume}{119}
  (\bibinfo{number}{15}): \bibinfo{pages}{152001}.
  \bibinfo{doi}{\doi{10.1103/PhysRevLett.119.152001}}.
\eprint{1705.10271}.

\bibtype{Article}%
\bibitem[Currie et al.(2017{\natexlab{b}})]{Currie:2016bfm}
\bibinfo{author}{Currie J}, \bibinfo{author}{Glover EWN} and
  \bibinfo{author}{Pires J} (\bibinfo{year}{2017}{\natexlab{b}}).
\bibinfo{title}{{Next-to-Next-to Leading Order QCD Predictions for Single Jet
  Inclusive Production at the LHC}}.
\bibinfo{journal}{{\em Phys. Rev. Lett.}} \bibinfo{volume}{118}
  (\bibinfo{number}{7}): \bibinfo{pages}{072002}.
  \bibinfo{doi}{\doi{10.1103/PhysRevLett.118.072002}}.
\eprint{1611.01460}.

\bibtype{Article}%
\bibitem[Czakon et al.(2025)]{Czakon:2025yti}
\bibinfo{author}{Czakon M}, \bibinfo{author}{Generet T}, \bibinfo{author}{Mitov
  A} and  \bibinfo{author}{Poncelet R} (\bibinfo{year}{2025}).
\bibinfo{title}{{Identified Hadron Production at Hadron Colliders in
  Next-to-Next-to-Leading-Order QCD}}.
\bibinfo{journal}{{\em Phys. Rev. Lett.}} \bibinfo{volume}{135}
  (\bibinfo{number}{17}): \bibinfo{pages}{17}.
  \bibinfo{doi}{\doi{10.1103/mhkl-vt96}}.
\eprint{2503.11489}.

\bibtype{Article}%
\bibitem[D'Agostini(1994)]{DAgostini:1993arp}
\bibinfo{author}{D'Agostini G} (\bibinfo{year}{1994}).
\bibinfo{title}{{On the use of the covariance matrix to fit correlated data}}.
\bibinfo{journal}{{\em Nucl. Instrum. Meth. A}} \bibinfo{volume}{346}:
  \bibinfo{pages}{306--311}. \bibinfo{doi}{\doi{10.1016/0168-9002(94)90719-6}}.

\bibtype{Article}%
\bibitem[de~Florian(2003)]{deFlorian:2002az}
\bibinfo{author}{de~Florian D} (\bibinfo{year}{2003}).
\bibinfo{title}{{Next-to-leading order QCD corrections to one hadron production
  in polarized pp collisions at RHIC}}.
\bibinfo{journal}{{\em Phys. Rev. D}} \bibinfo{volume}{67}:
  \bibinfo{pages}{054004}. \bibinfo{doi}{\doi{10.1103/PhysRevD.67.054004}}.
\eprint{hep-ph/0210442}.

\bibtype{Article}%
\bibitem[de~Florian and Rotstein~Habarnau(2013)]{deFlorian:2012wk}
\bibinfo{author}{de~Florian D} and  \bibinfo{author}{Rotstein~Habarnau Y}
  (\bibinfo{year}{2013}).
\bibinfo{title}{{Polarized semi-inclusive electroweak structure functions at
  next-to-leading-order}}.
\bibinfo{journal}{{\em Eur. Phys. J. C}} \bibinfo{volume}{73}
  (\bibinfo{number}{3}): \bibinfo{pages}{2356}.
  \bibinfo{doi}{\doi{10.1140/epjc/s10052-013-2356-3}}.
\eprint{1210.7203}.

\bibtype{Article}%
\bibitem[de~Florian et al.(2008)]{deFlorian:2008mr}
\bibinfo{author}{de~Florian D}, \bibinfo{author}{Sassot R},
  \bibinfo{author}{Stratmann M} and  \bibinfo{author}{Vogelsang W}
  (\bibinfo{year}{2008}).
\bibinfo{title}{{Global Analysis of Helicity Parton Densities and Their
  Uncertainties}}.
\bibinfo{journal}{{\em Phys. Rev. Lett.}} \bibinfo{volume}{101}:
  \bibinfo{pages}{072001}. \bibinfo{doi}{\doi{10.1103/PhysRevLett.101.072001}}.
\eprint{0804.0422}.

\bibtype{Article}%
\bibitem[de~Florian et al.(2009)]{deFlorian:2009vb}
\bibinfo{author}{de~Florian D}, \bibinfo{author}{Sassot R},
  \bibinfo{author}{Stratmann M} and  \bibinfo{author}{Vogelsang W}
  (\bibinfo{year}{2009}).
\bibinfo{title}{{Extraction of Spin-Dependent Parton Densities and Their
  Uncertainties}}.
\bibinfo{journal}{{\em Phys. Rev. D}} \bibinfo{volume}{80}:
  \bibinfo{pages}{034030}. \bibinfo{doi}{\doi{10.1103/PhysRevD.80.034030}}.
\eprint{0904.3821}.

\bibtype{Article}%
\bibitem[de~Florian et al.(2014)]{deFlorian:2014yva}
\bibinfo{author}{de~Florian D}, \bibinfo{author}{Sassot R},
  \bibinfo{author}{Stratmann M} and  \bibinfo{author}{Vogelsang W}
  (\bibinfo{year}{2014}).
\bibinfo{title}{{Evidence for polarization of gluons in the proton}}.
\bibinfo{journal}{{\em Phys. Rev. Lett.}} \bibinfo{volume}{113}
  (\bibinfo{number}{1}): \bibinfo{pages}{012001}.
  \bibinfo{doi}{\doi{10.1103/PhysRevLett.113.012001}}.
\eprint{1404.4293}.

\bibtype{Article}%
\bibitem[de~Florian et al.(2024{\natexlab{a}})]{deFlorian:2024hsu}
\bibinfo{author}{de~Florian D}, \bibinfo{author}{Conte LP} and
  \bibinfo{author}{Volonnino GF} (\bibinfo{year}{2024}{\natexlab{a}}).
\bibinfo{title}{{The polarized photon distribution function}}.
\bibinfo{journal}{{\em Eur. Phys. J. C}} \bibinfo{volume}{84}
  (\bibinfo{number}{9}): \bibinfo{pages}{905}.
  \bibinfo{doi}{\doi{10.1140/epjc/s10052-024-13294-4}}.
\eprint{2406.03414}.

\bibtype{Article}%
\bibitem[de~Florian et al.(2024{\natexlab{b}})]{deFlorian:2024utd}
\bibinfo{author}{de~Florian D}, \bibinfo{author}{Forte S} and
  \bibinfo{author}{Vogelsang W} (\bibinfo{year}{2024}{\natexlab{b}}).
\bibinfo{title}{{Higgs production at RHIC and the positivity of the gluon
  helicity distribution}}.
\bibinfo{journal}{{\em Phys. Rev. D}} \bibinfo{volume}{109}
  (\bibinfo{number}{7}): \bibinfo{pages}{074007}.
  \bibinfo{doi}{\doi{10.1103/PhysRevD.109.074007}}.
\eprint{2401.10814}.

\bibtype{Article}%
\bibitem[de~Groot et al.(1979{\natexlab{a}})]{deGroot:1979ugm}
\bibinfo{author}{de~Groot JGH} and  et al.
  (\bibinfo{year}{1979}{\natexlab{a}}).
\bibinfo{title}{{Comparison of Moments from the Valence Structure Function with
  QCD Predictions}}.
\bibinfo{journal}{{\em Phys. Lett. B}} \bibinfo{volume}{82}:
  \bibinfo{pages}{292--296}. \bibinfo{doi}{\doi{10.1016/0370-2693(79)90759-7}}.

\bibtype{Article}%
\bibitem[de~Groot et al.(1979{\natexlab{b}})]{deGroot:1978feq}
\bibinfo{author}{de~Groot JGH} and  et al.
  (\bibinfo{year}{1979}{\natexlab{b}}).
\bibinfo{title}{{Inclusive Interactions of High-Energy Neutrinos and
  anti-neutrinos in Iron}}.
\bibinfo{journal}{{\em Z. Phys. C}} \bibinfo{volume}{1}: \bibinfo{pages}{143}.
  \bibinfo{doi}{\doi{10.1007/BF01445406}}.

\bibtype{Article}%
\bibitem[Del~Debbio et al.(2022)]{DelDebbio:2021whr}
\bibinfo{author}{Del~Debbio L}, \bibinfo{author}{Giani T} and
  \bibinfo{author}{Wilson M} (\bibinfo{year}{2022}).
\bibinfo{title}{{Bayesian approach to inverse problems: an application to NNPDF
  closure testing}}.
\bibinfo{journal}{{\em Eur. Phys. J. C}} \bibinfo{volume}{82}
  (\bibinfo{number}{4}): \bibinfo{pages}{330}.
  \bibinfo{doi}{\doi{10.1140/epjc/s10052-022-10297-x}}.
\eprint{2111.05787}.

\bibtype{Article}%
\bibitem[Dennison(1927)]{Dennison:1927}
\bibinfo{author}{Dennison DM} (\bibinfo{year}{1927}).
\bibinfo{title}{A note on the specific heat of the hydrogen molecule}.
\bibinfo{journal}{{\em Proceedings of the Royal Society of London. Series A}}
  \bibinfo{volume}{115} (\bibinfo{number}{771}): \bibinfo{pages}{483--486}.

\bibtype{Article}%
\bibitem[Dharmawardane et al.(2006)]{CLAS:2006ozz}
\bibinfo{author}{Dharmawardane KV} and  et al. (\bibinfo{collaboration}{CLAS})
  (\bibinfo{year}{2006}).
\bibinfo{title}{{Measurement of the x- and Q**2-dependence of the asymmetry
  A(1) on the nucleon}}.
\bibinfo{journal}{{\em Phys. Lett. B}} \bibinfo{volume}{641}:
  \bibinfo{pages}{11--17}. \bibinfo{doi}{\doi{10.1016/j.physletb.2006.08.011}}.
\eprint{nucl-ex/0605028}.

\bibtype{Article}%
\bibitem[Dittmar et al.(2005)]{Dittmar:2005ed}
\bibinfo{author}{Dittmar M} and  et al. (\bibinfo{year}{2005}),
  \bibinfo{month}{11}.
\bibinfo{title}{{Working Group I: Parton distributions: Summary report for the
  HERA LHC Workshop Proceedings}} \eprint{hep-ph/0511119}.

\bibtype{Article}%
\bibitem[Dokshitzer(1977)]{Dokshitzer:1977sg}
\bibinfo{author}{Dokshitzer YL} (\bibinfo{year}{1977}).
\bibinfo{title}{{Calculation of the Structure Functions for Deep Inelastic
  Scattering and e+ e- Annihilation by Perturbation Theory in Quantum
  Chromodynamics.}}
\bibinfo{journal}{{\em Sov. Phys. JETP}} \bibinfo{volume}{46}:
  \bibinfo{pages}{641--653}.

\bibtype{Article}%
\bibitem[Estermann and Stern(1933)]{Estermann:1933a}
\bibinfo{author}{Estermann I} and  \bibinfo{author}{Stern O}
  (\bibinfo{year}{1933}).
\bibinfo{title}{{\"U}ber die magnetische ablenkung von wasserstoffmolek{\"u}len
  und das magnetische moment des protons. ii}.
\bibinfo{journal}{{\em Z. Phys.}} \bibinfo{volume}{85}:
  \bibinfo{pages}{17--24}. \bibinfo{doi}{\doi{10.1007/BF01330774}}.

\bibtype{Article}%
\bibitem[Estermann et al.(1933)]{Estermann:1933b}
\bibinfo{author}{Estermann I}, \bibinfo{author}{Frisch OR} and
  \bibinfo{author}{Stern O} (\bibinfo{year}{1933}).
\bibinfo{title}{Magnetic moment of the proton}.
\bibinfo{journal}{{\em Nature}} \bibinfo{volume}{132}: \bibinfo{pages}{169}.
  \bibinfo{doi}{\doi{10.1038/132169a0}}.

\bibtype{Article}%
\bibitem[Ethier et al.(2017)]{Ethier:2017zbq}
\bibinfo{author}{Ethier JJ}, \bibinfo{author}{Sato N} and
  \bibinfo{author}{Melnitchouk W} (\bibinfo{year}{2017}).
\bibinfo{title}{{First simultaneous extraction of spin-dependent parton
  distributions and fragmentation functions from a global QCD analysis}}.
\bibinfo{journal}{{\em Phys. Rev. Lett.}} \bibinfo{volume}{119}
  (\bibinfo{number}{13}): \bibinfo{pages}{132001}.
  \bibinfo{doi}{\doi{10.1103/PhysRevLett.119.132001}}.
\eprint{1705.05889}.

\bibtype{Article}%
\bibitem[Feynman(1969)]{Feynman:1969ej}
\bibinfo{author}{Feynman RP} (\bibinfo{year}{1969}).
\bibinfo{title}{{Very high-energy collisions of hadrons}}.
\bibinfo{journal}{{\em Phys. Rev. Lett.}} \bibinfo{volume}{23}:
  \bibinfo{pages}{1415--1417}.
  \bibinfo{doi}{\doi{10.1103/PhysRevLett.23.1415}}.

\bibtype{Article}%
\bibitem[Flay et al.(2016)]{JeffersonLabHallA:2016neg}
\bibinfo{author}{Flay D} and  et al. (\bibinfo{collaboration}{Jefferson Lab
  Hall A}) (\bibinfo{year}{2016}).
\bibinfo{title}{{Measurements of $d_{2}^{n}$ and $A_{1}^{n}$: Probing the
  neutron spin structure}}.
\bibinfo{journal}{{\em Phys. Rev. D}} \bibinfo{volume}{94}
  (\bibinfo{number}{5}): \bibinfo{pages}{052003}.
  \bibinfo{doi}{\doi{10.1103/PhysRevD.94.052003}}.
\eprint{1603.03612}.

\bibtype{Article}%
\bibitem[Forte et al.(1999)]{Forte:1998kd}
\bibinfo{author}{Forte S}, \bibinfo{author}{Altarelli G} and
  \bibinfo{author}{Ridolfi G} (\bibinfo{year}{1999}).
\bibinfo{title}{{Are parton distributions positive?}}
\bibinfo{journal}{{\em Nucl. Phys. B Proc. Suppl.}} \bibinfo{volume}{74}:
  \bibinfo{pages}{138--141}.
  \bibinfo{doi}{\doi{10.1016/S0920-5632(99)00150-4}}.
\eprint{hep-ph/9808462}.

\bibtype{Article}%
\bibitem[Forte et al.(2010)]{Forte:2010ta}
\bibinfo{author}{Forte S}, \bibinfo{author}{Laenen E}, \bibinfo{author}{Nason
  P} and  \bibinfo{author}{Rojo J} (\bibinfo{year}{2010}).
\bibinfo{title}{{Heavy quarks in deep-inelastic scattering}}.
\bibinfo{journal}{{\em Nucl. Phys. B}} \bibinfo{volume}{834}:
  \bibinfo{pages}{116--162}.
  \bibinfo{doi}{\doi{10.1016/j.nuclphysb.2010.03.014}}.
\eprint{1001.2312}.

\bibtype{Article}%
\bibitem[Friedman and Kendall(1972)]{Friedman:1972sy}
\bibinfo{author}{Friedman JI} and  \bibinfo{author}{Kendall HW}
  (\bibinfo{year}{1972}).
\bibinfo{title}{{Deep inelastic electron scattering}}.
\bibinfo{journal}{{\em Ann. Rev. Nucl. Part. Sci.}} \bibinfo{volume}{22}:
  \bibinfo{pages}{203--254}.
  \bibinfo{doi}{\doi{10.1146/annurev.ns.22.120172.001223}}.

\bibtype{Article}%
\bibitem[Frisch and Stern(1933)]{Frisch:1933a}
\bibinfo{author}{Frisch OR} and  \bibinfo{author}{Stern O}
  (\bibinfo{year}{1933}).
\bibinfo{title}{{\"U}ber die magnetische ablenkung von wasserstoffmolek{\"u}len
  und das magnetische moment des protons. i}.
\bibinfo{journal}{{\em Z. Phys.}} \bibinfo{volume}{85}: \bibinfo{pages}{4--16}.
  \bibinfo{doi}{\doi{10.1007/BF01330773}}.

\bibtype{Article}%
\bibitem[Gehrmann and Stirling(1995)]{Gehrmann:1994rb}
\bibinfo{author}{Gehrmann T} and  \bibinfo{author}{Stirling WJ}
  (\bibinfo{year}{1995}).
\bibinfo{title}{{Spin dependent parton distributions from polarized structure
  function data}}.
\bibinfo{journal}{{\em Z. Phys. C}} \bibinfo{volume}{65}:
  \bibinfo{pages}{461--470}. \bibinfo{doi}{\doi{10.1007/BF01556134}}.
\eprint{hep-ph/9406212}.

\bibtype{Article}%
\bibitem[Gehrmann and Stirling(1996)]{Gehrmann:1995ag}
\bibinfo{author}{Gehrmann T} and  \bibinfo{author}{Stirling WJ}
  (\bibinfo{year}{1996}).
\bibinfo{title}{{Polarized parton distributions in the nucleon}}.
\bibinfo{journal}{{\em Phys. Rev. D}} \bibinfo{volume}{53}:
  \bibinfo{pages}{6100--6109}. \bibinfo{doi}{\doi{10.1103/PhysRevD.53.6100}}.
\eprint{hep-ph/9512406}.

\bibtype{Article}%
\bibitem[Gell-Mann(1962)]{Gell-Mann:1962yej}
\bibinfo{author}{Gell-Mann M} (\bibinfo{year}{1962}).
\bibinfo{title}{{Symmetries of baryons and mesons}}.
\bibinfo{journal}{{\em Phys. Rev.}} \bibinfo{volume}{125}:
  \bibinfo{pages}{1067--1084}. \bibinfo{doi}{\doi{10.1103/PhysRev.125.1067}}.

\bibtype{Article}%
\bibitem[Gell-Mann(1964)]{Gell-Mann:1964ewy}
\bibinfo{author}{Gell-Mann M} (\bibinfo{year}{1964}).
\bibinfo{title}{{A Schematic Model of Baryons and Mesons}}.
\bibinfo{journal}{{\em Phys. Lett.}} \bibinfo{volume}{8}:
  \bibinfo{pages}{214--215}.
  \bibinfo{doi}{\doi{10.1016/S0031-9163(64)92001-3}}.

\bibtype{Article}%
\bibitem[Gluck et al.(1995)]{Gluck:1995yq}
\bibinfo{author}{Gluck M}, \bibinfo{author}{Reya E} and
  \bibinfo{author}{Vogelsang W} (\bibinfo{year}{1995}).
\bibinfo{title}{{Radiative parton model analysis of polarized deep inelastic
  lepton - nucleon scattering}}.
\bibinfo{journal}{{\em Phys. Lett. B}} \bibinfo{volume}{359}:
  \bibinfo{pages}{201--209}. \bibinfo{doi}{\doi{10.1016/0370-2693(95)01061-T}}.
\eprint{hep-ph/9507354}.

\bibtype{Article}%
\bibitem[Gluck et al.(1996)]{Gluck:1995yr}
\bibinfo{author}{Gluck M}, \bibinfo{author}{Reya E}, \bibinfo{author}{Stratmann
  M} and  \bibinfo{author}{Vogelsang W} (\bibinfo{year}{1996}).
\bibinfo{title}{{Next-to-leading order radiative parton model analysis of
  polarized deep inelastic lepton - nucleon scattering}}.
\bibinfo{journal}{{\em Phys. Rev. D}} \bibinfo{volume}{53}:
  \bibinfo{pages}{4775--4786}. \bibinfo{doi}{\doi{10.1103/PhysRevD.53.4775}}.
\eprint{hep-ph/9508347}.

\bibtype{Article}%
\bibitem[Goto et al.(2000)]{AsymmetryAnalysis:1999gsr}
\bibinfo{author}{Goto Y} and  et al. (\bibinfo{collaboration}{Asymmetry
  Analysis}) (\bibinfo{year}{2000}).
\bibinfo{title}{{Polarized parton distribution functions in the nucleon}}.
\bibinfo{journal}{{\em Phys. Rev. D}} \bibinfo{volume}{62}:
  \bibinfo{pages}{034017}. \bibinfo{doi}{\doi{10.1103/PhysRevD.62.034017}}.
\eprint{hep-ph/0001046}.

\bibtype{Article}%
\bibitem[Goudsmit and Uhlenbeck(1926)]{Goudschmidt:1926ea}
\bibinfo{author}{Goudsmit SA} and  \bibinfo{author}{Uhlenbeck GH}
  (\bibinfo{year}{1926}).
\bibinfo{title}{{Spinning electrons and the structure of spectra}}.
\bibinfo{journal}{{\em Nature}} \bibinfo{volume}{117}:
  \bibinfo{pages}{264--265}. \bibinfo{doi}{\doi{10.1038/117264a0}}.

\bibtype{Article}%
\bibitem[Goyal et al.(2024{\natexlab{a}})]{Goyal:2024tmo}
\bibinfo{author}{Goyal S}, \bibinfo{author}{Lee RN}, \bibinfo{author}{Moch SO},
  \bibinfo{author}{Pathak V}, \bibinfo{author}{Rana N} and
  \bibinfo{author}{Ravindran V} (\bibinfo{year}{2024}{\natexlab{a}}).
\bibinfo{title}{{Next-to-Next-to-Leading Order QCD Corrections to Polarized
  Semi-Inclusive Deep-Inelastic Scattering}}.
\bibinfo{journal}{{\em Phys. Rev. Lett.}} \bibinfo{volume}{133}:
  \bibinfo{pages}{211905}. \bibinfo{doi}{\doi{10.1103/PhysRevLett.133.211905}}.
\eprint{2404.09959}.

\bibtype{Article}%
\bibitem[Goyal et al.(2024{\natexlab{b}})]{Goyal:2023zdi}
\bibinfo{author}{Goyal S}, \bibinfo{author}{Moch SO}, \bibinfo{author}{Pathak
  V}, \bibinfo{author}{Rana N} and  \bibinfo{author}{Ravindran V}
  (\bibinfo{year}{2024}{\natexlab{b}}).
\bibinfo{title}{{Next-to-Next-to-Leading Order QCD Corrections to
  Semi-Inclusive Deep-Inelastic Scattering}}.
\bibinfo{journal}{{\em Phys. Rev. Lett.}} \bibinfo{volume}{132}
  (\bibinfo{number}{25}): \bibinfo{pages}{251902}.
  \bibinfo{doi}{\doi{10.1103/PhysRevLett.132.251902}}.
\eprint{2312.17711}.

\bibtype{Article}%
\bibitem[Goyal et al.(2025)]{Goyal:2024emo}
\bibinfo{author}{Goyal S}, \bibinfo{author}{Lee RN}, \bibinfo{author}{Moch SO},
  \bibinfo{author}{Pathak V}, \bibinfo{author}{Rana N} and
  \bibinfo{author}{Ravindran V} (\bibinfo{year}{2025}).
\bibinfo{title}{{NNLO QCD corrections to unpolarized and polarized SIDIS}}.
\bibinfo{journal}{{\em Phys. Rev. D}} \bibinfo{volume}{111}
  (\bibinfo{number}{9}): \bibinfo{pages}{094007}.
  \bibinfo{doi}{\doi{10.1103/PhysRevD.111.094007}}.
\eprint{2412.19309}.

\bibtype{Article}%
\bibitem[Goyal et al.(2026)]{Goyal:2026ccx}
\bibinfo{author}{Goyal S}, \bibinfo{author}{Moch SO}, \bibinfo{author}{Pathak
  V} and  \bibinfo{author}{Ravindran V} (\bibinfo{year}{2026}),
  \bibinfo{month}{3}.
\bibinfo{title}{{NNLO QCD corrections to unpolarized and polarized electroweak
  structure functions in semi-inclusive deep-inelastic scattering}}
  \eprint{2603.30012}.

\bibtype{Article}%
\bibitem[Gribov and Lipatov(1972)]{Gribov:1972ri}
\bibinfo{author}{Gribov VN} and  \bibinfo{author}{Lipatov LN}
  (\bibinfo{year}{1972}).
\bibinfo{title}{{Deep inelastic e p scattering in perturbation theory}}.
\bibinfo{journal}{{\em Sov. J. Nucl. Phys.}} \bibinfo{volume}{15}:
  \bibinfo{pages}{438--450}.

\bibtype{Article}%
\bibitem[Gross and Wilczek(1973{\natexlab{a}})]{Gross:1973ju}
\bibinfo{author}{Gross DJ} and  \bibinfo{author}{Wilczek F}
  (\bibinfo{year}{1973}{\natexlab{a}}).
\bibinfo{title}{{Asymptotically Free Gauge Theories - I}}.
\bibinfo{journal}{{\em Phys. Rev. D}} \bibinfo{volume}{8}:
  \bibinfo{pages}{3633--3652}. \bibinfo{doi}{\doi{10.1103/PhysRevD.8.3633}}.

\bibtype{Article}%
\bibitem[Gross and Wilczek(1973{\natexlab{b}})]{Gross:1973id}
\bibinfo{author}{Gross DJ} and  \bibinfo{author}{Wilczek F}
  (\bibinfo{year}{1973}{\natexlab{b}}).
\bibinfo{title}{{Ultraviolet Behavior of Nonabelian Gauge Theories}}.
\bibinfo{journal}{{\em Phys. Rev. Lett.}} \bibinfo{volume}{30}:
  \bibinfo{pages}{1343--1346}.
  \bibinfo{doi}{\doi{10.1103/PhysRevLett.30.1343}}.

\bibtype{Article}%
\bibitem[Gross and Wilczek(1974)]{Gross:1973zrg}
\bibinfo{author}{Gross DJ} and  \bibinfo{author}{Wilczek F}
  (\bibinfo{year}{1974}).
\bibinfo{title}{{Asymptotically Free Gauge Theories~2}}.
\bibinfo{journal}{{\em Phys. Rev. D}} \bibinfo{volume}{9}:
  \bibinfo{pages}{980--993}. \bibinfo{doi}{\doi{10.1103/PhysRevD.9.980}}.

\bibtype{Article}%
\bibitem[Harland-Lang et al.(2025)]{Harland-Lang:2024kvt}
\bibinfo{author}{Harland-Lang LA}, \bibinfo{author}{Cridge T} and
  \bibinfo{author}{Thorne RS} (\bibinfo{year}{2025}).
\bibinfo{title}{{A stress test of global PDF fits: closure testing the MSHT
  PDFs and a first direct comparison to the neural net approach}}.
\bibinfo{journal}{{\em Eur. Phys. J. C}} \bibinfo{volume}{85}
  (\bibinfo{number}{3}): \bibinfo{pages}{316}.
  \bibinfo{doi}{\doi{10.1140/epjc/s10052-025-13934-3}}.
\eprint{2407.07944}.

\bibtype{Article}%
\bibitem[Hekhorn and Stratmann(2018)]{Hekhorn:2018ywm}
\bibinfo{author}{Hekhorn F} and  \bibinfo{author}{Stratmann M}
  (\bibinfo{year}{2018}).
\bibinfo{title}{{Next-to-Leading Order QCD Corrections to Inclusive
  Heavy-Flavor Production in Polarized Deep-Inelastic Scattering}}.
\bibinfo{journal}{{\em Phys. Rev. D}} \bibinfo{volume}{98}
  (\bibinfo{number}{1}): \bibinfo{pages}{014018}.
  \bibinfo{doi}{\doi{10.1103/PhysRevD.98.014018}}.
\eprint{1805.09026}.

\bibtype{Article}%
\bibitem[Hekhorn et al.(2024)]{Hekhorn:2024tqm}
\bibinfo{author}{Hekhorn F}, \bibinfo{author}{Magni G}, \bibinfo{author}{Nocera
  ER}, \bibinfo{author}{Rabemananjara TR}, \bibinfo{author}{Rojo J},
  \bibinfo{author}{Schaus A} and  \bibinfo{author}{Stegeman R}
  (\bibinfo{year}{2024}).
\bibinfo{title}{{Heavy quarks in polarised deep-inelastic scattering at the
  electron-ion collider}}.
\bibinfo{journal}{{\em Eur. Phys. J. C}} \bibinfo{volume}{84}
  (\bibinfo{number}{2}): \bibinfo{pages}{189}.
  \bibinfo{doi}{\doi{10.1140/epjc/s10052-024-12524-z}}.
\eprint{2401.10127}.

\bibtype{Article}%
\bibitem[Hirai and Kumano(2009)]{Hirai:2008aj}
\bibinfo{author}{Hirai M} and  \bibinfo{author}{Kumano S}
  (\bibinfo{collaboration}{Asymmetry Analysis}) (\bibinfo{year}{2009}).
\bibinfo{title}{{Determination of gluon polarization from deep inelastic
  scattering and collider data}}.
\bibinfo{journal}{{\em Nucl. Phys. B}} \bibinfo{volume}{813}:
  \bibinfo{pages}{106--122}.
  \bibinfo{doi}{\doi{10.1016/j.nuclphysb.2008.12.026}}.
\eprint{0808.0413}.

\bibtype{Article}%
\bibitem[Hirai et al.(2004)]{Hirai:2003pm}
\bibinfo{author}{Hirai M}, \bibinfo{author}{Kumano S} and
  \bibinfo{author}{Saito N} (\bibinfo{collaboration}{Asymmetry Analysis})
  (\bibinfo{year}{2004}).
\bibinfo{title}{{Determination of polarized parton distribution functions and
  their uncertainties}}.
\bibinfo{journal}{{\em Phys. Rev. D}} \bibinfo{volume}{69}:
  \bibinfo{pages}{054021}. \bibinfo{doi}{\doi{10.1103/PhysRevD.69.054021}}.
\eprint{hep-ph/0312112}.

\bibtype{Article}%
\bibitem[Hunt-Smith et al.(2024)]{Hunt-Smith:2024khs}
\bibinfo{author}{Hunt-Smith NT}, \bibinfo{author}{Cocuzza C},
  \bibinfo{author}{Melnitchouk W}, \bibinfo{author}{Sato N},
  \bibinfo{author}{Thomas AW} and  \bibinfo{author}{White MJ}
  (\bibinfo{collaboration}{JAM}) (\bibinfo{year}{2024}).
\bibinfo{title}{{New Data-Driven Constraints on the Sign of Gluon Polarization
  in the Proton}}.
\bibinfo{journal}{{\em Phys. Rev. Lett.}} \bibinfo{volume}{133}
  (\bibinfo{number}{16}): \bibinfo{pages}{161901}.
  \bibinfo{doi}{\doi{10.1103/PhysRevLett.133.161901}}.
\eprint{2403.08117}.

\bibtype{Article}%
\bibitem[Jaffe and Manohar(1990)]{Jaffe:1989jz}
\bibinfo{author}{Jaffe RL} and  \bibinfo{author}{Manohar A}
  (\bibinfo{year}{1990}).
\bibinfo{title}{{The $g_1$ Problem: Fact and Fantasy on the Spin of the
  Proton}}.
\bibinfo{journal}{{\em Nucl. Phys. B}} \bibinfo{volume}{337}:
  \bibinfo{pages}{509--546}. \bibinfo{doi}{\doi{10.1016/0550-3213(90)90506-9}}.

\bibtype{Article}%
\bibitem[Jager et al.(2003)]{Jager:2002xm}
\bibinfo{author}{Jager B}, \bibinfo{author}{Schafer A},
  \bibinfo{author}{Stratmann M} and  \bibinfo{author}{Vogelsang W}
  (\bibinfo{year}{2003}).
\bibinfo{title}{{Next-to-leading order QCD corrections to high p(T) pion
  production in longitudinally polarized pp collisions}}.
\bibinfo{journal}{{\em Phys. Rev. D}} \bibinfo{volume}{67}:
  \bibinfo{pages}{054005}. \bibinfo{doi}{\doi{10.1103/PhysRevD.67.054005}}.
\eprint{hep-ph/0211007}.

\bibtype{Article}%
\bibitem[Jager et al.(2004)]{Jager:2004jh}
\bibinfo{author}{Jager B}, \bibinfo{author}{Stratmann M} and
  \bibinfo{author}{Vogelsang W} (\bibinfo{year}{2004}).
\bibinfo{title}{{Single inclusive jet production in polarized $p p$ collisions
  at $O(alpha^3_s)$}}.
\bibinfo{journal}{{\em Phys. Rev. D}} \bibinfo{volume}{70}:
  \bibinfo{pages}{034010}. \bibinfo{doi}{\doi{10.1103/PhysRevD.70.034010}}.
\eprint{hep-ph/0404057}.

\bibtype{Article}%
\bibitem[Je{\v{z}}o et al.(2026)]{Jezo:2026adf}
\bibinfo{author}{Je{\v{z}}o T}, \bibinfo{author}{Nocera ER},
  \bibinfo{author}{Rabemananjara TR}, \bibinfo{author}{Schwan C},
  \bibinfo{author}{Sharma T} and  \bibinfo{author}{Wissmann J}
  (\bibinfo{year}{2026}), \bibinfo{month}{6}.
\bibinfo{title}{{PineAPPLv1: fast and flexible theory predictions for present
  and future colliders}} \eprint{2606.17134}.

\bibtype{Article}%
\bibitem[Ji et al.(2021)]{Ji:2020ena}
\bibinfo{author}{Ji X}, \bibinfo{author}{Yuan F} and  \bibinfo{author}{Zhao Y}
  (\bibinfo{year}{2021}).
\bibinfo{title}{{What we know and what we don{\textquoteright}t know about the
  proton spin after 30 years}}.
\bibinfo{journal}{{\em Nature Rev. Phys.}} \bibinfo{volume}{3}
  (\bibinfo{number}{1}): \bibinfo{pages}{27--38}.
  \bibinfo{doi}{\doi{10.1038/s42254-020-00248-4}}.
\eprint{2009.01291}.

\bibtype{Article}%
\bibitem[Kang and Soffer(2011)]{Kang:2011qz}
\bibinfo{author}{Kang ZB} and  \bibinfo{author}{Soffer J}
  (\bibinfo{year}{2011}).
\bibinfo{title}{{General positivity bounds for spin observables in particle
  inclusive production}}.
\bibinfo{journal}{{\em Phys. Rev. D}} \bibinfo{volume}{83}:
  \bibinfo{pages}{114020}. \bibinfo{doi}{\doi{10.1103/PhysRevD.83.114020}}.
\eprint{1104.2920}.

\bibtype{Article}%
\bibitem[Karpie et al.(2024)]{Karpie:2023nyg}
\bibinfo{author}{Karpie J}, \bibinfo{author}{Whitehill RM},
  \bibinfo{author}{Melnitchouk W}, \bibinfo{author}{Monahan C},
  \bibinfo{author}{Orginos K}, \bibinfo{author}{Qiu JW},
  \bibinfo{author}{Richards DG}, \bibinfo{author}{Sato N} and
  \bibinfo{author}{Zafeiropoulos S} (\bibinfo{collaboration}{Jefferson Lab
  Angular Momentum, HadStruc}) (\bibinfo{year}{2024}).
\bibinfo{title}{{Gluon helicity from global analysis of experimental data and
  lattice QCD Ioffe time distributions}}.
\bibinfo{journal}{{\em Phys. Rev. D}} \bibinfo{volume}{109}
  (\bibinfo{number}{3}): \bibinfo{pages}{036031}.
  \bibinfo{doi}{\doi{10.1103/PhysRevD.109.036031}}.
\eprint{2310.18179}.

\bibtype{Article}%
\bibitem[Kawamura et al.(2012)]{Kawamura:2012cr}
\bibinfo{author}{Kawamura H}, \bibinfo{author}{Lo~Presti NA},
  \bibinfo{author}{Moch S} and  \bibinfo{author}{Vogt A}
  (\bibinfo{year}{2012}).
\bibinfo{title}{{On the next-to-next-to-leading order QCD corrections to
  heavy-quark production in deep-inelastic scattering}}.
\bibinfo{journal}{{\em Nucl. Phys. B}} \bibinfo{volume}{864}:
  \bibinfo{pages}{399--468}.
  \bibinfo{doi}{\doi{10.1016/j.nuclphysb.2012.07.001}}.
\eprint{1205.5727}.

\bibtype{Article}%
\bibitem[Khanpour et al.(2026)]{Khanpour:2026erj}
\bibinfo{author}{Khanpour H}, \bibinfo{author}{Soleymaninia M},
  \bibinfo{author}{Azizi M}, \bibinfo{author}{Klasen M},
  \bibinfo{author}{Hashamipour H}, \bibinfo{author}{Salajegheh M} and
  \bibinfo{author}{Mei{\ss}ner UG} (\bibinfo{collaboration}{HAPS})
  (\bibinfo{year}{2026}).
\bibinfo{title}{{Toward precision helicity PDFs from global DIS and SIDIS fits
  with projected EIC measurements}}.
\bibinfo{journal}{{\em Phys. Rev. D}} \bibinfo{volume}{113}
  (\bibinfo{number}{11}): \bibinfo{pages}{114010}.
  \bibinfo{doi}{\doi{10.1103/wc54-mnfh}}.
\eprint{2602.17298}.

\bibtype{Article}%
\bibitem[Kovchegov and Tawabutr(2020)]{Kovchegov:2020hgb}
\bibinfo{author}{Kovchegov YV} and  \bibinfo{author}{Tawabutr Y}
  (\bibinfo{year}{2020}).
\bibinfo{title}{{Helicity at Small $x$: Oscillations Generated by Bringing Back
  the Quarks}}.
\bibinfo{journal}{{\em JHEP}} \bibinfo{volume}{08}: \bibinfo{pages}{014}.
  \bibinfo{doi}{\doi{10.1007/JHEP08(2020)014}}.
\eprint{2005.07285}.

\bibtype{Article}%
\bibitem[Kovchegov et al.(2016)]{Kovchegov:2015pbl}
\bibinfo{author}{Kovchegov YV}, \bibinfo{author}{Pitonyak D} and
  \bibinfo{author}{Sievert MD} (\bibinfo{year}{2016}).
\bibinfo{title}{{Helicity Evolution at Small-x}}.
\bibinfo{journal}{{\em JHEP}} \bibinfo{volume}{01}: \bibinfo{pages}{072}.
  \bibinfo{doi}{\doi{10.1007/JHEP01(2016)072}}.
\bibinfo{note}{[Erratum: JHEP 10, 148 (2016)]}, \eprint{1511.06737}.

\bibtype{Article}%
\bibitem[Kovchegov et al.(2017{\natexlab{a}})]{Kovchegov:2017lsr}
\bibinfo{author}{Kovchegov YV}, \bibinfo{author}{Pitonyak D} and
  \bibinfo{author}{Sievert MD} (\bibinfo{year}{2017}{\natexlab{a}}).
\bibinfo{title}{{Small-$x$ Asymptotics of the Gluon Helicity Distribution}}.
\bibinfo{journal}{{\em JHEP}} \bibinfo{volume}{10}: \bibinfo{pages}{198}.
  \bibinfo{doi}{\doi{10.1007/JHEP10(2017)198}}.
\eprint{1706.04236}.

\bibtype{Article}%
\bibitem[Kovchegov et al.(2017{\natexlab{b}})]{Kovchegov:2016weo}
\bibinfo{author}{Kovchegov YV}, \bibinfo{author}{Pitonyak D} and
  \bibinfo{author}{Sievert MD} (\bibinfo{year}{2017}{\natexlab{b}}).
\bibinfo{title}{{Small-$x$ asymptotics of the quark helicity distribution}}.
\bibinfo{journal}{{\em Phys. Rev. Lett.}} \bibinfo{volume}{118}
  (\bibinfo{number}{5}): \bibinfo{pages}{052001}.
  \bibinfo{doi}{\doi{10.1103/PhysRevLett.118.052001}}.
\eprint{1610.06188}.

\bibtype{Article}%
\bibitem[Kovchegov et al.(2017{\natexlab{c}})]{Kovchegov:2017jxc}
\bibinfo{author}{Kovchegov YV}, \bibinfo{author}{Pitonyak D} and
  \bibinfo{author}{Sievert MD} (\bibinfo{year}{2017}{\natexlab{c}}).
\bibinfo{title}{{Small-$x$ Asymptotics of the Quark Helicity Distribution:
  Analytic Results}}.
\bibinfo{journal}{{\em Phys. Lett. B}} \bibinfo{volume}{772}:
  \bibinfo{pages}{136--140}.
  \bibinfo{doi}{\doi{10.1016/j.physletb.2017.06.032}}.
\eprint{1703.05809}.

\bibtype{Article}%
\bibitem[Kovchegov et al.(2022)]{Kovchegov:2021lvz}
\bibinfo{author}{Kovchegov YV}, \bibinfo{author}{Tarasov A} and
  \bibinfo{author}{Tawabutr Y} (\bibinfo{year}{2022}).
\bibinfo{title}{{Helicity evolution at small x: the single-logarithmic
  contribution}}.
\bibinfo{journal}{{\em JHEP}} \bibinfo{volume}{03}: \bibinfo{pages}{184}.
  \bibinfo{doi}{\doi{10.1007/JHEP03(2022)184}}.
\eprint{2104.11765}.

\bibtype{Article}%
\bibitem[Leader and Anselmino(1988)]{Leader:1988vd}
\bibinfo{author}{Leader E} and  \bibinfo{author}{Anselmino M}
  (\bibinfo{year}{1988}).
\bibinfo{title}{{A crisis in the parton model: Where, oh where is the
  proton{\textquoteright}s spin?}}
\bibinfo{journal}{{\em Z. Phys. C}} \bibinfo{volume}{41}: \bibinfo{pages}{239}.
  \bibinfo{doi}{\doi{10.1007/BF01566922}}.

\bibtype{Article}%
\bibitem[Leader and Lorc{\'e}(2014)]{Leader:2013jra}
\bibinfo{author}{Leader E} and  \bibinfo{author}{Lorc{\'e} C}
  (\bibinfo{year}{2014}).
\bibinfo{title}{{The angular momentum controversy: What{\textquoteright}s it
  all about and does it matter?}}
\bibinfo{journal}{{\em Phys. Rept.}} \bibinfo{volume}{541}
  (\bibinfo{number}{3}): \bibinfo{pages}{163--248}.
  \bibinfo{doi}{\doi{10.1016/j.physrep.2014.02.010}}.
\eprint{1309.4235}.

\bibtype{Article}%
\bibitem[Leader et al.(1998)]{Leader:1998qv}
\bibinfo{author}{Leader E}, \bibinfo{author}{Sidorov AV} and
  \bibinfo{author}{Stamenov DB} (\bibinfo{year}{1998}).
\bibinfo{title}{{Polarized parton densities in the nucleon}}.
\bibinfo{journal}{{\em Phys. Rev. D}} \bibinfo{volume}{58}:
  \bibinfo{pages}{114028}. \bibinfo{doi}{\doi{10.1103/PhysRevD.58.114028}}.
\eprint{hep-ph/9807251}.

\bibtype{Article}%
\bibitem[Leader et al.(2002)]{Leader:2001kh}
\bibinfo{author}{Leader E}, \bibinfo{author}{Sidorov AV} and
  \bibinfo{author}{Stamenov DB} (\bibinfo{year}{2002}).
\bibinfo{title}{{A New evaluation of polarized parton densities in the
  nucleon}}.
\bibinfo{journal}{{\em Eur. Phys. J. C}} \bibinfo{volume}{23}:
  \bibinfo{pages}{479--485}. \bibinfo{doi}{\doi{10.1007/s100520200901}}.
\eprint{hep-ph/0111267}.

\bibtype{Article}%
\bibitem[Leader et al.(2006)]{Leader:2005ci}
\bibinfo{author}{Leader E}, \bibinfo{author}{Sidorov AV} and
  \bibinfo{author}{Stamenov DB} (\bibinfo{year}{2006}).
\bibinfo{title}{{Longitudinal polarized parton densities updated}}.
\bibinfo{journal}{{\em Phys. Rev. D}} \bibinfo{volume}{73}:
  \bibinfo{pages}{034023}. \bibinfo{doi}{\doi{10.1103/PhysRevD.73.034023}}.
\eprint{hep-ph/0512114}.

\bibtype{Article}%
\bibitem[Leader et al.(2007)]{Leader:2006xc}
\bibinfo{author}{Leader E}, \bibinfo{author}{Sidorov AV} and
  \bibinfo{author}{Stamenov DB} (\bibinfo{year}{2007}).
\bibinfo{title}{{Impact of CLAS and COMPASS data on Polarized Parton Densities
  and Higher Twist}}.
\bibinfo{journal}{{\em Phys. Rev. D}} \bibinfo{volume}{75}:
  \bibinfo{pages}{074027}. \bibinfo{doi}{\doi{10.1103/PhysRevD.75.074027}}.
\eprint{hep-ph/0612360}.

\bibtype{Article}%
\bibitem[Leader et al.(2010)]{Leader:2010rb}
\bibinfo{author}{Leader E}, \bibinfo{author}{Sidorov AV} and
  \bibinfo{author}{Stamenov DB} (\bibinfo{year}{2010}).
\bibinfo{title}{{Determination of Polarized PDFs from a QCD Analysis of
  Inclusive and Semi-inclusive Deep Inelastic Scattering Data}}.
\bibinfo{journal}{{\em Phys. Rev. D}} \bibinfo{volume}{82}:
  \bibinfo{pages}{114018}. \bibinfo{doi}{\doi{10.1103/PhysRevD.82.114018}}.
\eprint{1010.0574}.

\bibtype{Article}%
\bibitem[Lin et al.(2018)]{Lin:2017snn}
\bibinfo{author}{Lin HW} and  et al. (\bibinfo{year}{2018}).
\bibinfo{title}{{Parton distributions and lattice QCD calculations: a community
  white paper}}.
\bibinfo{journal}{{\em Prog. Part. Nucl. Phys.}} \bibinfo{volume}{100}:
  \bibinfo{pages}{107--160}. \bibinfo{doi}{\doi{10.1016/j.ppnp.2018.01.007}}.
\eprint{1711.07916}.

\bibtype{Article}%
\bibitem[Lipatov(1974)]{Lipatov:1974qm}
\bibinfo{author}{Lipatov LN} (\bibinfo{year}{1974}).
\bibinfo{title}{{The parton model and perturbation theory}}.
\bibinfo{journal}{{\em Yad. Fiz.}} \bibinfo{volume}{20}:
  \bibinfo{pages}{181--198}.

\bibtype{Article}%
\bibitem[Mitov et al.(2006)]{Mitov:2006ic}
\bibinfo{author}{Mitov A}, \bibinfo{author}{Moch S} and  \bibinfo{author}{Vogt
  A} (\bibinfo{year}{2006}).
\bibinfo{title}{{Next-to-Next-to-Leading Order Evolution of Non-Singlet
  Fragmentation Functions}}.
\bibinfo{journal}{{\em Phys. Lett. B}} \bibinfo{volume}{638}:
  \bibinfo{pages}{61--67}. \bibinfo{doi}{\doi{10.1016/j.physletb.2006.05.005}}.
\eprint{hep-ph/0604053}.

\bibtype{Article}%
\bibitem[Moch and Vogt(2008)]{Moch:2007tx}
\bibinfo{author}{Moch S} and  \bibinfo{author}{Vogt A} (\bibinfo{year}{2008}).
\bibinfo{title}{{On third-order timelike splitting functions and top-mediated
  Higgs decay into hadrons}}.
\bibinfo{journal}{{\em Phys. Lett. B}} \bibinfo{volume}{659}:
  \bibinfo{pages}{290--296}.
  \bibinfo{doi}{\doi{10.1016/j.physletb.2007.10.069}}.
\eprint{0709.3899}.

\bibtype{Article}%
\bibitem[Moch et al.(2004)]{Moch:2004pa}
\bibinfo{author}{Moch S}, \bibinfo{author}{Vermaseren JAM} and
  \bibinfo{author}{Vogt A} (\bibinfo{year}{2004}).
\bibinfo{title}{{The Three loop splitting functions in QCD: The Nonsinglet
  case}}.
\bibinfo{journal}{{\em Nucl. Phys. B}} \bibinfo{volume}{688}:
  \bibinfo{pages}{101--134}.
  \bibinfo{doi}{\doi{10.1016/j.nuclphysb.2004.03.030}}.
\eprint{hep-ph/0403192}.

\bibtype{Article}%
\bibitem[Moch et al.(2014)]{Moch:2014sna}
\bibinfo{author}{Moch S}, \bibinfo{author}{Vermaseren JAM} and
  \bibinfo{author}{Vogt A} (\bibinfo{year}{2014}).
\bibinfo{title}{{The Three-Loop Splitting Functions in QCD: The
  Helicity-Dependent Case}}.
\bibinfo{journal}{{\em Nucl. Phys. B}} \bibinfo{volume}{889}:
  \bibinfo{pages}{351--400}.
  \bibinfo{doi}{\doi{10.1016/j.nuclphysb.2014.10.016}}.
\eprint{1409.5131}.

\bibtype{Article}%
\bibitem[Moch et al.(2015)]{Moch:2015usa}
\bibinfo{author}{Moch S}, \bibinfo{author}{Vermaseren JAM} and
  \bibinfo{author}{Vogt A} (\bibinfo{year}{2015}).
\bibinfo{title}{{On {\ensuremath{\gamma}}5 in higher-order QCD calculations and
  the NNLO evolution of the polarized valence distribution}}.
\bibinfo{journal}{{\em Phys. Lett. B}} \bibinfo{volume}{748}:
  \bibinfo{pages}{432--438}.
  \bibinfo{doi}{\doi{10.1016/j.physletb.2015.07.027}}.
\eprint{1506.04517}.

\bibtype{Article}%
\bibitem[Nocera(2014)]{Nocera:2014rea}
\bibinfo{author}{Nocera ER} (\bibinfo{year}{2014}).
\bibinfo{title}{{Flavor asymmetry of the polarized nucleon sea}}.
\bibinfo{journal}{{\em PoS}} \bibinfo{volume}{DIS2014}: \bibinfo{pages}{204}.
  \bibinfo{doi}{\doi{10.22323/1.203.0204}}.

\bibtype{Article}%
\bibitem[Nocera(2015)]{Nocera:2014uea}
\bibinfo{author}{Nocera ER} (\bibinfo{year}{2015}).
\bibinfo{title}{{Small- and large-$x$ nucleon spin structure from a global QCD
  analysis of polarized Parton Distribution Functions}}.
\bibinfo{journal}{{\em Phys. Lett. B}} \bibinfo{volume}{742}:
  \bibinfo{pages}{117--125}.
  \bibinfo{doi}{\doi{10.1016/j.physletb.2015.01.021}}.
\eprint{1410.7290}.

\bibtype{Article}%
\bibitem[Nocera et al.(2014)]{Nocera:2014gqa}
\bibinfo{author}{Nocera ER}, \bibinfo{author}{Ball RD}, \bibinfo{author}{Forte
  S}, \bibinfo{author}{Ridolfi G} and  \bibinfo{author}{Rojo J}
  (\bibinfo{collaboration}{NNPDF}) (\bibinfo{year}{2014}).
\bibinfo{title}{{A first unbiased global determination of polarized PDFs and
  their uncertainties}}.
\bibinfo{journal}{{\em Nucl. Phys. B}} \bibinfo{volume}{887}:
  \bibinfo{pages}{276--308}.
  \bibinfo{doi}{\doi{10.1016/j.nuclphysb.2014.08.008}}.
\eprint{1406.5539}.

\bibtype{Article}%
\bibitem[Prok et al.(2014)]{CLAS:2014qtg}
\bibinfo{author}{Prok Y} and  et al. (\bibinfo{collaboration}{CLAS})
  (\bibinfo{year}{2014}).
\bibinfo{title}{{Precision measurements of $g_1$ of the proton and the deuteron
  with 6 GeV electrons}}.
\bibinfo{journal}{{\em Phys. Rev. C}} \bibinfo{volume}{90}
  (\bibinfo{number}{2}): \bibinfo{pages}{025212}.
  \bibinfo{doi}{\doi{10.1103/PhysRevC.90.025212}}.
\eprint{1404.6231}.

\bibtype{Article}%
\bibitem[Pumplin et al.(2002)]{Pumplin:2002vw}
\bibinfo{author}{Pumplin J}, \bibinfo{author}{Stump DR},
  \bibinfo{author}{Huston J}, \bibinfo{author}{Lai HL},
  \bibinfo{author}{Nadolsky PM} and  \bibinfo{author}{Tung WK}
  (\bibinfo{year}{2002}).
\bibinfo{title}{{New generation of parton distributions with uncertainties from
  global QCD analysis}}.
\bibinfo{journal}{{\em JHEP}} \bibinfo{volume}{07}: \bibinfo{pages}{012}.
  \bibinfo{doi}{\doi{10.1088/1126-6708/2002/07/012}}.
\eprint{hep-ph/0201195}.

\bibtype{Article}%
\bibitem[Regge(1959)]{Regge:1959mz}
\bibinfo{author}{Regge T} (\bibinfo{year}{1959}).
\bibinfo{title}{{Introduction to complex orbital momenta}}.
\bibinfo{journal}{{\em Nuovo Cim.}} \bibinfo{volume}{14}: \bibinfo{pages}{951}.
  \bibinfo{doi}{\doi{10.1007/BF02728177}}.

\bibtype{Book}%
\bibitem[Roberts(1994)]{Roberts:1990ww}
\bibinfo{author}{Roberts RG} (\bibinfo{year}{1994}), \bibinfo{month}{2}.
\bibinfo{title}{{The Structure of the proton: Deep inelastic scattering}},
  \bibinfo{series}{Cambridge Monographs on Mathematical Physics},
  \bibinfo{publisher}{Cambridge University Press}.
\bibinfo{comment}{ISBN} \bibinfo{isbn}{978-0-521-44944-1, 978-1-139-24244-8}.
\bibinfo{doi}{\doi{10.1017/CBO9780511564062}}.

\bibtype{Article}%
\bibitem[Salam and Rojo(2009)]{Salam:2008qg}
\bibinfo{author}{Salam GP} and  \bibinfo{author}{Rojo J}
  (\bibinfo{year}{2009}).
\bibinfo{title}{{A Higher Order Perturbative Parton Evolution Toolkit
  (HOPPET)}}.
\bibinfo{journal}{{\em Comput. Phys. Commun.}} \bibinfo{volume}{180}:
  \bibinfo{pages}{120--156}. \bibinfo{doi}{\doi{10.1016/j.cpc.2008.08.010}}.
\eprint{0804.3755}.

\bibtype{Article}%
\bibitem[Sato et al.(2016)]{Sato:2016tuz}
\bibinfo{author}{Sato N}, \bibinfo{author}{Melnitchouk W},
  \bibinfo{author}{Kuhn SE}, \bibinfo{author}{Ethier JJ} and
  \bibinfo{author}{Accardi A} (\bibinfo{collaboration}{Jefferson Lab Angular
  Momentum}) (\bibinfo{year}{2016}).
\bibinfo{title}{{Iterative Monte Carlo analysis of spin-dependent parton
  distributions}}.
\bibinfo{journal}{{\em Phys. Rev. D}} \bibinfo{volume}{93}
  (\bibinfo{number}{7}): \bibinfo{pages}{074005}.
  \bibinfo{doi}{\doi{10.1103/PhysRevD.93.074005}}.
\eprint{1601.07782}.

\bibtype{Misc}%
\bibitem[Schwan et al.(2025)]{christopher_schwan_2025_15635174}
\bibinfo{author}{Schwan C}, \bibinfo{author}{Rabemananjara TR},
  \bibinfo{author}{Candido A}, \bibinfo{author}{Hekhorn F},
  \bibinfo{author}{Sharma T}, \bibinfo{author}{Carrazza S},
  \bibinfo{author}{Barontini A}, \bibinfo{author}{Wissmann J} and
  \bibinfo{author}{Cruz-Martinez JM} (\bibinfo{year}{2025}),
  \bibinfo{month}{Jun.}
\bibinfo{title}{Nnpdf/pineappl: v1.0.0}.
\bibinfo{doi}{\doi{10.5281/zenodo.15635174}}.
\bibinfo{url}{\url{https://doi.org/10.5281/zenodo.15635174}}.

\bibtype{Mastersthesis}%
\bibitem[Simolo(2006)]{Simolo:2006iw}
\bibinfo{author}{Simolo C} (\bibinfo{year}{2006}).
\bibinfo{title}{{Extraction of characteristic constants in QCD with
  perturbative and nonperturbative methods}}.
\bibinfo{type}{Other thesis}.
\eprint{0807.1501}, \bibinfo{type}{Other thesis}.

\bibtype{Article}%
\bibitem[Soffer(2003)]{Soffer:2003qj}
\bibinfo{author}{Soffer J} (\bibinfo{year}{2003}).
\bibinfo{title}{{Positivity constraints on initial spin observables in
  inclusive reactions}}.
\bibinfo{journal}{{\em Phys. Rev. Lett.}} \bibinfo{volume}{91}:
  \bibinfo{pages}{092005}. \bibinfo{doi}{\doi{10.1103/PhysRevLett.91.092005}}.
\eprint{hep-ph/0305222}.

\bibtype{Article}%
\bibitem[Stratmann and Vogelsang(2001)]{Stratmann:2001pb}
\bibinfo{author}{Stratmann M} and  \bibinfo{author}{Vogelsang W}
  (\bibinfo{year}{2001}).
\bibinfo{title}{{Towards a global analysis of polarized parton distributions}}.
\bibinfo{journal}{{\em Phys. Rev. D}} \bibinfo{volume}{64}:
  \bibinfo{pages}{114007}. \bibinfo{doi}{\doi{10.1103/PhysRevD.64.114007}}.
\eprint{hep-ph/0107064}.

\bibtype{Article}%
\bibitem[Stuart(2010)]{Stuart_2010}
\bibinfo{author}{Stuart AM} (\bibinfo{year}{2010}).
\bibinfo{title}{Inverse problems: A bayesian perspective}.
\bibinfo{journal}{{\em Acta Numerica}} \bibinfo{volume}{19}:
  \bibinfo{pages}{451–559}. \bibinfo{doi}{\doi{10.1017/S0962492910000061}}.

\bibtype{Article}%
\bibitem['t~Hooft(1973)]{tHooft:1973mfk}
\bibinfo{author}{'t~Hooft G} (\bibinfo{year}{1973}).
\bibinfo{title}{{Dimensional regularization and the renormalization group}}.
\bibinfo{journal}{{\em Nucl. Phys. B}} \bibinfo{volume}{61}:
  \bibinfo{pages}{455--468}. \bibinfo{doi}{\doi{10.1016/0550-3213(73)90376-3}}.

\bibtype{Article}%
\bibitem['t~Hooft and Veltman(1972)]{tHooft:1972tcz}
\bibinfo{author}{'t~Hooft G} and  \bibinfo{author}{Veltman MJG}
  (\bibinfo{year}{1972}).
\bibinfo{title}{{Regularization and Renormalization of Gauge Fields}}.
\bibinfo{journal}{{\em Nucl. Phys. B}} \bibinfo{volume}{44}:
  \bibinfo{pages}{189--213}. \bibinfo{doi}{\doi{10.1016/0550-3213(72)90279-9}}.

\bibtype{Article}%
\bibitem[Taghavi-Shahri et al.(2016)]{Taghavi-Shahri:2016idw}
\bibinfo{author}{Taghavi-Shahri F}, \bibinfo{author}{Khanpour H},
  \bibinfo{author}{Atashbar~Tehrani S} and  \bibinfo{author}{Alizadeh~Yazdi Z}
  (\bibinfo{year}{2016}).
\bibinfo{title}{{Next-to-next-to-leading order QCD analysis of spin-dependent
  parton distribution functions and their uncertainties: Jacobi polynomials
  approach}}.
\bibinfo{journal}{{\em Phys. Rev. D}} \bibinfo{volume}{93}
  (\bibinfo{number}{11}): \bibinfo{pages}{114024}.
  \bibinfo{doi}{\doi{10.1103/PhysRevD.93.114024}}.
\eprint{1603.03157}.

\bibtype{Article}%
\bibitem[Uhlenbeck and Goudsmit(1925)]{Uhlenbeck:1925pqz}
\bibinfo{author}{Uhlenbeck G} and  \bibinfo{author}{Goudsmit S}
  (\bibinfo{year}{1925}).
\bibinfo{title}{{Ersetzung der Hypothese vom unmechanischen Zwang durch eine
  Forderung bez{\"u}glich des inneren Verhaltens jedes einzelnen Elektrons}}.
\bibinfo{journal}{{\em Naturwiss.}} \bibinfo{volume}{13}
  (\bibinfo{number}{47}): \bibinfo{pages}{953--954}.
  \bibinfo{doi}{\doi{10.1007/bf01558878}}.

\bibtype{Article}%
\bibitem[van Neerven and Zijlstra(1991)]{vanNeerven:1991nn}
\bibinfo{author}{van Neerven WL} and  \bibinfo{author}{Zijlstra EB}
  (\bibinfo{year}{1991}).
\bibinfo{title}{{Order alpha-s**2 contributions to the deep inelastic Wilson
  coefficient}}.
\bibinfo{journal}{{\em Phys. Lett. B}} \bibinfo{volume}{272}:
  \bibinfo{pages}{127--133}. \bibinfo{doi}{\doi{10.1016/0370-2693(91)91024-P}}.

\bibtype{Article}%
\bibitem[Vermaseren et al.(2005)]{Vermaseren:2005qc}
\bibinfo{author}{Vermaseren JAM}, \bibinfo{author}{Vogt A} and
  \bibinfo{author}{Moch S} (\bibinfo{year}{2005}).
\bibinfo{title}{{The Third-order QCD corrections to deep-inelastic scattering
  by photon exchange}}.
\bibinfo{journal}{{\em Nucl. Phys. B}} \bibinfo{volume}{724}:
  \bibinfo{pages}{3--182}.
  \bibinfo{doi}{\doi{10.1016/j.nuclphysb.2005.06.020}}.
\eprint{hep-ph/0504242}.

\bibtype{Article}%
\bibitem[Vogt(2005)]{Vogt:2004ns}
\bibinfo{author}{Vogt A} (\bibinfo{year}{2005}).
\bibinfo{title}{{Efficient evolution of unpolarized and polarized parton
  distributions with QCD-PEGASUS}}.
\bibinfo{journal}{{\em Comput. Phys. Commun.}} \bibinfo{volume}{170}:
  \bibinfo{pages}{65--92}. \bibinfo{doi}{\doi{10.1016/j.cpc.2005.03.103}}.
\eprint{hep-ph/0408244}.

\bibtype{Article}%
\bibitem[Vogt et al.(2004)]{Vogt:2004mw}
\bibinfo{author}{Vogt A}, \bibinfo{author}{Moch S} and
  \bibinfo{author}{Vermaseren JAM} (\bibinfo{year}{2004}).
\bibinfo{title}{{The Three-loop splitting functions in QCD: The Singlet case}}.
\bibinfo{journal}{{\em Nucl. Phys. B}} \bibinfo{volume}{691}:
  \bibinfo{pages}{129--181}.
  \bibinfo{doi}{\doi{10.1016/j.nuclphysb.2004.04.024}}.
\eprint{hep-ph/0404111}.

\bibtype{Article}%
\bibitem[Weinberg(1973)]{Weinberg:1973un}
\bibinfo{author}{Weinberg S} (\bibinfo{year}{1973}).
\bibinfo{title}{{Nonabelian Gauge Theories of the Strong Interactions}}.
\bibinfo{journal}{{\em Phys. Rev. Lett.}} \bibinfo{volume}{31}:
  \bibinfo{pages}{494--497}. \bibinfo{doi}{\doi{10.1103/PhysRevLett.31.494}}.

\bibtype{Article}%
\bibitem[Zheng et al.(2004)]{JeffersonLabHallA:2004tea}
\bibinfo{author}{Zheng X} and  et al. (\bibinfo{collaboration}{Jefferson Lab
  Hall A}) (\bibinfo{year}{2004}).
\bibinfo{title}{{Precision measurement of the neutron spin asymmetries and
  spin-dependent structure functions in the valence quark region}}.
\bibinfo{journal}{{\em Phys. Rev. C}} \bibinfo{volume}{70}:
  \bibinfo{pages}{065207}. \bibinfo{doi}{\doi{10.1103/PhysRevC.70.065207}}.
\eprint{nucl-ex/0405006}.

\bibtype{Article}%
\bibitem[Zhou et al.(2021)]{Zhou:2021llj}
\bibinfo{author}{Zhou Y}, \bibinfo{author}{Cocuzza C},
  \bibinfo{author}{Delcarro F}, \bibinfo{author}{Melnitchouk W},
  \bibinfo{author}{Metz A} and  \bibinfo{author}{Sato N}
  (\bibinfo{collaboration}{Jefferson Lab Angular Momentum (JAM)})
  (\bibinfo{year}{2021}).
\bibinfo{title}{{Revisiting quark and gluon polarization in the proton at the
  EIC}}.
\bibinfo{journal}{{\em Phys. Rev. D}} \bibinfo{volume}{104}
  (\bibinfo{number}{3}): \bibinfo{pages}{034028}.
  \bibinfo{doi}{\doi{10.1103/PhysRevD.104.034028}}.
\eprint{2105.04434}.

\bibtype{Article}%
\bibitem[Zhou et al.(2022)]{Zhou:2022wzm}
\bibinfo{author}{Zhou Y}, \bibinfo{author}{Sato N} and
  \bibinfo{author}{Melnitchouk W} (\bibinfo{collaboration}{Jefferson Lab
  Angular Momentum (JAM)}) (\bibinfo{year}{2022}).
\bibinfo{title}{{How well do we know the gluon polarization in the proton?}}
\bibinfo{journal}{{\em Phys. Rev. D}} \bibinfo{volume}{105}
  (\bibinfo{number}{7}): \bibinfo{pages}{074022}.
  \bibinfo{doi}{\doi{10.1103/PhysRevD.105.074022}}.
\eprint{2201.02075}.

\bibtype{Article}%
\bibitem[Zijlstra and van Neerven(1991)]{Zijlstra:1991qc}
\bibinfo{author}{Zijlstra EB} and  \bibinfo{author}{van Neerven WL}
  (\bibinfo{year}{1991}).
\bibinfo{title}{{Contribution of the second order gluonic Wilson coefficient to
  the deep inelastic structure function}}.
\bibinfo{journal}{{\em Phys. Lett. B}} \bibinfo{volume}{273}:
  \bibinfo{pages}{476--482}. \bibinfo{doi}{\doi{10.1016/0370-2693(91)90301-6}}.

\bibtype{Article}%
\bibitem[Zijlstra and van Neerven(1992{\natexlab{a}})]{Zijlstra:1992kj}
\bibinfo{author}{Zijlstra EB} and  \bibinfo{author}{van Neerven WL}
  (\bibinfo{year}{1992}{\natexlab{a}}).
\bibinfo{title}{{Order alpha-s**2 correction to the structure function F3 (x,
  Q**2) in deep inelastic neutrino - hadron scattering}}.
\bibinfo{journal}{{\em Phys. Lett. B}} \bibinfo{volume}{297}:
  \bibinfo{pages}{377--384}. \bibinfo{doi}{\doi{10.1016/0370-2693(92)91277-G}}.

\bibtype{Article}%
\bibitem[Zijlstra and van Neerven(1992{\natexlab{b}})]{Zijlstra:1992qd}
\bibinfo{author}{Zijlstra EB} and  \bibinfo{author}{van Neerven WL}
  (\bibinfo{year}{1992}{\natexlab{b}}).
\bibinfo{title}{{Order alpha-s**2 QCD corrections to the deep inelastic proton
  structure functions F2 and F(L)}}.
\bibinfo{journal}{{\em Nucl. Phys. B}} \bibinfo{volume}{383}:
  \bibinfo{pages}{525--574}. \bibinfo{doi}{\doi{10.1016/0550-3213(92)90087-R}}.

\bibtype{Article}%
\bibitem[Zijlstra and van Neerven(1994)]{Zijlstra:1993sh}
\bibinfo{author}{Zijlstra EB} and  \bibinfo{author}{van Neerven WL}
  (\bibinfo{year}{1994}).
\bibinfo{title}{{Order-$\alpha_s^2$ corrections to the polarized structure
  function $g_1 (x,Q^2)$}}.
\bibinfo{journal}{{\em Nucl. Phys. B}} \bibinfo{volume}{417}:
  \bibinfo{pages}{61--100}. \bibinfo{doi}{\doi{10.1016/0550-3213(94)90538-X}}.
\bibinfo{note}{[Erratum: Nucl.Phys.B 426, 245 (1994), Erratum: Nucl.Phys.B 773,
  105--106 (2007), Erratum: Nucl.Phys.B 501, 599--599 (1997)]}.

\bibtype{Article}%
\bibitem[Zweig(1964{\natexlab{a}})]{Zweig:1964ruk}
\bibinfo{author}{Zweig G} (\bibinfo{year}{1964}{\natexlab{a}}),
  \bibinfo{month}{1}.
\bibinfo{title}{{An SU(3) model for strong interaction symmetry and its
  breaking. Version 1}} \bibinfo{doi}{\doi{10.17181/CERN-TH-401}}.

\bibtype{inbook}%
\bibitem[Zweig(1964{\natexlab{b}})]{Zweig:1964jf}
\bibinfo{author}{Zweig G} (\bibinfo{year}{1964}{\natexlab{b}}),
  \bibinfo{month}{2}.
\bibinfo{title}{{An SU(3) model for strong interaction symmetry and its
  breaking. Version 2}}.
 \bibinfo{pages}{22--101}.
\bibinfo{doi}{\doi{10.17181/CERN-TH-412}}.

\end{thebibliography*}

\end{document}